%% file: submit.tex
\documentclass[journal=jacsat,manuscript=article]{achemso}

\usepackage[version=3]{mhchem} 
\usepackage{graphicx}%
\usepackage{dcolumn}%
\usepackage{bm}%
\usepackage{multirow}
\usepackage{xcolor}
\usepackage{soul}
\usepackage{xr}
\usepackage{import}
\usepackage{chemfig}
\usepackage{tikz}
\usetikzlibrary{decorations}

\newcommand{\set}[1]{{\it Set#1}}
\newcommand{\kl}{$D_{\rm KL}$ }
\newcommand{\js}{$D_{\rm JS}$ }

\makeatletter
\newcommand*{\addFileDependency}[1]{%
  \typeout{(#1)}
  \@addtofilelist{#1}
  \IfFileExists{#1}{}{\typeout{No file #1.}}
}
\makeatother

\pgfdeclaredecoration{ddbond}{initial}{
  \state{initial}[width=4pt]{
    \pgfpathlineto{\pgfpoint{4pt}{0pt}}
    \pgfpathmoveto{\pgfpoint{2pt}{2pt}}
    \pgfpathlineto{\pgfpoint{4pt}{2pt}}
    \pgfpathmoveto{\pgfpoint{4pt}{0pt}}
  }
  \state{final}{
    \pgfpathlineto{\pgfpointdecoratedpathlast}
  }
}
\tikzset{lddbond/.style={decorate, decoration=ddbond}}

\setchemfig{atom sep=2em}

\title[Artificial DB]{Augmenting chemical databases for atomistic
  machine learning by sampling conformational space}

\author{Luis Itza Vazquez-Salazar}%
\email{l.i.vazquez-salazar@thphys.uni-heidelberg.de}
\affiliation{Department of Chemistry, University of Basel, Basel,
  Switzerland}
\altaffiliation{Current address: Institute for Theoretical Physics, Heidelberg University, Heidelberg, Germany}
\author{Markus Meuwly} \email{m.meuwly@unibas.ch}
\affiliation{Department of Chemistry, University of Basel, Basel,
  Switzerland}

\date{\today}

\begin{document}

\begin{abstract}
Machine learning (ML) has become a standard tool for the exploration
of chemical space. Much of the performance of such models depends on the chosen database for a given task. Here, this aspect is investigated for "chemical tasks" including the prediction of hybridization, oxidation, substituent effects, and aromaticity, starting from an initial "restricted" database (iRD). Choosing molecules for augmenting this iRD, including increasing numbers of conformations generated at different temperatures, and retraining the models can improve predictions of the models on the selected "tasks". Addition
of a small percentage of conformers (1\% ) obtained at 300 K improves the performance in almost all cases. On the other hand, and in
line with previous studies, redundancy and highly deformed structures in the augmentation set compromise prediction quality. Energy and bond
distributions were evaluated by means of Kullback-Leibler (\kl) and
Jensen-Shannon (\js) divergence and Wasserstein distance
($W_{1}$). The findings of this work
provide a baseline for the rational augmentation of chemical
databases or the creation of synthetic databases.
\end{abstract}

\section{Introduction}
Chemical space (CS) as the set of all possible molecules or
materials\cite{kirkpatrick2004chemical,von2013first,von2020exploring,huang:2021,coley2021defining,medina2022chemical}
is extraordinarily large. It has been theorized that the total number
of possible substances\cite{restrepo2022chemical,gorse2006diversity}
is about $10^{200}$. This large size makes the exploration of CS a big
challenge but also an important opportunity for scientific and
technological advancement. In this regard, computational simulations
have been consolidated as a powerful tool for this task. With the rise
of machine learning (ML) methods, obtaining high-quality predictions
of chemical properties at a low computational cost has become easier
than ever. Consequently, the exploration of CS has progressed in the
direction of computational compound
design\cite{sandonas2023freedom,fallani2023enabling}.\\

\noindent
Nevertheless, for an ML method to perform adequately on a range of -
potentially chemically diverse systems, it requires a sufficiently
broad corpus of data that adequately covers the CS to be probed and
described. In chemistry, generating such reference data incurs a high
computational cost with associated environmental costs
\cite{MM.rev:2023}, besides being limited by the size of the molecular
systems of interest. To address the problems associated with the generation of reference data, it has been proposed\cite{kulichenko2024data} to incorporate data from atoms-in-molecules fragments\cite{huang2020quantum}  (amons) or external chemical databases, which help to explore CS. Another viable alternative is using information from conformational space represented by a
potential energy surface (PES). It has
been proposed that the chemical information contained in a chemical
bond and, consequently, in the conformational space provides valuable
information that can help to study CS \cite{shaik2013one}. In
particular, for ML methods, it was previously found that the
exploration of CS can be improved by adding adequate information from
the configurational space represented by the
PES\cite{vazquezsalazar2021}.\\

\noindent
Although adding samples from conformational space is a convenient way
to improve the ability of a model to explore CS, there is no clear
guidance on {\it how} this should be done. Currently, this addition of
samples is made by obtaining hundreds or thousands of conformers for a
few molecules (e.g. QM7-X \cite{hoja2021qm7}) or for a large number of
molecules (e.g. ANI-1\cite{smith2017ani}). However, such an approach
generates data redundancies and the prediction capability may deteriorates
as a consequence\cite{vazquezsalazar2021}. Besides that, such an
approach is only feasible if sufficient computational resources are
available. Furthermore, data redundancy in a data set leads to the
well-known problem of ``dataset
imbalance''\cite{quinonero2008dataset}. In cheminformatics, efforts
have been made to address this problem \cite{banerjee2018prediction,
  hemmerich2020cover, korkmaz2020deep}, though mostly in the context
of classification tasks. Unfortunately, for atomistic machine learning
and to the best of our knowledge, there is only one example of studies
that addresses the question of chemical and conformational diversity for
ML-based models.\cite{shenoy2023role}\\

\noindent
The present work has a twofold aim. Firstly, to understand from a
chemical perspective, how a chemical database can be improved by
adding samples from conformational space because it has been recently
found that the addition of conformers leads to improvement on the
prediction of chemical properties\cite{hamakawa2025understanding}. For
this, the influence of simulation temperature and number of samples
will be evaluated. In addition, the question of ``dataset imbalance''
in a chemical dataset will be considered by explicitly biasing the
initial dataset and then adding conformers to improve the
initially biased datasets in view of a particular chemical task. The starting datasets were created to
explore different chemical aspects and, therefore, were generated with
specific and separate biases. As a difference to earlier efforts\cite{shenoy2023role}, the focus here is on specific
chemical aspects of the databases, while chemical structure diversity was not extensively evaluated.\\

\noindent
The second goal of the present work is to determine and quantify
whether extending conformational space covered during sampling can
compensate for a lack of exploration of chemical space in a reference
database used for prediction. Therefore, "new chemistry" was added to
the restricted databases by sampling the conformational space of one
or many molecules that contained features of interest in the target
database. In the following, this will be referred to as
''Structure-based addition".\\

\noindent
This article is structured as follows. First, the
construction of the artificial databases, data augmentation
strategies, and ML method set-up are described in the methods
section. Next, the results of the different aspects of the data
augmentation are discussed. Finally, some conclusions from the
different strategies evaluated are drawn.\\

\section{Methods}

\subsection{Machine Learning}
The machine learning model employed was
PhysNet\cite{unke2019physnet}. This model belongs to the general class
of graph neural networks\cite{reiser2022graph,corso2024graph},
examples of such NNs include, but are not limited to,
SchNet\cite{schnet:2018}, PaiNN\cite{schuett2021painn},
Nequip\cite{batzner2022nequip} or MACE\cite{batatia2022mace}, to name
a few. All these approaches have proven their outstanding performance
in predicting quantum chemical properties. Therefore, in this work,
PhysNet is used to represent those NNs. In this work, the modified
version of PhysNet\cite{vazquez2022uncertainty} to allow uncertainty
quantification (UQ) based on Deep Evidential Regression
(DER)\cite{soleimany2021evidential} was used. In such an approach it is
assumed that the energies are normally distributed
$P(E)=\mathcal{N}(\mu,\sigma^{2})$. The corresponding prior
distribution is a Normal-Inverse Gamma (NIG) distribution, described
by four parameters ($\gamma$, $\nu$, $\alpha$,
$\beta$).\cite{amini2020} The loss function to be optimized is a
dual-objective loss $\mathcal{L}(x)$ with two terms: the first term
maximizes model fitting, and the second penalizes incorrect
predictions:
\begin{equation}
    \mathcal{L}(x) = \mathcal{L}^{\rm NLL}(x)
    +\lambda(\mathcal{L}^{\rm R}(x) - \varepsilon)
    \label{eq:loss_funct}
\end{equation}

\noindent
The first term of eq. \ref{eq:loss_funct} is the negative log-likelihood
(NLL) of the model evidence that can be represented as a Student-$t$
distribution
\begin{equation}
    \mathcal{L}^{\rm NLL}(x) = \frac{1}{2} \log
    \left(\frac{\pi}{\nu}\right) - \alpha \log(\Omega) + (\alpha +
    \frac{1}{2}) \log((x-\gamma)^{2}\nu + \Omega) + \log
    \left(\frac{\Gamma(\alpha)}{\Gamma(\alpha+\frac{1}{2})}\right)
    \label{eq:loss_nll}
\end{equation}
where $\Omega = 2\beta(1+\nu)$ and $x$ is the value predicted by the
neural network.\cite{amini2020} The second term in Equation
\ref{eq:loss_funct}, $\mathcal{L}^{\rm R}(x)$, corresponds to a
regularizer that minimizes the evidence for incorrect predictions
(Equation \ref{eq:loss_ref}).
\begin{equation}
    \mathcal{L}^{\rm R}(x) = |x-\gamma|\cdot(2\nu + \alpha)
    \label{eq:loss_ref}
\end{equation}

\noindent
For all trainings in the present work, the hyperparameter $\lambda$ in
Equation \ref{eq:loss_funct}, governing the neural network's
confidence, was set to 0.2. Unless otherwise specified, other
hyperparameters (number of modules, number of radial basis function,
dimensionality of feature space, and others) remained unchanged from
those used previously.\cite{unke2019physnet,vazquez2022uncertainty} For training the
NNs, a standard 8:1:1 split for training, validation, and test sets
was employed. The training procedure was run over 1000 epochs with a
batch size of 32 using the ADAM optimizer\cite{kingma2014adam}. A
validation step for the model was done every five epochs. Three models
with different starting seeds (28, 42, and 64) were obtained for each
augmented database. Model performance for the restricted databases
(i.e. before adding new points) was assessed on the test set; see
Table \ref{sitab:init_perf}. After adding new data, the constructed
models were re-evaluated on the target databases as outlined in Table
\ref{tab:compo}.\\

\subsection{Databases, Data Augmentation and Tasks}
\begin{figure}
    \centering
    \includegraphics[width=0.8\textwidth, height=0.6\textheight]{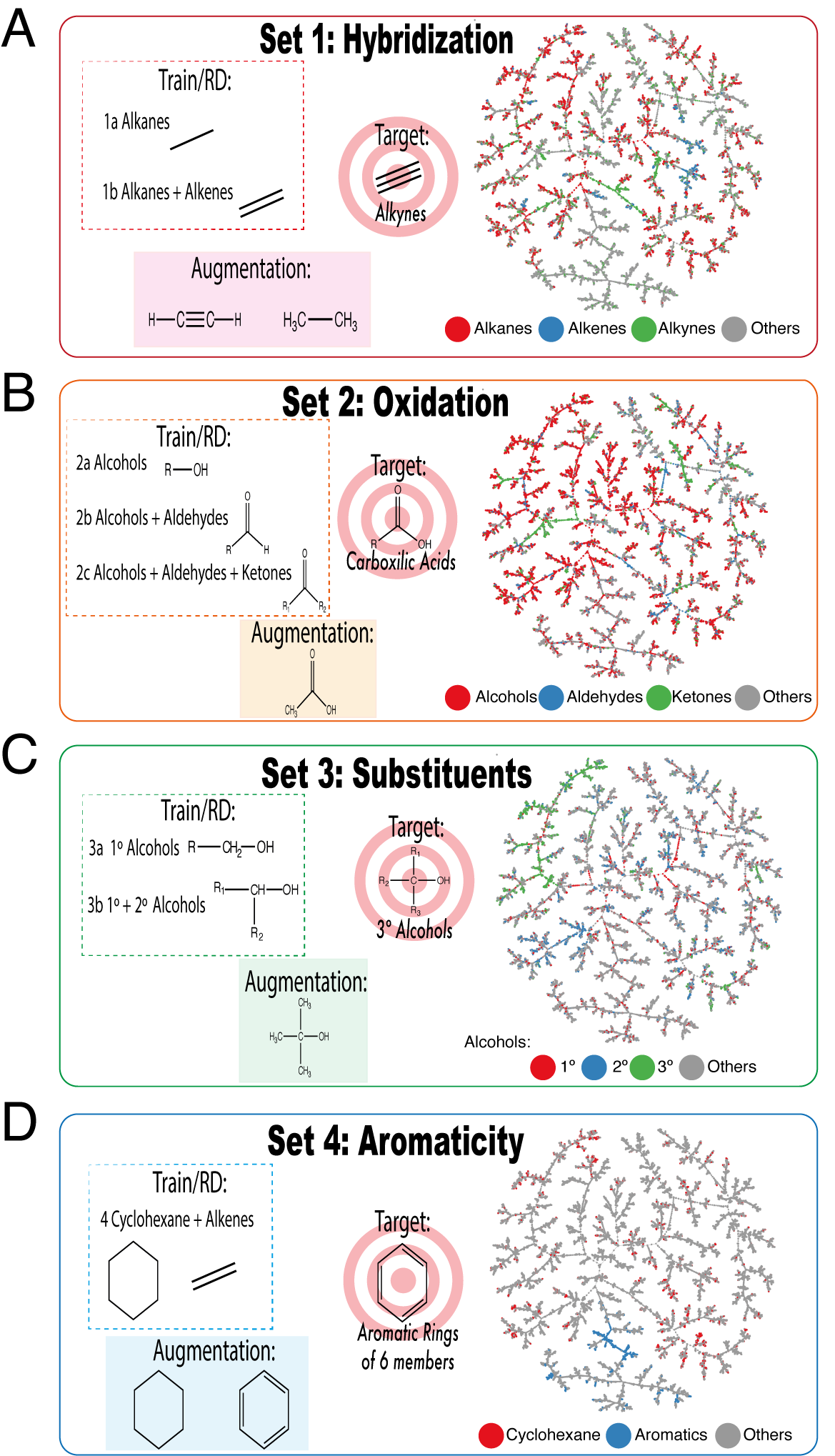}
    \caption{\textbf{Restricted Databases.} Summary of the RDs used in
      this work. Each panel shows the chemical structures of the RDs
      used for training, together with the target structures and the
      molecules used for data augmentation. On the right side of each
      panel is the TMAP representation of the QM9 databases. The
      molecules with moieties of interest are highlighted if the
      sample does not present the fragment of interest is not coloured
      (grey). Panel A shows the molecules in the first set constituted
      by different hybridization of the C-C bond. Panel B shows
      different oxidation states of organic molecules; it is important
      to mention that QM9 does not have recognizable carboxylic
      acids. Panel C shows alcohol molecules with different numbers of
      substituents. Finally, panel D shows molecules with cyclohexane
      and aromatic rings with six atoms.}
    \label{fig:tmap_bias}
\end{figure}

Four databases covering different chemical aspects, henceforth
chemically restricted or restricted databases (RD), were constructed
to study the impact of the augmentation of RDs with conformers to
later evaluate the generalizability of the model on predicting
structures outside the initial training dataset. The chemical target
properties considered were hybridization, oxidation, chirality, and
aromaticity, see Table \ref{tab:compo} and Figure \ref{fig:tmap_bias}
for a summary.\\

\noindent
Construction of the RDs started by extracting molecules from the QM9
database\cite{ramakrishnan2014quantum}, comprising solely molecules
composed of carbon, nitrogen, oxygen, and fluorine. Each molecule in
QM9 is limited to a maximum of nine heavy atoms. To ensure data
quality, molecules failing the geometry consistency check adopted
within QM9\cite{ramakrishnan2014quantum} were excluded from the
dataset. This yielded a ``curated'' version with 130'219 molecules,
down from the initial 130,831. The parent database was filtered to
create the restricted database by using \textit{FragmentMatcher}
within the RDKit software package.\cite{landrum2013rdkit} The process
involved considering the SMILES representations of molecules in
curated QM9 for selection, alongside the generation of SMARTS patterns
to identify functional groups of interest, with additional SMARTS
patterns to exclude certain groups. Using the same strategy, the
target datasets were constructed.\\

\noindent
{\it Set1} was created to understand changes in {\bf carbon atom
  hybridization} (Figure \ref{fig:tmap_bias}A). It consists of two
subsets: {\it Set1a} containing only molecules with single
\chemfig{C-C} bonds (sp$^{3}$), excluding double (\chemfig{C = C},
\chemfig[atom sep=2em]{C-[,,,,lddbond={+}]C}, \chemfig{C=N},
\chemfig{C=O}, \chemfig{N=N}) and triple (\chemfig{C~C},
\chemfig{C~N}) bonds. {\it Set1b} included molecules with
\chemfig{C=C} bonds (sp$^2$), but excluding triple bonds. The target
was to predict \chemfig{C~C} bonds (sp$^{1}$). In this case, ethane
and acetylene were selected for the structure-based augmentation
strategy because these molecules are considered extreme examples of
\chemfig{C-C} bonding (Figure \ref{fig:tmap_bias}A and
\ref{sifig:bonds}).\\

\noindent 
{\it Set2} examined changes in the {\bf oxidation state} of organic
molecules as quantified by the loss of electron density around the
C-atom attached to an oxygen\cite{Klein} (Figure
\ref{fig:tmap_bias}B). For this RD, the task was to infer the energy
of molecules with an oxidation state of $+2$ (\chemfig{ROHC=O},
carboxylic acids) from a database that contains compounds with
oxidation states of $-2$ (\chemfig{R-OH}, alcohols) or 0
(\chemfig{R-CH=O}, aldehydes or \chemfig{R_{1}-R_{2}C=O},
ketones). Following the classification based on oxidation states, {\it
  Set2} was split into three subsets: {\it Set2a} contains only
alcohols, {\it Set2b} contains {\it Set2a} and aldehydes, and {\it
  Set2c} is based on {\it Set2b} and ketones. It must be mentioned
that for the QM9 databases, no molecules with the SMARTS fragment of
carboxylic acid ([CX3](=O)[OX2H1]) were detected by RDKit. Therefore,
compounds with carboxylic acids (target database) were obtained from
the PC9 database\cite{glavatskikh2019dataset} by filtering samples
featuring carboxylic acids (\chemfig{ROHC=O}) in that database. The
resulting structures were optimized at the level of theory used for
QM9 (B3LYP/6-311G(2df,p)) using Gaussian16\cite{gaussian16}. It was
checked that all molecules correspond to a stationary point by
assuring the absence of imaginary frequencies. For {\it Set2}, the
structure-based augmentation was done using formic acid because it
represents the minimum example of a carboxylic acid (Figure
\ref{fig:tmap_bias}B).\\

\begin{table}
\caption{Composition of the initial restricted datasets used in this
  work. The first column identifies the "chemical task" to
  be inferred by the Neural Network model. The size of the subset
  column refers to the total number of molecules used for training,
  validation and testing.}
\begin{tabular}{c|c|c|c}
Set/Task                & Composition of Subset          & Target Molecules                   & Size of subset \\ \hline
\multirow{2}{*}{1 Hybridization} & Alkanes (\set{1a})                      & \multirow{2}{*}{Alkynes}           & \multirow{2}{*}{31250}                   \\
                   & Alkanes + Alkenes (\set{1b})              &                                    &                                          \\ \hline
\multirow{3}{*}{2 Oxidation} & Alcohols (\set{2a})                       & \multirow{3}{*}{Carboxilic Acids}  & \multirow{3}{*}{31250}                   \\
                   & Alcohols + Aldehydes (\set{2b})          &                                    &                                          \\
                   & Alcohols + Aldehydes + Ketones (\set{2c}) &                                    &                                          \\ \hline
\multirow{2}{*}{3 Chirality} & Primary Alcohols  (\set{3a})             & \multirow{2}{*}{Tertiary Alcohols} & 10816                                    \\
                   & Secondary Alcohols (\set{3b})            &                                    & 25695                                    \\ \hline
4  Aromaticity  & Alkenes + Cyclohexane (\set{4})         & Aromatic rings    & 15673  \\
 & & of  6 atoms & \\ 
\end{tabular}
\label{tab:compo}
\end{table}

\noindent
{\it Set3} was biased towards exploring the impact of {\bf
  substituents} on the carbon atom with an attached -OH group, see
Figure \ref{fig:tmap_bias}C). Specifically, the model's ability to
infer {\bf chirality} from molecules lacking this property was of
interest. Alcohols were selected for constructing the RD as they can
be differentiated based on the number of alkyl groups attached to the
carbon in the $\alpha$-position. The set was divided into two subsets:
{\it Set3a} consisted of primary alcohols (\chemfig{RH_2C-OH}), and
{\it Set3b} consisted of {\it Set3a} complemented by secondary
(\chemfig{R_2HC-OH}) alcohols. The target compounds to be predicted
were tertiary alcohols (\chemfig{R_3C-OH}). In this case,
conformations derived from \textit{tert}-butanol, the minimum example
of a tertiary alcohol (Figure \ref{fig:tmap_bias}C), were used for the
structure-based addition.\\

\noindent
{\it Set4} was geared towards recognizing the concept of
\textbf{aromaticity} in chemistry. For this purpose, the RD
exclusively consisted of molecules containing cyclohexane and alkenes,
see Figure \ref{fig:tmap_bias}D. The alkenes from \textit{Set1} were
reused and complemented by compounds in QM9 that contain a cyclohexane
ring. In this case, the target dataset comprised compounds with an
aromatic ring of six members. As for {\it Set1}, two molecules were
used for augmentation based on structure: cyclohexane and benzene
represent extreme cases of double bonds in a six-atom carbon ring
(Figure \ref{fig:tmap_bias}D).\\

\subsection{Sample Generation}
For the purpose of this work, samples from the conformational space of
one or two representative molecules covering the target functional
property were generated, see Figure \ref{fig:tmap_bias}. Normal Mode
Sampling (NMS) was used to generate conformational samples for
augmentation of the iRBs. For NMS, the vibrational normal mode vectors
$\boldsymbol{Q} = {\boldsymbol{q}_i}$ for mode $i$ were obtained from
a normal mode analysis of a molecule in its equilibrium conformation,
$\boldsymbol{x_\mathrm{eq}}$. New conformations were generated by
displacing atom coordinates away from $\boldsymbol{x_\mathrm{eq}}$ by
randomly scaled normal mode coordinates $i = [1..N_{f}]$ by a factor
\begin{equation}
    R_{i} = \pm \sqrt{\dfrac{3c_{i}N_{a}k_{b}T}{K_{i}}}
\label{eq:nms}
\end{equation}
In equation \ref{eq:nms}, $N_{a}$ is the number of atoms, $k_{b}$ is
the Boltzmann constant, $K_{i}$ are the force constants obtained from
the normal model analysis, and $c_{i}$ are pseudo-random numbers in
the range of [0,1], and $T$ is the temperature in K. The sign in
expression \ref{eq:nms} is randomly defined by a Bernoulli
distribution with $P=0.5$. \\

\noindent
In this work, two aspects of ``structure-based addition'' were
evaluated. First, the {\it effect of temperature} was evaluated by
generating samples with NMS at different temperatures; $T \in [300,
  500, 1000, 2000]$ K. In each case, 1000 samples were generated and
added to the initial RDs. The second aspect studied was the {\it
  effect of the number of added samples}. For this, different numbers
of conformers of the selected molecules were generated by NMS at 300
K. The number of added samples was determined as a percentage of the
total number of molecules in the initial databases, see Table
\ref{sitab:number_of_samples}. These percentages were 1, 5, 10, and 25
\%. Consistent with QM9, for all NMS-generated structures single-point
energy calculations at the B3LYP/6-311G(2df,p) level were carried out
using the Gaussian16 program.\cite{gaussian16}\\

\subsection{Distributional Analysis}
Following the methodology described
previously\cite{vazquezsalazar2021}, the structural properties of the
RDs were characterized by using Gaussian kernel
density\cite{diwekar2015probability} estimation of the bond
distributions. For this, distributions of \chemfig{C-C} ,
\chemfig{C-O} and \chemfig{C-H} bonds were considered. For {\it Set3}
the distribution of \chemfig{O-H} bonds was also included. The target
and test distribution of bond distances were compared by way of the
Kullback-Leibler (KL) divergence\cite{cover2006elements}
\begin{equation}
D_{\rm KL}(p\parallel q) = \int_{r_{min}}^{r_{max}} p(x) log\left(
\frac{p(x)}{q(x)} \right) dx
\label{eq:kl}
\end{equation}
If the database $p(x)$ contains more information than the target set
$q(x)$, $D_{\rm KL} (p||q)>0$, and if particular information is
missing, $D_{\rm KL} (p||q)<0$. Notice that the integration limits are
the minimum ($r_{min}$) and maximum ($r_{max}$) distances present in
the database. This means that the values of the distributions $p(x)$
and $q(x)$ are not normalized and the KL divergence is negative in
regions where $p(x)<q(x)$, if those regions have a larger area than
the positive regions, the value of $D_{\rm KL} (p||q)$ would be
negative.\\

\noindent
Another useful metric to compare two distributions is the Jensen-Shannon (JS)
divergence\cite{lin1991divergence,nielsen2019jensen}
\begin{align}
D_{\rm JS}(p\parallel q) = \frac{1}{2}D_{\rm KL}(p\parallel
\frac{p+q}{2})+ \frac{1}{2}D_{\rm KL}(q\parallel \frac{p+q}{2}) \\ =
\frac{1}{2} \int_{r_{min}}^{r_{max}} \left[
p(x)\log\left(\frac{2p(x)}{p(x)+q(x)}\right) +
q(x)\log\left(\frac{2q(x)}{q(x)+p(x)}\right)\right] dx.
\label{eq:js}
\end{align}
This is a symmetrized version of the KL divergence and quantifies the
total divergence from the mean distribution, as it returns the
averaged sum of the divergence between each distribution and the
arithmetic mean of the distributions\cite{nielsen2019jensen}. The \kl
and \js metrics contain complementary information about the
distributions to compare: \kl is more sensitive to local changes as it
quantifies how much $p(x)$ underestimates $q(x)$, but not the
opposite. On the other hand, \js provides a more balanced
interpretation. Therefore, \js measures overall changes in the test
and target distributions.\\

\noindent
For comparing energy distributions between the initial, augmented, and
target distributions of samples, the Wasserstein or Earth mover's
distance, as computed by SciPy\cite{virtanen2020scipy}, was used. This
quantity is defined as\cite{villani2009wasserstein}:
\begin{equation}
    W_{n}(p,q) = \left(\inf_{\psi \in \Psi(p,q)} \int |x-y|^{n} d\psi(x,y)\right)^{1/n}
    \label{eq:wass_gen}
\end{equation}
where $x$ and $y$ are points of distribution $p(x)$ and $q(y)$, and
$\Psi(p(x),q(y))$ is the set of all possible joint probability
distributions between $p(x)$ and $q(y)$. This can be interpreted as
the minimal effort for moving a proportion of mass ($\psi(x,y)$) over
a distance ($|x-y|^{n}$) to reconfigure the distribution $p(x)$ into
$p(y)$\cite{weng2019gan,panaretos2019statistical}. In this work, the
distributions of energies ($p(E)$) are 1-dimensional. Therefore,
$W_{1}$ was considered as
\begin{equation}
    W_{1}(p,q) = \int_{-\infty}^{\infty} |p(E)-q(E)| dE
    \label{eq:wass_1d}
\end{equation}
which is equivalent to Eq. \ref{eq:wass_gen}, for a proof of this see
Ref.\cite{ramdas2017wasserstein}. In equation \ref{eq:wass_1d},
choices for $p(E)$ and $q(E)$ are the distributions of energies of the
henceforth initial (iRD) and augmented RD (aRD), or the target
databases regardless of order because Eq. \ref{eq:wass_1d} is
symmetric and follows the triangle
inequality\cite{panaretos2019statistical}. In general, a small value
of $W_1$ indicates that the distributions compared are close in shape
and position, while large values of $W_1$ imply the distributions are
less similar.\\

\subsection{Fraction of Improved/Worsened Predictions}
Usual metrics consider average changes in the prediction of the
samples in the test set. Nevertheless, individual changes in the
energies after modifications of the training database are important in
this context because these provide information about how the model
responds to additions/changes in specific parts of CS. Such changes
were quantified by considering the fraction of molecules for which the
prediction errors ($\mathcal{E}_{i}$) increase ($f_{\uparrow}$) or
decrease ($f_{\downarrow}$). The fraction
\begin{equation}
    f_{\uparrow}=\dfrac{\sum_{i}
      \eta_{i}(\mathcal{E}_{i}^{\alpha},\mathcal{E}_{i}^{0})}{n_{\rm
        total}}
    \label{eq:inc}
\end{equation}
$n_{\rm total}$ is the total number of samples in the target dataset,
$\eta_{i}$ is defined as
\begin{equation*}
    \eta_{i}(\mathcal{E}_{i}^{\alpha},\mathcal{E}_{i}^{0}) = \left\{
    \begin{array}{ll}
         1 & {\rm If} \ |\mathcal{E}_{i}^{\alpha}| > |\mathcal{E}_{i}^{0}| \\
         0 & {\rm If} \ |\mathcal{E}_{i}^{\alpha}| \leq |\mathcal{E}_{i}^{0}|
    \end{array}
    \right.
\end{equation*}
where $\mathcal{E}^{0}_{i}$ is the error in prediction by the initial
RD and $\mathcal{E}^{\alpha}_{i}$ is the error for the enriched
dataset for the condition $\alpha$ which is the temperature of
sampling or percentage of samples added to the iRDs.  Conversely,
$f_{\downarrow} = 1-f_{\uparrow}$ is the fraction of molecules for
which the absolute error decreases
(i.e. $|\mathcal{E}_{i}^{\alpha}|<|\mathcal{E}_{i}^{0}|$). The values
of $f_{\uparrow,\downarrow}$ clarify for which percentage of the
molecules in the target DBs the energy prediction
improves/deteriorates. Furthermore, $f_{\uparrow,\downarrow}$ quantify
whether observed changes in other metrics, such as MAE, result from
variations in predictions of energy for the majority or minority of
molecules in the target database.\\

\section{Results and Discussion}
In the following, results for structure-based data augmentation applied to four typical and concrete chemical questions
- hybridization, oxidation, substitution effects, aromaticity - are
presented and discussed. First, the impact of temperature on sample
generation was assessed, followed by effects based on increased
numbers of samples. For both ``structure-based augmentations'' the
samples were added to the iRDs, and new NNs were trained for the
aRDs. \\

\subsection{The Effect of Temperature}
Key to the present work is the notion that sampling different regions
of conformational space may help to cover parts of chemical space not
accessed by the iRDs. Therefore, adding samples from conformational
space might compensate for missing or underrepresented chemical
species. In consequence, determining which regions of conformational
space provide information for improving predictions for the task at
hand is a primary challenge in data augmentation.\\

\noindent
Increasing the temperature at which samples were generated is a first
rational way that leads to structures perturbed away from the minimum
energy conformation. For example, a sufficiently stretched double bond
will adopt a distribution of the electrons that is more reminiscent of
a single bond; see Figure \ref{sifig:bonds}. For this, conformations
of a chosen molecule, hereafter "samples", were added to the iRDs, the
NN was retrained and the effect of the new conformational space
covered was evaluated on the target data set.\\

\begin{figure}
    \centering
    \includegraphics[width=0.9\textwidth]{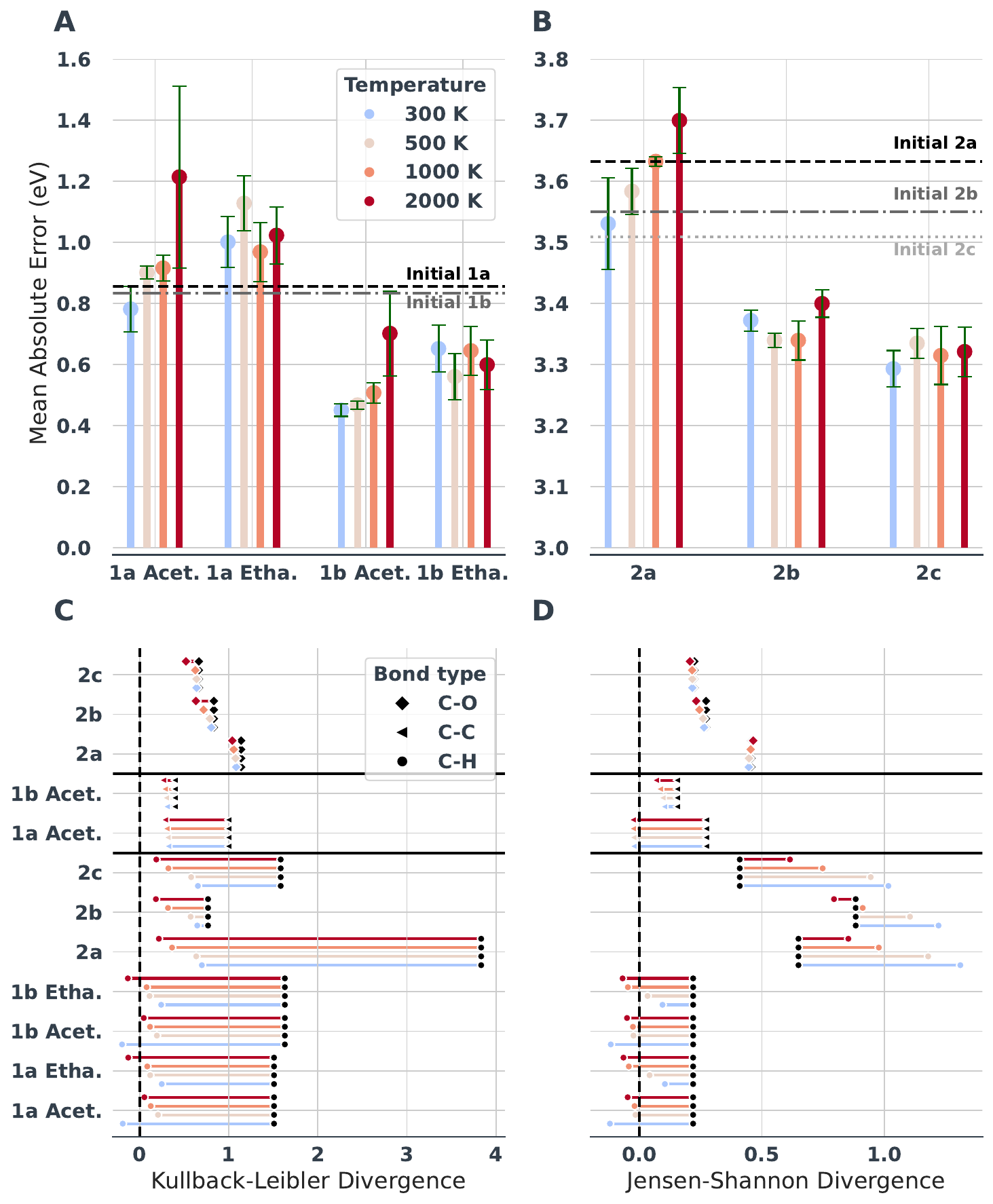}
    \caption{\textbf{Results for Temperature effect \set{1} \&
        \set{2}} A and B.  Change in the Mean Absolute Error (MAE) for
      the target dataset of the restricted databases 1 and 2 depending
      on $T$ used for NMS of representative structure(s). The results
      show the mean over three models initialized with different
      seeds. Green bars represent the standard deviation of the
      MAEs. The performance of the model in the target dataset before
      adding samples is shown as horizontal dotted
      lines. C. Kullback-Leibler divergence for different bond
      distributions (C-C, C-H, and C-O). The black circle indicates
      the initial value, and the final point is the value after the
      samples were added. Some values were omitted for
      clarity. D. Similar to C but for the Jensen-Shannon divergence
      (c.f. Equation \ref{eq:js}).}
    \label{fig:mae_enhanced_temp_12}
\end{figure}

\begin{figure}
    \centering
    \includegraphics[width=0.9\textwidth]{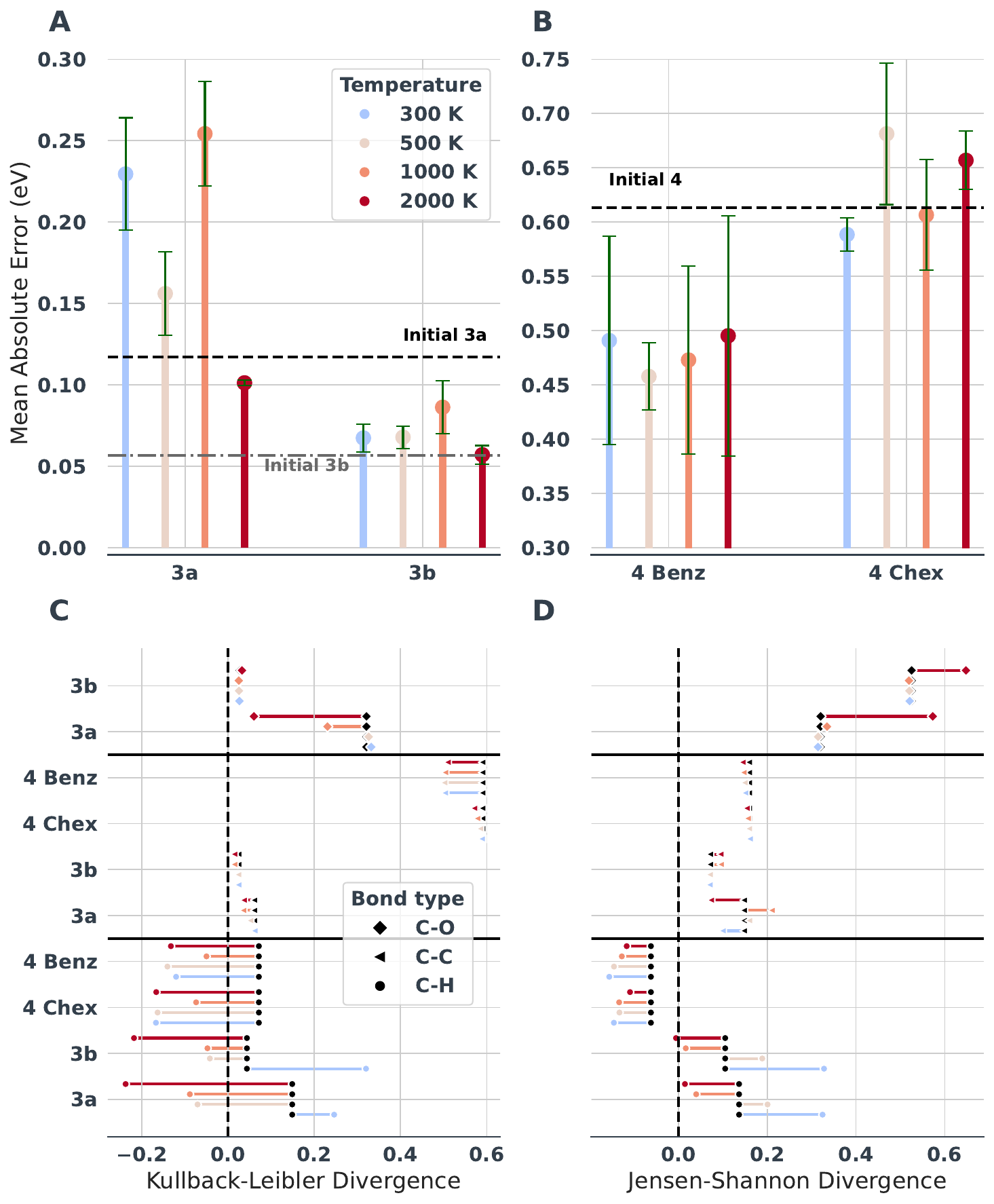}
    \caption{\textbf{Results for Temperature effect, \set{3} \&
        \set{4}} A and B.  Change in the Mean Absolute Error (MAE) for
      the target dataset of the restricted databases 3 and 4 depending
      on $T$ used for NMS of representative structure(s). The results
      show the mean over three models initialized with different
      seeds. Green bars represent the standard deviation of the
      MAEs. The performance of the model in the target dataset before
      adding samples is shown as horizontal dotted
      lines. C. Kullback-Leibler divergence for different bond
      distributions (C-C, C-H, and C-O). The black circle indicates
      the initial value, and the final point is the value after the
      samples were added. Some values were omitted for
      clarity. D. Similar to C but for the Jensen-Shannon divergence
      (c.f. Equation \ref{eq:js}).}
    \label{fig:mae_enhanced_temp_34}
\end{figure}

\noindent
\textit{Mean Absolute Error:} First, the overall performance of the
aRDs compared with results from models trained on the iRD was assessed by calculating the Mean Absolute Error (MAE) between
the energies of the molecules in the target dataset and those
predicted by the different models trained with the aRDs, see Figures
\ref{fig:mae_enhanced_temp_12}A/B and
\ref{fig:mae_enhanced_temp_34}A/B. For \textit{Set1}, the effect of
including samples generated at elevated temperature increases the MAE
proportionally to $T$ and is further supported by the error
distributions (Figure \ref{sifig:violin_set1_temp}). The MAE for the
iRD \textit{Set1b} is consistently smaller ($\approx 0.45-0.7$ eV)
than for \textit{Set1a} ($\approx 0.8-1.2$ eV) (Figure
\ref{fig:mae_enhanced_temp_12}A). This is in line with what was
observed for the models before augmentation. For the aRDs, it is
observed that \textit{Set1a} is more sensitive to augmentation
(broader error distributions) than \textit{Set1b} (Figure
\ref{sifig:violin_set1_temp}).\\

\noindent
Complementary to the effect of the composition of the RD, the effect
of the representative molecule used to generate the added samples was
observed for \textit{Set1a/b}: addition of acetylene conformers leads
to better performance when considering the MAE with the best
performing model as \textit{Set1b}-Acet at 300 K. This reduces the MAE
to $\approx 0.4$ eV, a reduction of $\approx 50\%$. At the same time,
the worst performance is observed for \textit{Set1a}-Acet at 2000 K,
increasing its MAE by approximately $0.4$ eV. In contrast, the addition of ethane samples leads to an increase in the MAE for \set{1a} while a reduction for \set{1b}. In both cases with ethane, the effect of the temperature is meaningless.\\

\noindent
For \textit{Set2}, the effect of adding formic acid
(FA) conformations is more noticeable in going from \textit{Set2a} to \textit{Set2b} and \textit{Set2c}. Improvements in the MAE are
$\approx 10$ \% with variations of $\approx 0.3$ eV between the models
augmented with FA conformations. Overall, the MAE distributions remain largely unchanged,
without significant variations (Figure
\ref{sifig:violin_set2_temp}). \set{2a} contains alcohols but no oxidized compounds, whereas \set{2b} and \set{2c} feature \chemfig{C=O} bonds. Thus, even without augmentation \set{2b} and \set{2c} perform better on the "task", see dashed lines in Figure \ref{fig:mae_enhanced_temp_12}. Although there is some slight improvement when adding samples of FA generated at 300 K to \set{2a} the effect vanishes for higher-temperature samples. Contrary to that, a clear improvement irrespective of the temperature at which augmentation samples were generated are found for \set{2b} and \set{2c}. One reason for this is that the distributional overlap of augmented \set{2b} and \set{2c} with the target data set increases which is confirmed by considering changes in \kl (Figure \ref{fig:mae_enhanced_temp_12}C) and $W_1$, see Figure \ref{sifig:wd_temp}B. \\

\noindent
Results for \textit{Set3} exhibited a negative effect in the MAE upon
adding 1000 samples generated at different temperatures. This effect
is more evident for \set{3a} than for \set{3b} showing larger changes
in the error distributions; see Figure
\ref{sifig:violin_set3_temp}. In both subsets of \set{3a}, the lowest
MAE is reached at the highest temperature, in clear contrast with
\textit{Set1} and \textit{Set2}, indicating that adding more disturbed
structures is beneficial for \set{3}. For \textit{Set4} the MAE varies
by $\approx \pm 0.1$ eV ($\approx 15$ \%) from the initial value. It
is observed that the variations are relatively stable ($\pm 0.05$ eV)
as a function of the temperature of sampling (Figure
\ref{fig:mae_enhanced_temp_34}B). Adding benzene conformations
improves model performance, whereas the opposite is observed when
cyclohexane conformers are added. (Figure
\ref{sifig:violin_set4_temp}).\\

\noindent
{\it Energy Distributions:} The temperature effect on the energy
distributions is overall insignificant. Considering changes in the
Wasserstein distance depending on the temperature at which samples
were generated, see Figure \ref{sifig:wd_temp}, all values for $W_1$
are comparable except for \set{3}. On the other hand, the effect of
augmenting the databases the effects are beneficial for \set{1a}-Etha,
\set{2}, and \set{4}, detrimental for \set{1b}-Acet and \set{3}, and
neutral for \set{1a}-Acet, and \set{1b}-Etha.\\

\begin{figure}
    \centering
    \includegraphics[width=0.7\textwidth]{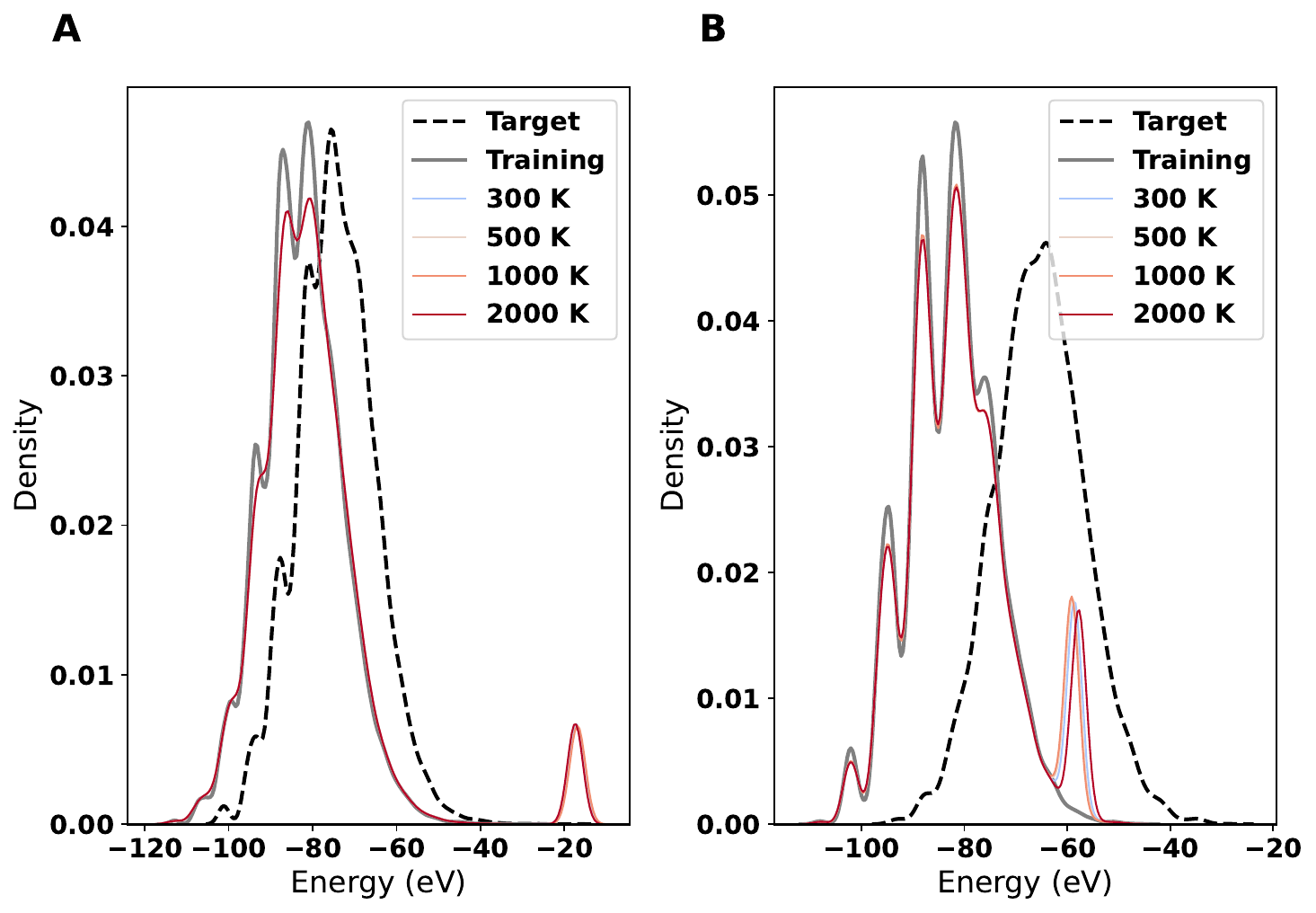}
    \caption{\textbf{Energy distributions of iRDs,aRDs and target
        databases}. A. Energy distribution for \set{1a}-Acet showing
      a case with small shifts and changes. B. Energy distribution for
      \set{4}-Benz showing a different centre of mass between the
      target and iRD. }
    \label{fig:energy_dist_extremes}
\end{figure}

\noindent
In all cases, the energy distributions, $P(E)$, of the aDBs were
bimodal, see Figure \ref{fig:energy_dist_extremes}. However, DER
assumes Gaussian distributed energies which makes learning bimodal
distributions challenging as had been recently also found for
uncertainty quantification of reactive potential energy surfaces
\cite{vazquez2025outlier}. The emergence of a second peak in $P(E)$
for the aDBs is a consequence of adding samples with high energy. In
general, the intensity of the new peak in $P(E)$ is independent of the
temperature of sampling, except for \set{3}. Thus, two extreme cases
can be distinguished. In the first (Figure
\ref{fig:energy_dist_extremes}A), the added samples are at high
energy. This potentially leads to
improvements in the generalizability of the models trained with the
aRDs. However, this improvement might occur in regions of chemical
space not part of the evaluated ``tasks''. For the second case (Figure
\ref{fig:energy_dist_extremes}B), $P(E)$ of the target DBs and the
iRDs overlap little because the centers of mass of both distributions
are far apart. Therefore, adding high-energy samples is
beneficial. Specifically, for (\set{4}), the addition of benzene
conformations generates a new peak around $-60$ eV, located near the
centre of mass of the target database. A complete description of the
energy distributions $P(E)$ (Figures
\ref{sifig:ener_dist_set1_temp}-\ref{sifig:ener_dist_set4_temp}) is
given in the SI.\\

\noindent
{\it Distribution of Geometries:} Next, the effect of sampling
temperature on the coverage of the target atom-atom separation
distributions was analyzed. For this, the Kullback-Leibler (\kl)
(c.f. Equation \ref{eq:kl}) and Jensen-Shannon (\js) (c.f. Equation
\ref{eq:js}) divergence between the target databases and the iRDs and
aRDs (Figure \ref{fig:mae_enhanced_temp_12}C/D and
\ref{fig:mae_enhanced_temp_34}C/D) were analyzed. In cases for which
the difference between the initial bond distribution of the iRD and
aRD was rather small, results are not reported. Starting with
\textit{Set1a/b}, \kl values for the \chemfig{C-H} bond distances show
largest values of \kl at low temperatures, except for \set{1a/b}-Acet
at 300 K. It must be noticed that for \set{1} the initial value of \kl 
is smaller for \set{1a} than for \set{1b}.  The largest reduction in
\kl is observed at high temperatures for \set{1a}-Etha (Figure
\ref{fig:mae_enhanced_temp_12} C). At the same time, \kl increases for
\set{1a/b}-Acet regardless of the composition of the iRD.  For
\chemfig{C-C} bonds, it is noticed that for \textit{Set1a}-Acet \kl
reduces more than for \textit{Set1b}-Acet because the iRD
\textit{Set1b} contains \chemfig{C-C} single and \chemfig{C=C} double
bonds. This reduces the difference between the target and aRDs for
\set{1b} compared to \set{1a}. This effect is reinforced by
augmentation with acetylene.  The previous findings are corroborated
by the results of the \js divergence (Figure
\ref{fig:mae_enhanced_temp_12}D).\\

\noindent
Moving to \textit{Set2}, the largest changes in \kl for \chemfig{C-H}
distances are found for \textit{Set2a}, followed by \textit{Set2c},
and \textit{Set2b}. Complementary, the largest differences in \kl
values are obtained at high temperatures. For \set{2}, the values of
the \js divergence displays the opposite trend than \kl. A possible
explanation for this is that the overlap of the distributions in
high-probability regions (e.g. near the centre of mass of the
distribution) for both distributions improves, leading to small values
of \kl. However, \js increases because the distributions of aRDs are
broadened (e.g. the appearance of samples at regions uncovered by the
target distribution), reducing the total overlap of the distributions.
The \kl values for the \chemfig{C-O} distributions follow the trend
observed for \chemfig{C-H} but with a considerably smaller
magnitude.\\

\noindent
Changes in \kl and \js for \textit{Set3} and \textit{Set4} are shown
in Figure \ref{fig:mae_enhanced_temp_34}C/D. Unlike previous
datasets, no clear trend is observed for them. For \set{3}, the values
of \kl for \chemfig{C-H} increase at low temperature to later
decrease, reaching a minimum at the highest temperature for both
subsets. A similar trend is observed for \js of \chemfig{C-H} for
\set{3a/b}. Values of \kl and \js for \set{3a/b} for \chemfig{C-C}
bond show negligible changes, while the same values for \chemfig{C-O}
display discernible differences only for \set{3a}. However, for this
set, \kl values are reduced. In contrast, \js values increase,
indicating that the samples obtained at high temperatures correspond
to conformations far from the average distribution of the test and
target that do not help to improve the prediction. Lastly, for \set{4}
\kl and \js for \chemfig{C-H} decrease regardless of the molecules used for augmentation and the temperature at which conformers were generated. The values of \kl for
\chemfig{C-C} of \set{4}-Benz reduce slightly while the \js
value does not change. \\

\begin{figure}
    \centering
    \includegraphics[width=0.9\textwidth]{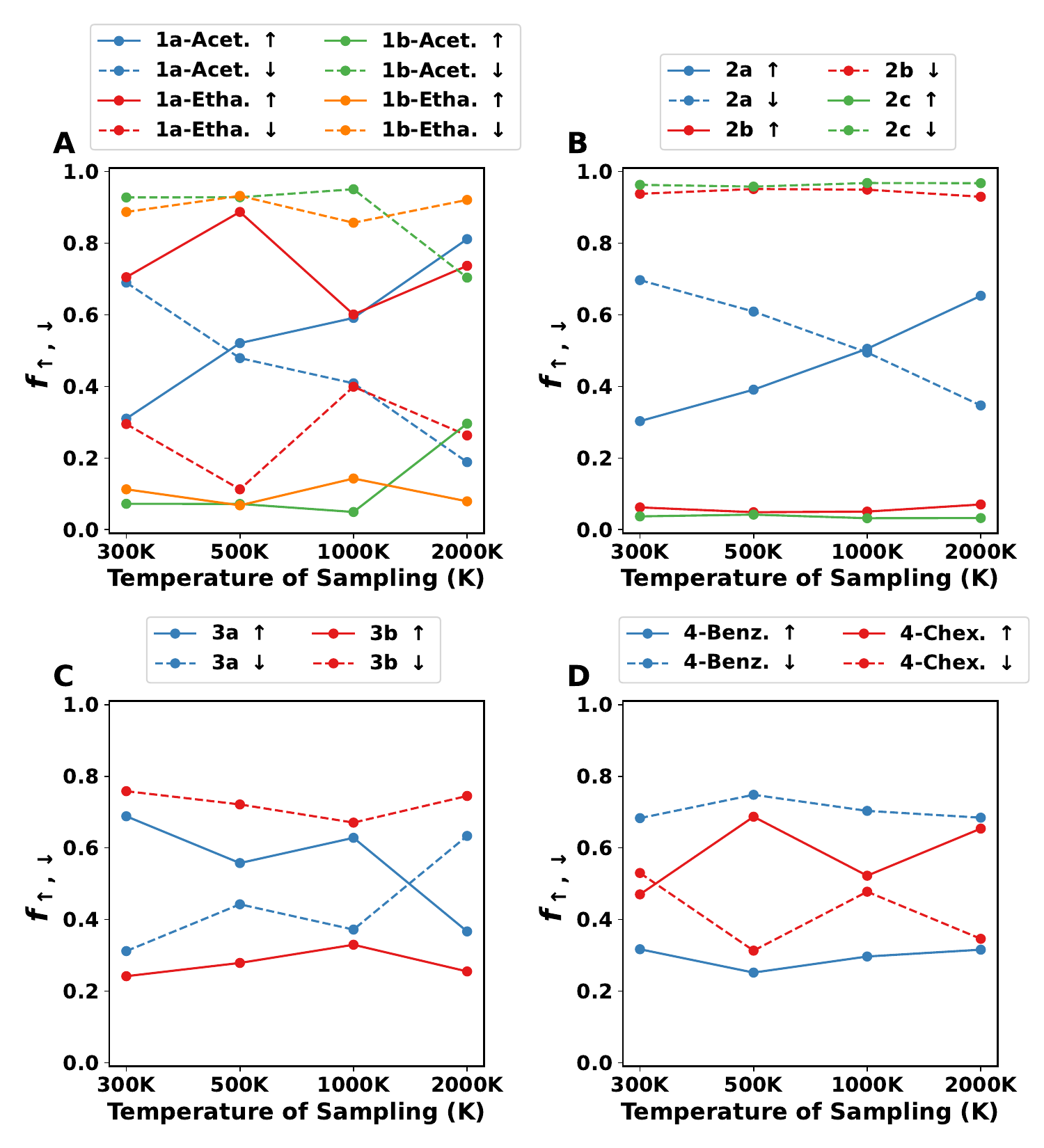}
    \caption{\textbf{Fraction of improved/worsened predictions} Values of $f_{\uparrow/\downarrow}$ (c.f. Eq. \ref{eq:inc}) are shown as a function of the temperature used to generate the conformers of the selected molecules for each of the iRD studied in this work. Panel A shows the results for \set{1}, panel B display results for \set{2}, panel C for \set{3} and panel D for \set{4}}
    \label{fig:inc_dec_temp}
\end{figure}

\noindent
\textit{Fraction of Improved/Worsened Predictions:} The quantities
described so far correspond to averages of the predicted quantity or
changes over all samples in the test set. Next, the fraction of
molecules in the target DBs for which the prediction errors increase
($f_{\uparrow}$) or decrease ($f_{\downarrow}$), see methods
section. Values of $f_{\uparrow}$ and $f_{\downarrow}$ for all RDs are
shown in Figure \ref{fig:inc_dec_temp}. Results for \set{1} depend on
the composition of the iRDs with values of $f_{\downarrow} > 0.8$ for
\set{1b} regardless of the temperature of sampling while for \set{1a}
the largest value was $f_{\downarrow} = 0.7$. In addition, \set{1a} is
more sensitive to the molecule used for augmentation of the RDs. For
\set{1a}-Acet, $f_{\uparrow}$ increases linearly with temperature
whereas for \set{1a}-Etha $0.6 < f_{\uparrow} < 0.8$. For
\textit{Set2} $f_{\downarrow}$ increases with temperature for
\textit{Set2a} (Figure \ref{fig:inc_dec_temp}B) in line with the
modest increase in MAE shown in Figure \ref{fig:mae_enhanced_temp_12}B
whereas for subsets \textit{Set2b} and \textit{Set2c} $f_{\downarrow}
\sim 90$\% regardless of the temperature at which added samples were
generated.\\

\noindent
Moving to {\it Set3}, the values of $f_{\downarrow}$ for
\textit{Set3a} and \textit{Set3b} exhibit contrasting trends.  For
\textit{Set3a}, $f_{\uparrow} \sim 60$ \% for all temperatures except
at 2000 K, for which it drops to $\sim 40$ \%. Conversely, for
\textit{Set3b}, $f_{\downarrow} \sim 70$ \% across all sampling
temperatures (Figure \ref{fig:inc_dec_temp}C). This suggests that the new information introduced by the conformers (e.g., tertiary alcohols) differs more from the content in iRDs for \set{3a} (primary alcohols) than from \set{3b} (a mix of primary and secondary alcohols). Additionally, predicting more complex chemical environments, such as chiral carbon atoms, likely requires a chemically more diverse dataset, so that the model can learn such environment. As a result, augmentation of \set{3b} provides a better framework to describe chirality than augmented \set{3a}. The findings for
\textit{Set4}-Benz conformers reveal an improvement in prediction for
approximately 70\% of the molecules in the target set
(i.e. $f\downarrow< 0.7$), irrespective of the sampling temperature
(Figure \ref{fig:inc_dec_temp}D). Conversely, for \textit{Set4}-Chex,
the fraction $f_{\uparrow}$ oscillates between 40\% to 60\%,
increasing with the sample temperature. Complementary to this, a
discussion of the distribution of changes in the predicted energy
($\Delta E_{\rm pred} = E^{\rm Pred}_{0}-E^{\rm Pred}_{T}$) and
respective figures (Figure
\ref{sifig:dist_err_set1_temp}-\ref{sifig:dist_err_set4_temp}) is
given in the SI. \\

\noindent
In summary, for most iRDs, augmentation leads to performance improvements with largest reductions of the MAE for low temperatures (i.e. 300 K). The only exception is
\set{3} which displays better prediction performance as the sampling
temperature increases. A chemical interpretation may be that the
samples in the target DBs are more perturbed by the presence of
several substituents on the carbon atom of the alcohol. Hence, generating samples at higher temperatures leads to stronger deformations which ultimately improves the trained model. The target DB of \textit{Set2} proves to be the most
challenging to predict. This can be attributed to the fact that the
target set was taken from a different parent DB, which contains
chemical groups not covered by the iRD (e.g. the nitro group is not
present in QM9). Therefore, the addition of conformers of FA is not
sufficient to improve the predictions, as observed by the minimal
changes in MAE of predictions. Another possible explanation is that most of the added conformations involve changes in \chemfig{C-H} bonds, which have minimal impact on improving oxidation predictions in organic molecules, as \chemfig{C-H} stretching does not significantly contribute to describing this property. In contrast, \textit{Set1} exhibits the most significant
changes in the mean absolute error (MAE), with \textit{Set1a/b}-Acet
yielding the best results.  Lastly, the results of \textit{Set4}-Benz
have the best performance independent of the sampling temperature.\\

\subsection{The Effect of the Number of Samples} 
Next, the impact of the {\it number of samples added} to augment the
iRDs for predicting the target dataset was considered. For this,
samples generated at $T = 300$ K were chosen as this temperature led
to the most significant decrease in the MAE for most iRDs. The number
of structures used for augmentation ranged from 1 \% to 25 \% with
respect to the initial number of samples in the iRDs.\\

\begin{figure}
    \centering
    \includegraphics[scale=0.6]{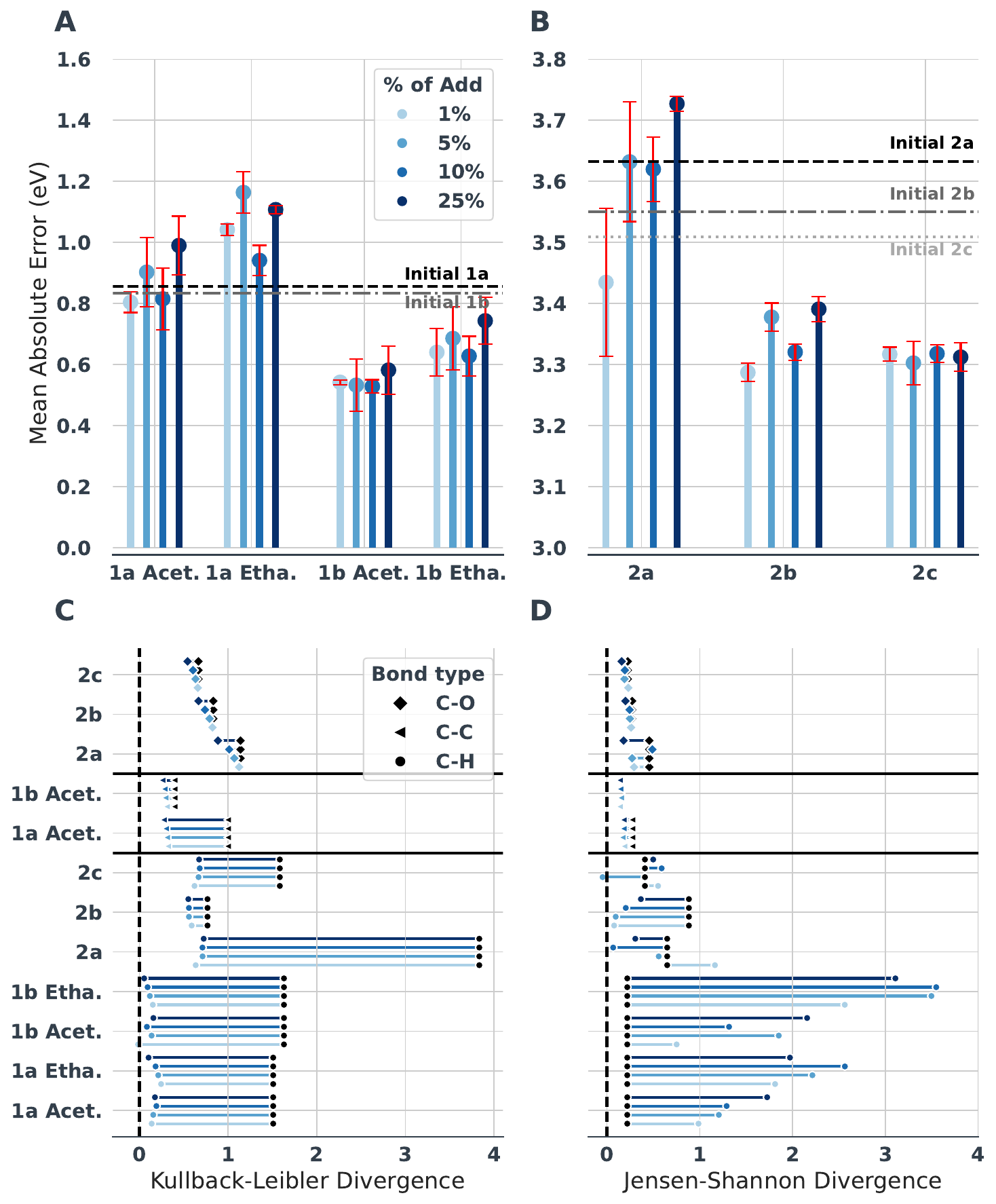}
    \caption{\textbf{Results for the number of samples added. \set{1}
        \& \set{2}} Panels A and B: Change in the Mean Absolute Error (MAE)
      for the target dataset of the restricted databases 1 and 2
      depending on percentage added used for NMS of representative
      structure(s). The results show the mean over three models
      initialized with different seeds. The error bars represent the
      standard deviation of the MAE over the different values. In each
      of the panels, the performance of the model in the target
      dataset before adding samples is shown on horizontal dotted
      lines. Panel C: Kullback-Leibler divergence for different bond
      distributions (C-C, C-H, and C-O). In all cases, the black dot
      indicates the initial value, and the final point is the value
      after the samples have been added. Some values were omitted for
      clarity. Panel D: As for panel C but for the Jensen-Shannon divergence
      (c.f. Equation \ref{eq:js}).}
    \label{fig:mae_enhanced_size}
\end{figure}

\noindent
\textit{Mean Absolute Error:} Figure \ref{fig:mae_enhanced_size}
illustrates the impact on the mean absolute error (MAE) for models
trained with iRDs and subsequently augmented with varying sample
sizes. Again, the effect is inconsistent across all datasets and does
not remain constant with the number of added samples to the training
databases. \textit{Set1} have different results for both subsets, with
better results for \textit{Set1b}. Regarding the molecule used for
augmentation, it is noticed that \set{1a/b}-Acet yield larger
improvements than \set{1a/b}-Etha (Figure
\ref{fig:mae_enhanced_size}A). These observations are confirmed by the
error distributions in Figure \ref{sifig:violin_set1_large}, which
display more compact distributions for \textit{Set1b} than
\textit{Set1a}.  Additionally, \set{a/b}-Acet shows error
distributions less spread out than \set{1a/b}-Etha.\\

\noindent
Continuing with \textit{Set2}, changes in MAE with respect to the
initial value remain minor (Figure \ref{fig:mae_enhanced_size}). The
most significant improvements were observed for \textit{Set2c},
followed by \textit{Set2b}, and \textit{Set2a} because, from a chemical perspective, the number of oxidized compounds increases. This is also consistent with what was observed for the
temperature effect. Changes in the MAE with respect to the number of
samples show the largest oscillations in inverse order of the
improvements in this quantity whereas changes in the error
distributions across different subsets are marginal (Figure
\ref{sifig:violin_set2_large}).\\

\begin{figure}
    \centering
    \includegraphics[width=0.9\textwidth]{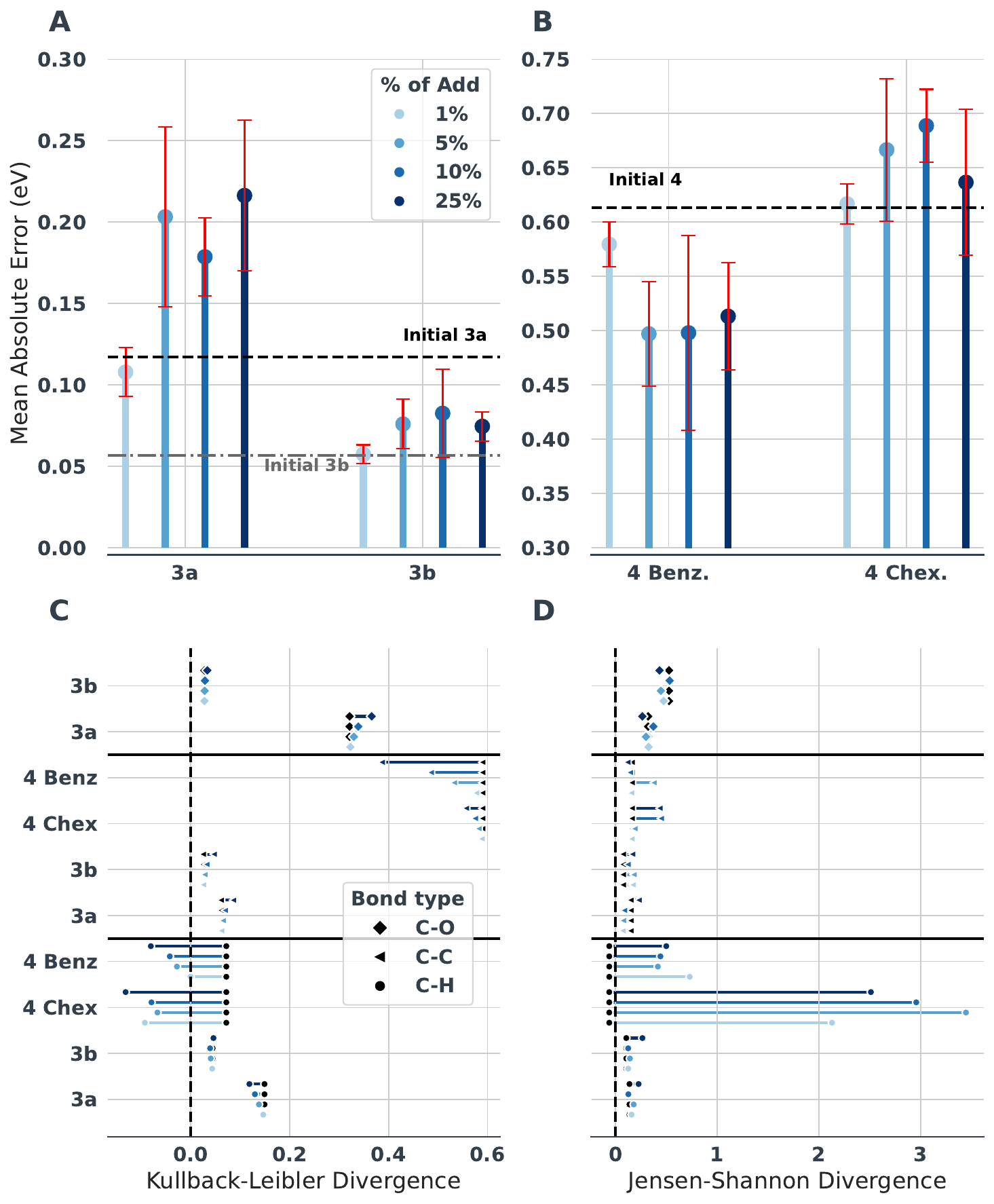}
    \caption{ \textbf{Results for the number of samples added. \set{3}
        \& \set{4}} A and B.  Change in the Mean Absolute Error (MAE)
      for the target dataset of the restricted databases 3 and 4
      depending on percentage added used for NMS of representative
      structure(s). The results show the mean over three models
      initialized with different seeds. The error bars represent the
      standard deviation of the MAE over the different values. In each
      of the panels, the performance of the model in the target
      dataset before adding samples is shown on horizontal dotted
      lines. C. Kullback-Leibler divergence for different bond
      distributions (C-C, C-H, and C-O). In all cases, the black dot
      indicates the initial value, and the final point is the value
      after the samples have been added. Some values were omitted for
      clarity. D. Similar to C but for the Jensen-Shannon divergence
      (c.f. Equation \ref{eq:js}).}
    \label{fig:mae_enhanced_samp_34}
\end{figure}

\noindent
In the case of \textit{Set3}, augmentation yields negative effects for
both subsets with a slight yet continuous increase with the number of
added samples (Figure \ref{fig:mae_enhanced_samp_34}A). The increase
in the MAE is accompanied by shifts in the error distributions for
\textit{Set3} (Figure \ref{sifig:violin_set3_large}) with substantial
displacements in the distribution's centre of mass for \set{3a} and an
expansion of the distribution tails for \set{3b}. The results for
\textit{Set4} depend on the molecule used for the augmentation (Figure
\ref{fig:mae_enhanced_samp_34}B) with overall changes of $\sim
0.1$ eV.  Reductions in MAE dependent on the number of samples added
are observed for \set{4}-Benz. In contrast, increases in MAE are
observed for \set{4}-Chex. This leads to significant changes in the
error distributions for \set{4}-Benz while they are negligible for
\set{4}-Chex (Figure \ref{sifig:violin_set4_large}).\\

\noindent
\textit{Energy distributions:} Changes in the energy distribution of
iRDs, aRDs, and target sets (Figures \ref{sifig:ener_dist_set1_large}
to \ref{sifig:ener_dist_set4_large}) revealed shifts in the
distributions of energies of the aRDs towards smaller energies. In
addition, a second peak in $P(E)$ of aRDs emerges at high energies,
but the intensity of the new peak does not correlate with the number
of added samples. The values of $W_1$ reveal an irregular pattern with
respect to the percentage of added samples; see Figure
\ref{sifig:wd_perc}. For \set{1}, the composition of the iRD and the
molecule used for augmentation have different effects on the magnitude
of $W_{1}$. For \set{1a}, $W_1$ increases with the number of added
samples, although the largest value of $W_1$ is observed at 1\% for
\set{1a}-Acet and at 5\% for \set{1a}-Etha. In the former case, the
values of $W_1$ are larger than the initial number after 10\%
addition, while for the latter the values of $W_1$ are smaller, except
for 5\% addition. Conversely, for \set{1b}, the values of $W_1$ are
constantly larger than their initial value and decrease as the number
of samples added increases. Contrary to what was observed for
\set{1a}, an exception is noticed for \set{1b}-Acet at 10\% which
shows the largest value of $W_1$ for \set{1}.\\

\noindent
Continuing with \set{2}, the trends are more regular, whereby $W_{1}$
decreases with respect to their initial values.  Complementary, the
value of $W_1$ reduces as the number of added samples increases.
Nevertheless, it should be noted that the value of $W_1$ depends on
the iRDs' composition. $W_1$ is largest for 1\% addition (\set{2b})
and 5\% for the rest.  Next, \set{3} results show clear trends. In all
cases, the value of the $W_1$ for the aRDs is larger than its initial
value. The value of $W_1$ for \set{3} reduces as the number of
samples increases, with the largest value of WD at 1\%.  Lastly,
\set{4} shows a reduction of $W_1$ in both cases and at all addition
percentages. Contrary to \set{2} and \set{3}, the value of $W_1$ for
\set{4} increases with the increase in the percentage of samples
added. These observations indicate that adding a large number of
samples leads to problems for prediction because the ensuing
redundancy ``confuses'' the trained model. These observations are in
line with a previous study\cite{vazquez2022uncertainty} where it was
found that conflicting similar information leads to problems in
prediction.\\

\noindent
\textit{Distribution of Geometries:} The relevant \kl and \js
divergences for bond lengths after the addition of samples from
conformational space are shown in Figures \ref{fig:mae_enhanced_size}
C/D and \ref{fig:mae_enhanced_temp_34} C/D.  For \textit{Set1}, it is
clear that the values of the \kl distance for \chemfig{C-H} bonds are
reduced for all percentages of sampling addition and irrespective of
the molecule used for augmentation. In this case, the values of the
\js distance gives us a clearer picture of the changes in the bond
distribution, which shows increases as the number of samples
grows. This difference in values between \kl and \js indicates that
the distributions have a better overlap at high-probability values
while diverging on low-probability areas, which is a natural
consequence of adding more disturbed samples.  Continuing our
discussion, the values of \kl and \js for the \chemfig{C-C} bonds show
modest reductions, which are only considerable for \set{1}-Acet. Those
reductions are more evident for \set{1a} than for \set{1b} with more
marked changes for \kl divergence than \js, which can be attributed
to the fact that the addition of conformations of conformers from
acetylene has a larger impact on reducing the distance in \set{1a}
because the samples are added to regions which were initially loosely
cover.  However, changes in the value of \js, are negligible.\\

\noindent
Moving to the results of \set{2} for \chemfig{C-H} bonds, the \kl
divergence is reduced in all cases, with the more noticeable changes
at low addition percentages. On the contrary, values of \js follow an
irregular pattern with reductions for \set{2a}, with except of 1 \%
addition, and \set{2b}. Lastly, \set{2c} show small increases except
for 5 \% addition.  It is interesting to note that \set{2b} has the
lowest values of initial \kl divergence and, at the same time, the
largest value of \js among those in \set{2} . A possible explanation
might be that the amount of \chemfig{C-H} bonds from aldehydes plays a
considerable role at bond distances similar to those of the target
distribution of carboxylic acids. For \chemfig{C-H} bonds changes in
\kl for \set{3a} are more visible than for \set{3b}, specifically at
the largest percentages of augmentation whereas \kl and \js for
\chemfig{C-C} of \set{3} do not change noticeably. Contrary to that,
\set{4} displays clear reductions of \kl value for \chemfig{C-H} and
for both subsets of \set{4} the percentage of addition is
evident. Contrary to this, the values of \js for \chemfig{C-H} in
\set{3} increase irregularly for \set{4}-Chex while \set{4}-Benz has a
clear dependence. As in the case of the effect of temperature, values
of \js and \kl for \chemfig{C-C} of \set{4} do not display meaningful
changes to the initial values.\\

\noindent
\textit{Fraction of Improved/Worsened Predictions:} The fractions
$f_{\uparrow}$ (Equation \ref{eq:inc}) and $f_{\downarrow}$ were also
analyzed, see Figure \ref{fig:inc_dec_large}. For \textit{Set1},
notably, \set{1a} exhibits larger values of $f_{\uparrow}$ compared to
\set{1b}, in line with results from the MAE values. Furthermore,
\set{1a/b}-Acet displays larger values of $f_{\downarrow}$ compared
to \set{1a/b}-Etha (Figure \ref{fig:inc_dec_large}A). Regarding the
impact of the number of added samples, an oscillatory pattern is
observed across all aRDs of \set{1}.  \set{1a}-Etha augmentation show
high values of $f_{\uparrow}$ ($ \sim 0.8$) reaching a maximum at 90\%
at 5\% augmentation. Conversely, \set{1a}-Acet displays larger values
of $f_{\downarrow}$ oscillating between 70\% and 30\%.  Results for
\set{1b} are more consistent and independent of the number of added
samples.  \set{1b}-Acet maintains a constant $f_{\downarrow}$ value of
around 90\% meanwhile, \set{1b}-Etha fluctuates between 90\% and 70\%
for all levels of addition.\\

\noindent
Moving on to \textit{Set2}, the findings align with the observations
of the previous section on temperature effect. Specifically, for \set{2a} the value of $f_{\uparrow}$ increases with the
number of added samples. At the same time, \set{2b} and \set{2c}
maintain constant values of $f_{\downarrow}$ exceeding 90\% regardless
of the sample size (see Figure \ref{fig:inc_dec_large}B). Concerning
\textit{Set3}, a consistent opposite trend between \set{3a} and
\set{3b} is evident (see Figure \ref{fig:inc_dec_large}C). For
\set{3a}, $f_{\uparrow}$ increases with the number of added samples,
whereas \set{3b} maintains a high value ($>70$\%) of $f_{\downarrow}$
regardless of database enrichment. Sets \set{4}-Benz and
\set{4}-Chex, exhibit opposing trends, with \set{4}-Benz showing a
$f_{\downarrow}$ value of approximately 60\% (see Figure
\ref{fig:inc_dec_large}D). Contrariwise, \set{4}-Chex demonstrates
$f_{\uparrow}$ values close to 60\%. Complementary discussion of the distribution of changes in the predicted energy and Figures \ref{sifig:dist_err_set1_large}-\ref{sifig:dist_err_set4_large} can be found in the SI.\\

\noindent
In summary, this section studied the effect of the number of
conformers of the representative molecule added to the iRDs. Here, the results highlight again that improved performance can be achieved by adding a
small fraction of samples to the iRDs. Therefore, adding a large
number of conformers either harms or has negligible effects
on prediction accuracy. This is in line with previous observations
\cite{vazquezsalazar2021}.  Once again, \textit{Set2} emerges as the
most challenging to predict, with marginal reductions of the MAE of
prediction regardless of the number of samples added.  On the other
hand, \textit{Set1a/b}-Acet yields the largest reductions of MAE. \\

\section{Summary and Conclusions}
This study investigated the augmentation of initially restricted
chemical databases by adding samples from the conformation space of
representative structures of the target databases. The iRDs were designed to cover various chemical aspects,
including hybridization, oxidation, chirality, and aromaticity. 
The
performance assessment of the addition was focused on mean absolute
error, the fraction of samples with increased/decreased absolute error
in the target dataset, changes in $E_{\rm pred}$, and the chemical
structures of samples exhibiting significant changes in $E_{\rm
  pred}$. In addition, analysis of the changes in the distribution of
energies between target and augmented distribution using WD and the
distributions of bond distances \textit{via} the \kl and \js
divergences were studied. The iRDs were augmented by generating
samples from normal mode sampling of a representative molecule
corresponding to the targeted chemical aspect. The temperature of
sampling and the number of samples added to the iRDs were examined. The results indicate that, in general, adding samples from a single
molecule had minimal effects on most of the DBs.  Nevertheless, it was
possible to observe the impact of the temperature of sampling used to
generate samples in the prediction.  Then, the influence of
temperature was found to slightly degrade prediction accuracy across
most databases, with optimal results achieved at 300 K.  On the other
hand, the analysis of the effect of the number of samples added to the
iRDs in the prediction showed that addition of smaller sample sizes (1 \%) yielded better
performance. This suggests that redundancy and highly disturbed
structure addition adversely affect prediction quality. \\

\noindent
Analysis of the overlap between distributions of energy from target DB
and iRD provides a rational basis for model improvement. As an
analogy, umbrella sampling simulations for determining free energies
from atomistic simulations require energy distributions from
neighbouring sampling windows to
overlap\cite{torrie1977nonphysical,kastner2011umbrella} Similar to
that, the $P(E)$ for the target DB and the iRD need to overlap for
meaningful model performance on the task. If the two distributions do
not overlap, the augmentation procedure needs to ensure that such an
overlap is generated. This overlap could be optimized by including $W_1$ into the loss function with a corresponding hyperparameter, as it is common in generative
ML\cite{dukler2019wasserstein}.\\

\noindent
One major
  finding of the present work is the fact that the addition of
  measured amounts (1\% ) of judiciously chosen molecules can improve
  the performance in the prediction of energies in view of particular
  "chemical tasks". Ideally, such augmentation would occur in an
  orthogonal fashion within chemical space relative to the iRD. In
  other words, a more direct approach to "designing" improved aRDs for a
  given task will consist of constructing feature vectors that are
  orthogonal to the feature vectors from a trained model on an iRD
  based on the target database, akin to the
  well-known Gram-Schmidt orthogonalization for vector spaces. In a
  next step, these newly generated feature vectors must be translated
  back to chemical space through another ML model (e.g. variational
  autoencoder\cite{kingma2019introduction} or generative adversarial networks\cite{Schwalbe-Koda2020}). Such an approach requires an improved understanding of how feature vectors and the underlying chemistry are related. This task falls in the domain of explainable AI, on which
  recent progress for potential energy surfaces has been made \cite
  {esders2025analyzing}.\\

\noindent
The results of this work show that incorporating samples from
conformational space can improve property prediction. However,
results also indicate that adding a single moiety fails to fully
address the distribution shift issue across different
databases. Therefore, alternative methods which cover wider ranges of chemical space need to be explored. Achieving this will help to rebalance initially biased databases for a particular chemical task. Some general recommendations
can be drawn from the observations here. First, new samples must be
generated at low temperatures to capture relevant regions of
conformational space that help improve prediction, as the molecules
wished to predict are in equilibrium. In addition, it should be
noticed that a small number of samples can yield a significant impact
on the prediction, while the largest number introduces redundant
information, which makes prediction harder. Results of this work
provide a baseline for the creation of synthetic
databases\cite{gardner2023synthetic} or the inversion of the data
generation process\cite{reizinger2024cross}.\\

\section*{Acknowledgment}
The authors gratefully acknowledge financial support from the Swiss
National Science Foundation through grants $200020\_219779$ (MM),
$200021\_215088$ (MM), the NCCR-MUST (MM), and the
University of Basel. LIVS acknowledges funding from the Swiss National Science
Foundation (Grant P500PN\_222297) to develop the last stages of this
work. The authors acknowledge Eric Boittier and Julia Nguyen for helpful
discussions in the initial stage of the project.

\bibliography{refs}%

\appendix

\import{./}{si.tex}

\end{document}

%% file: si.tex
\renewcommand{\thepage}{S\arabic{page}}
\renewcommand{\thetable}{S\arabic{table}}
\renewcommand{\thefigure}{S\arabic{figure}}
\renewcommand{\theequation}{S\arabic{equation}}

\setcounter{page}{1}
\setcounter{figure}{0}

\section{Supplementary Discussion}

\subsection{Analysis of energy distributions}
\subsubsection{The effect of temperature}

\begin{figure}
    \centering
    \includegraphics[width=0.9\textwidth]{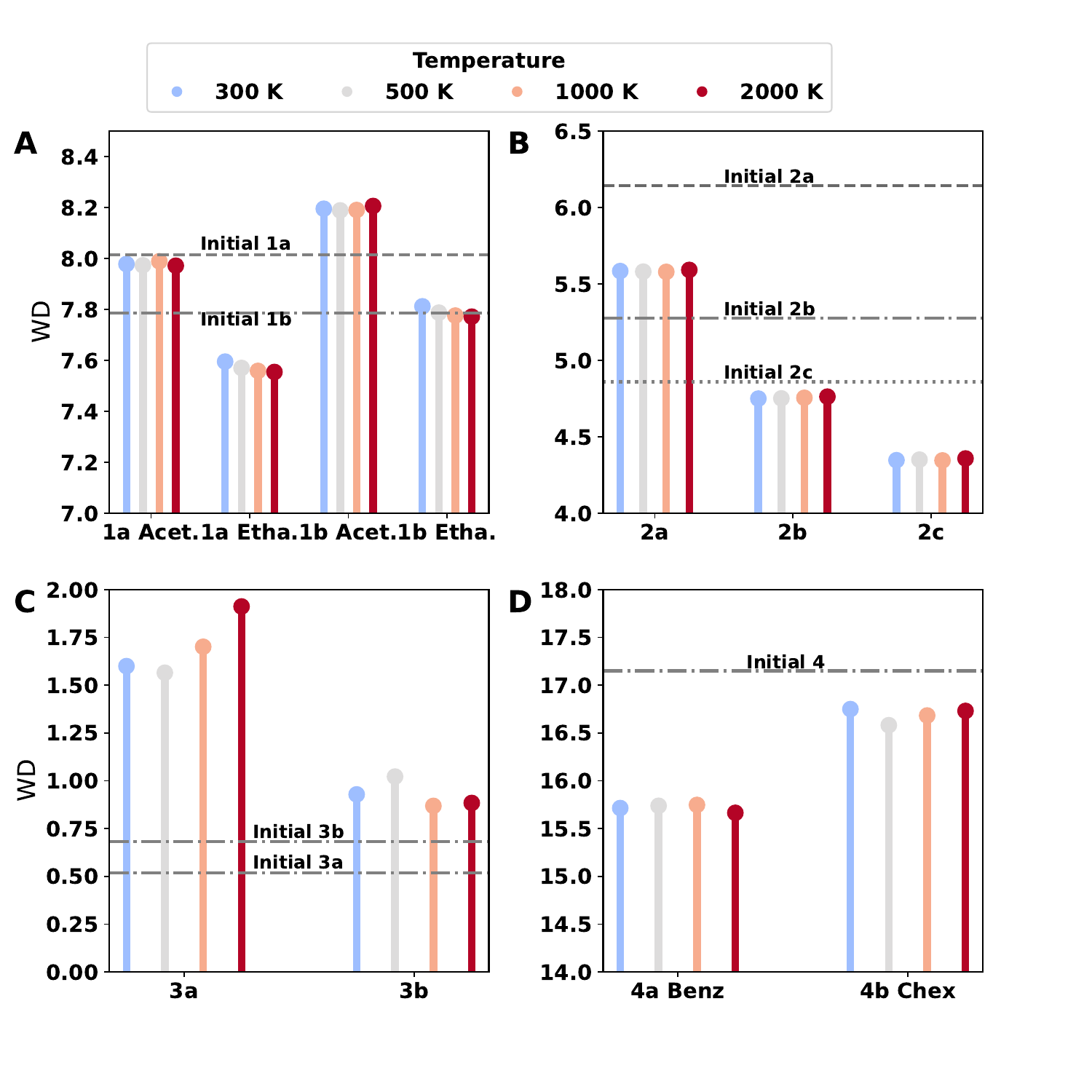}
    \caption{Wasserstein distance between the enhanced and target energy distributions for all the restricted databases studied in this work. A grey line(s) represents the initial distance between training and target distribution in each panel. The scale of the different axes is not uniform to better exemplify the changes in the distances. }
    \label{sifig:wd_temp}
\end{figure}

\begin{figure}[h]
    \centering
    \includegraphics[width=0.9\textwidth]{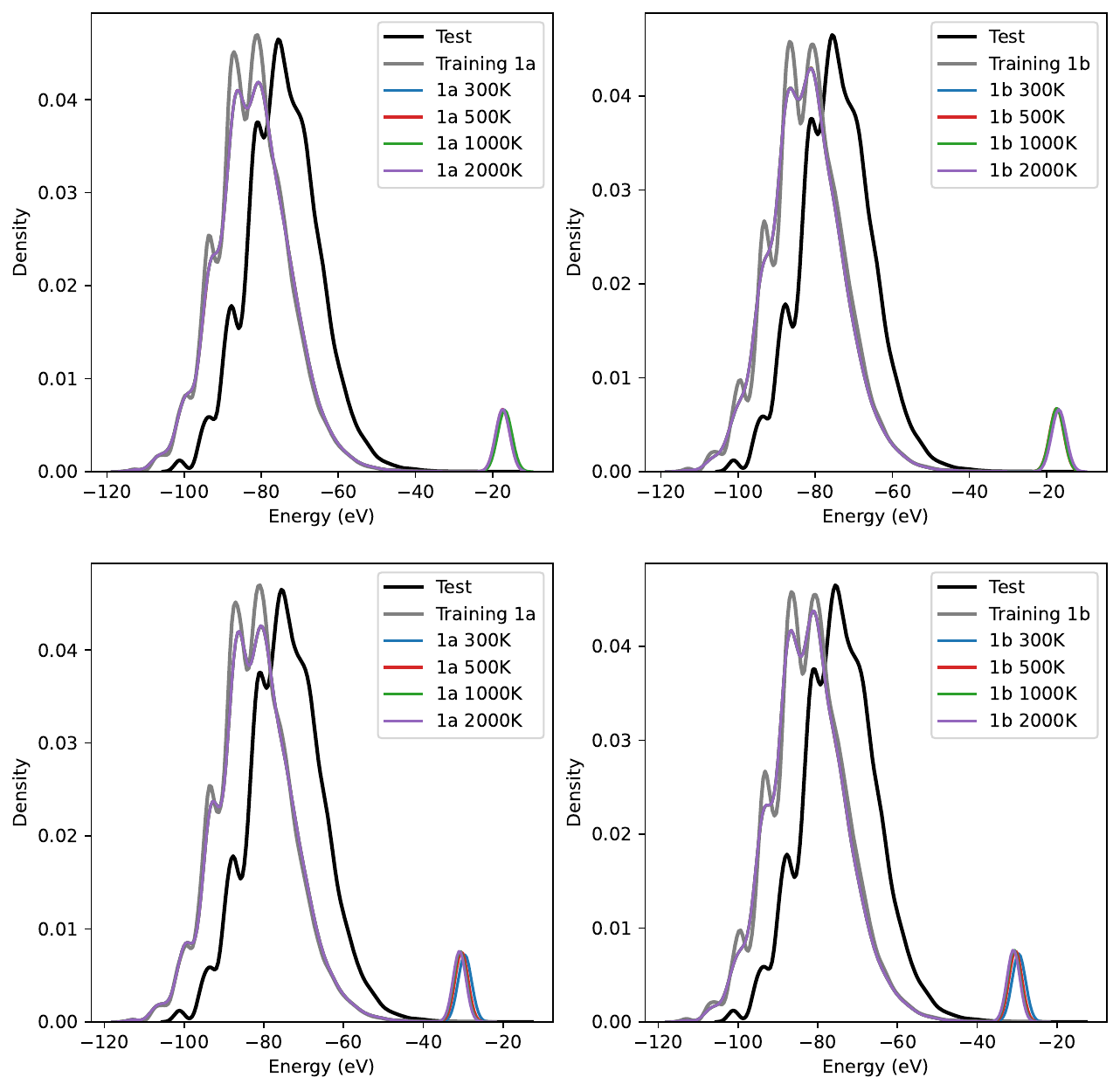}
    \caption{Energy distribution for the testing, initial training
      dataset and the enhanced datasets by temperature for \textit{Set1}.}
    \label{sifig:ener_dist_set1_temp}
\end{figure}

\begin{figure}
    \centering
    \includegraphics[width=0.9\textwidth]{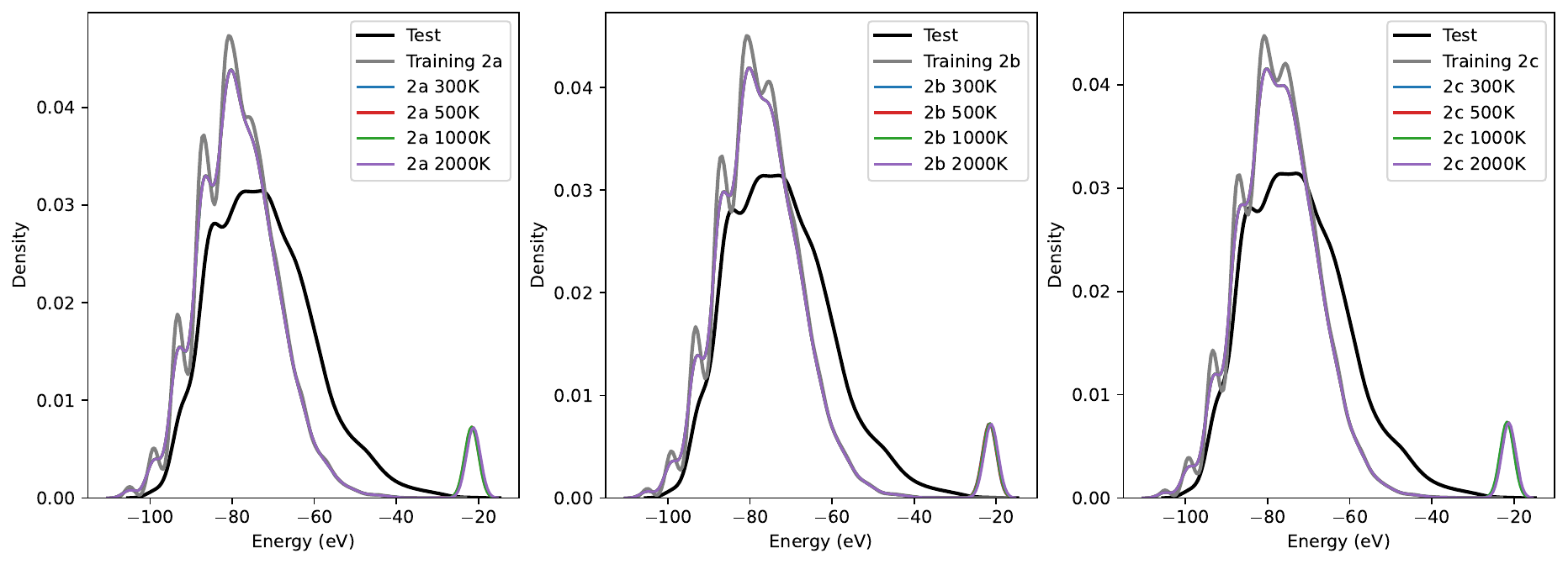}
    \caption{Energy distribution for the testing, initial training
      dataset and the enhanced datasets by temperature for \textit{Set2}.}
    \label{sifig:ener_dist_set2_temp}
\end{figure}

Energy distributions of the
iRDs, aRDs and target databases were compared. This analysis provides
information on how the added structures reduce the distributional
differences between the target and aDBs. Starting with {\it Set1} and
{\it Set2}, data augmentation resulted in bimodal energy
distributions, see Figures \ref{sifig:ener_dist_set1_temp} and
\ref{sifig:ener_dist_set2_temp} with peaks at high energies ($> -40$
eV). A quantitative picture of the changes in the energy distributions
can be obtained from the Wasserstein distance $W_1$, see Equation
\ref{eq:wass_1d} and Figure \ref{sifig:wd_temp}. For {\it Set1a/b},
the effect of the molecule used for augmenting the RDs heavily depends
on the composition of the iRDs. Adding acetylene conformations
increased the value of $W_1$ by 0.2 units for \textit{Set1b}, but no
noticeable change was found for \textit{Set1a}. With ethane used for
augmentation, on the other hand, $W_1$ reduces on average by 0.4 units
for \set{1a} whereas the reductions are negligible for \set{1b}.  A
possible explanation for these observations is that the ethane samples
of high energy are closer to the target distribution for
\textit{Set1} because of the contraction/elongation of the \chemfig{C-C} bond. This
explanation is consistent with the reduction in $W_1$ distance and the
position of the second peak of the energy distributions of the aDBs
for \textit{Set1}-Etha.\\

\begin{figure}
    \centering
    \includegraphics[width=0.9\textwidth]{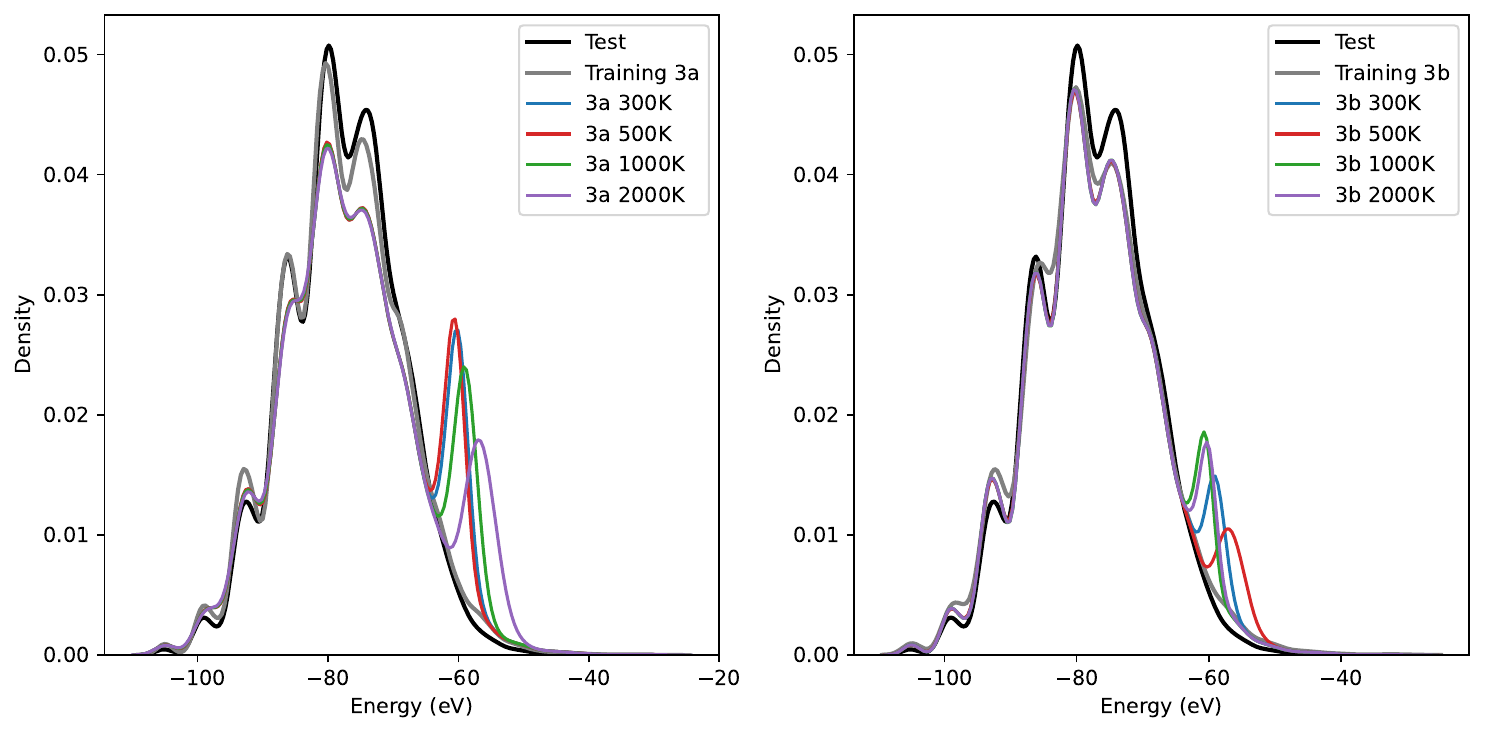}
    \caption{Energy distribution for the testing, initial training
      dataset and the enhanced datasets by temperature for \textit{Set3}.}
    \label{sifig:ener_dist_set3_temp}
\end{figure}

\noindent
For \textit{Set2}, the energy distributions changed little, such as
the appearance of a second peak at $\approx -20$ eV and small changes
near the centre of mass of the distribution (Figure
\ref{sifig:ener_dist_set2_temp}). Nevertheless, the value of $W_1$ for
\set{2} reduces as a function of the composition of the iRDs. The energy distribution $P(E)$ for {\it Set3},
see Figure \ref{sifig:ener_dist_set3_temp}, is bimodal with a second
peak near $-60$ eV, which shifts as a function of temperature. For
both subsets of \textit{Set 3}, the value of $W_1$ for the aDBs
increases by $\approx 1$ unit for \set{3a} and $\approx 0.25$ unit for
\set{3b} to the value of $W_1$ for iDBs. For \set{4}-Benz/Chex $W_1$ decreases. As a general finding it is noted that the values of $W_{1}$ depend insignificantly on the sampling temperature.\\

\subsection{Analysis of differences in energy prediction.}
\subsubsection{The effect of temperature}
\begin{figure}
    \centering
    \includegraphics[width=0.9\textwidth]{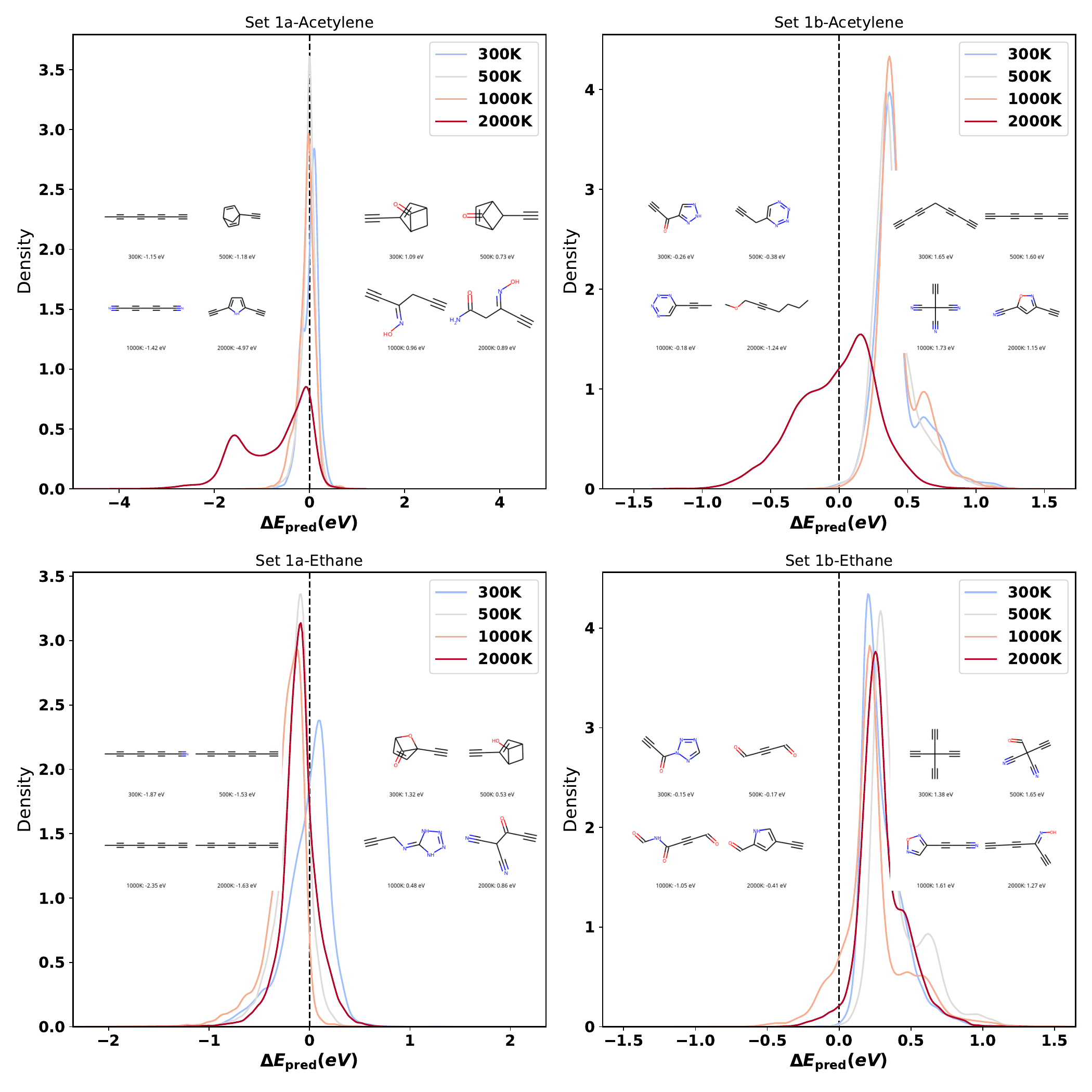}
    \caption{Distribution of change in predicted energy to the
      temperature ($\Delta E = E_{0}- E_{T}$, here
      $T\in\{300,500,1000,2000\}$K ) for the datasets of \textit{Set1}. Each
      panel shows the molecule with the largest decrease or increase
      in $\Delta E$ for the different temperatures}
    \label{sifig:dist_err_set1_temp}
\end{figure}

\noindent
Complementary to the analysis of $f_{\uparrow/\downarrow}$ in the main text, the distribution of changes
in the predicted energy ($\Delta E_{\rm pred} = E^{\rm
  Pred}_{0}-E^{\rm Pred}_{T}$) was determined for \set{1}; see Figure
\ref{sifig:dist_err_set1_temp}. For \set{1b} the center of mass of
$P(\Delta E_{\rm pred})$ shifts to positive values. This indicates
that for most of the molecules the predicted energy decreases. On the
contrary for \set{1a}-Acet $P(\Delta E_{\rm pred})$ is centered around
0 or shifted by a few eV to positive values with the notable exception
of $T=2000$ K for which the distribution becomes bimodal. For
\set{1a}-Etha, the centre of mass of $P(\Delta E_{\rm pred})$ shifts to
negative values, implying that the predicted energy increases in
comparison with the initial values.  Lastly, we also identify the
molecules which suffer the largest increases or decreases in $\Delta
E_{\rm pred}$. Samples with the largest changes in $\Delta E_{\rm
  pred}$ after augmentation feature multiple triple
bonds. Interestingly, opposite effects were observed for \set{1a} and
\set{1b}. In the first, the predicted energy for molecules with more
triple bonds increases (i.e.  $\Delta E_{\rm pred} < 0$), while for the
second it is reduced ($\Delta E_{\rm pred} > 0$). \\
 
\begin{figure}
    \centering
    \includegraphics[width=\textwidth]{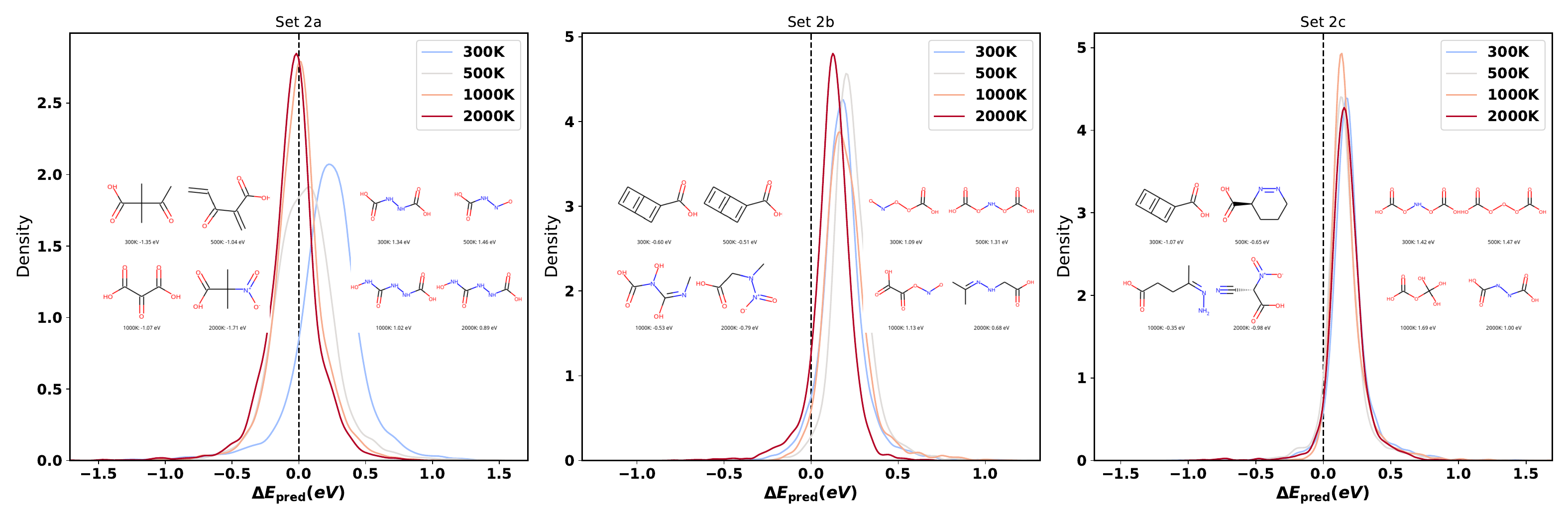}
    \caption{Distribution of change in predicted energy to the
      temperature ($\Delta E = E_{0}- E_{T}$, here
      $T\in\{300,500,1000,2000\}$K ) for the datasets of \textit{Set2}. Each
      panel shows the molecule with the largest decrease or increase
      in $\Delta E$ for the different temperatures.}
    \label{sifig:dist_err_set2_temp}
\end{figure}

\noindent
 Analysis of the $P(\Delta E_{\rm pred})$ (Figure
\ref{sifig:dist_err_set2_temp}) shows that $E_{\rm pred}$ reduces for
\textit{Set2a} at 300 and 500 K, with a slight shift of the maximum of
$P(\Delta E_{\rm pred})$ at higher temperatures. Conversely, both
subsets \textit{Set2b} and \textit{Set2c} shift the center of mass of
$P(\Delta E_{\rm pred})$ towards positive values of $\Delta E_{\rm
  pred}$ across all temperatures. The molecular structure of the
samples with largest decreases in predicted energy contains
\chemfig{N-O}, \chemfig{N-N} or \chemfig{O-O} bonds. On the other
hand, the most significant decreases in $E_{\rm pred}$ feature
\chemfig{C=C}, \chemfig{C=N} and a six-carbon ring.\\

\begin{figure}
    \centering
    \includegraphics[width=0.9\textwidth]{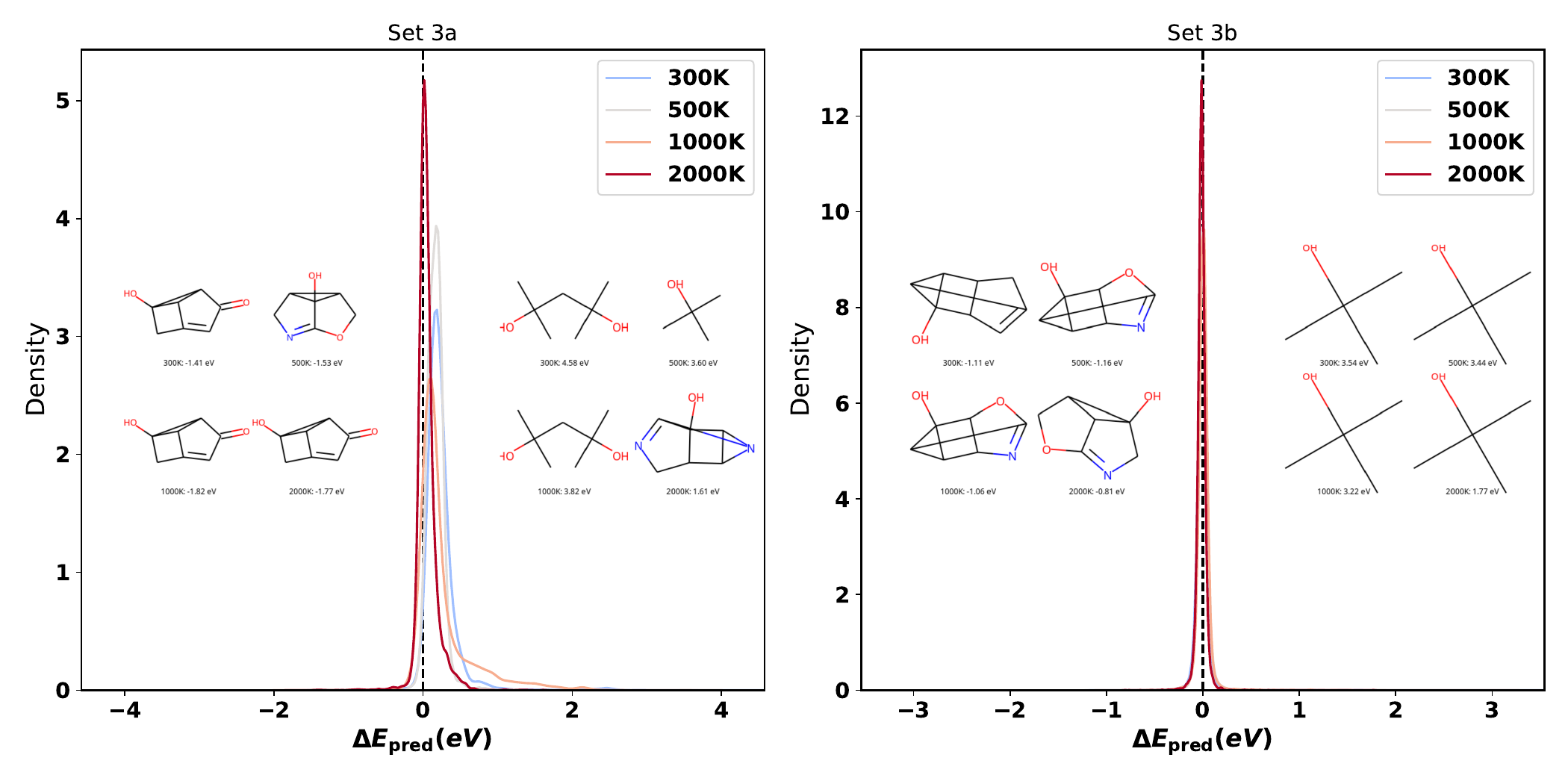}
    \caption{Distribution of change in predicted energy to the
      temperature ($\Delta E = E_{0}- E_{T}$, here
      $T\in\{300,500,1000,2000\}$K ) for the datasets of \textit{Set3}. Each
      panel shows the molecule with the largest decrease or increase
      in $\Delta E$ for the different temperatures.}
    \label{sifig:dist_err_set3_temp}
\end{figure}

\noindent
 The distributions $P(
\Delta E_{\rm pred})$ for both subsets are sharply peaked around zero
at all temperatures (Figure \ref{sifig:dist_err_set3_temp}). However,
for \textit{Set3a}, the RD with the best performance, there is a
slight displacement of the distribution towards positive values with
pronounced tails.  Molecules with large negative $\Delta E_{\rm
  pred}$, (-1.4 to -1.8 eV for \set{3a} and -1.1 to -0.8 for \set{3b})
in \textit{Set3} typically comprise structures with multiple bridged
rings, whereas those with large positive values exhibit simpler
structures.\\

\begin{figure}
    \centering
    \includegraphics[width=0.9\textwidth]{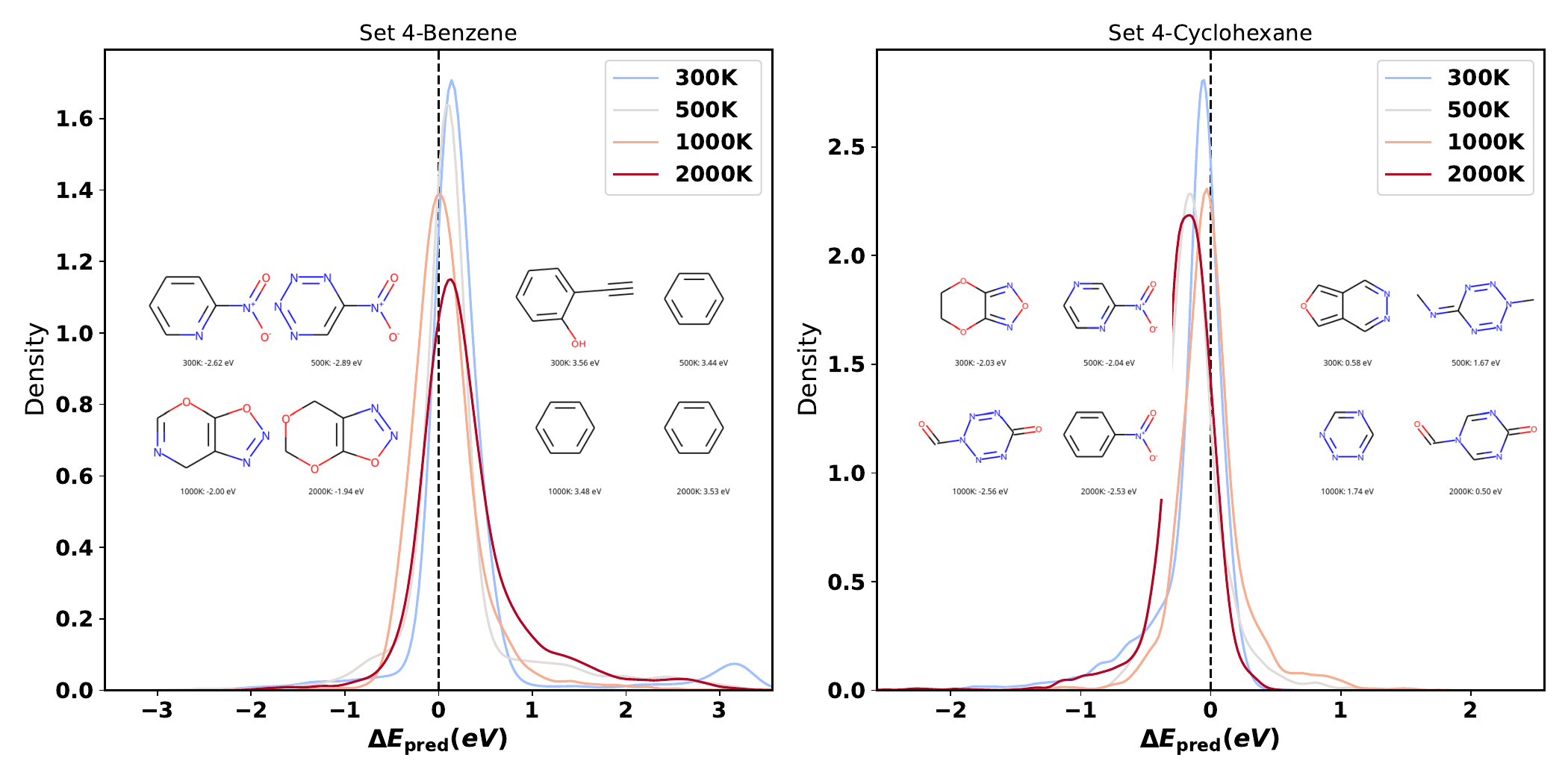}
    \caption{Distribution of change in predicted energy to the
      temperature ($\Delta E = E_{0}- E_{T}$, here
      $T\in\{300,500,1000,2000\}$K ) for the datasets of \textit{Set4}. Each
      panel shows the molecule with the largest decrease or increase
      in $\Delta E$ for the different temperatures.}
    \label{sifig:dist_err_set4_temp}
\end{figure}

\noindent
Lastly, the distribution of $\Delta E_{\rm pred}$ highlights the opposing trends
observed for \textit{Set4} with the different molecules (benzene and
cyclohexane) used for the augmentation (Figure
\ref{sifig:dist_err_set4_temp}).  The distribution of $P(\Delta E_{\rm
  pred})$ for \set{4}-Benz.  shifts towards positive values with tails
extending up to 3 eV.  Conversely, for \set{4}-Chex., $P(\Delta E_{\rm
  pred})$ shifts to negative values of $\Delta E_{\rm pred}$. The
effect of temperature is reflected in changes in the width of
$P(\Delta E_{\rm pred})$ and its tails, which grow as the temperature
increases.  The structure of the molecules with the largest increases
of $E_{\rm pred}$ in \textit{Set4} contain multiple heteroatoms
organized in bicycles or feature the presence of the nitro group.
Meanwhile, structures which reduce $E_{\rm pred}$ are usually single
aromatic rings.\\

\subsubsection{The effect of the number of samples}

\begin{figure}
    \centering
    \includegraphics[width=0.9\textwidth]{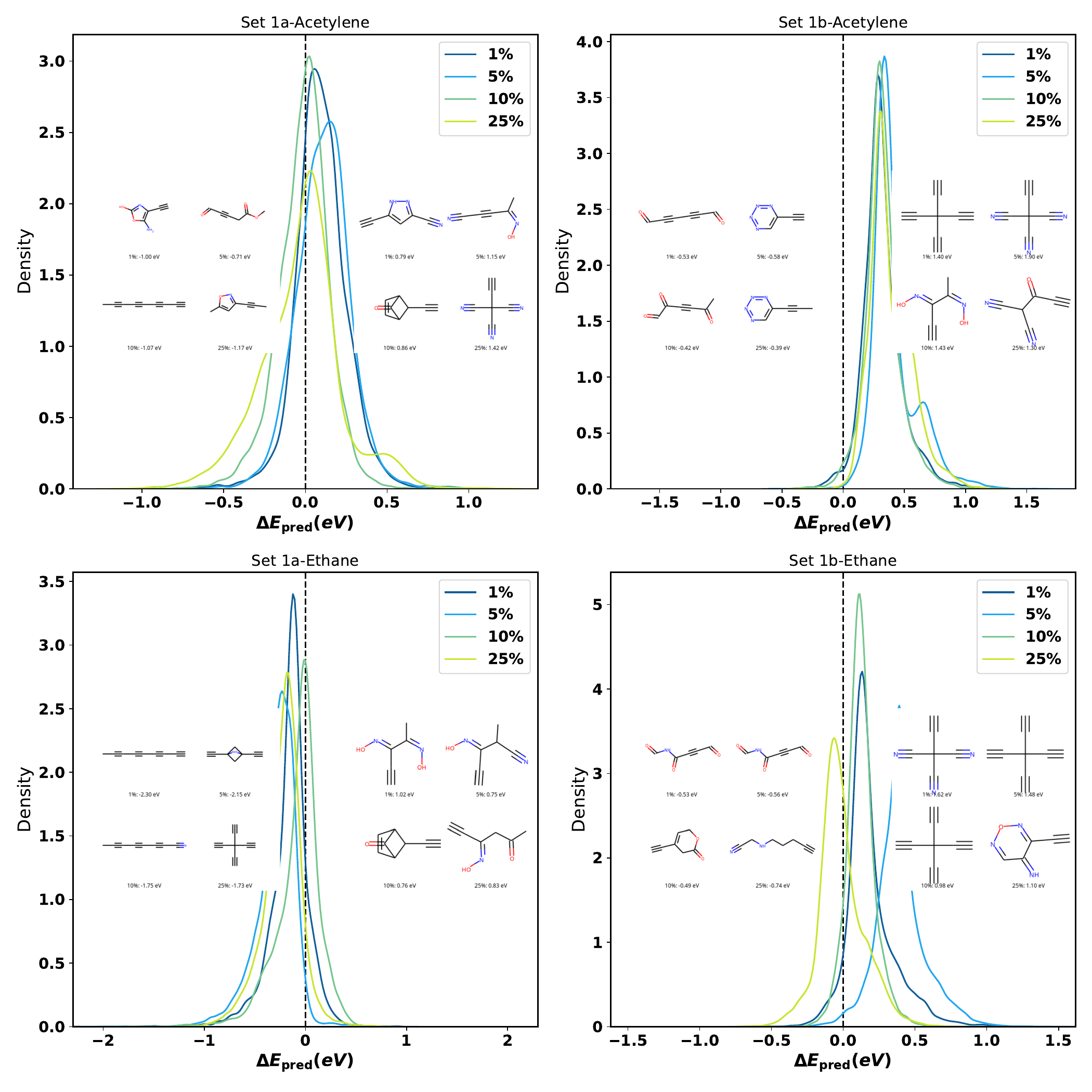}
    \caption{Distribution of change in predicted energy to the
      percentages of samples added ($\Delta E = E_{0}- E_{i}$, here $i
      \in [1,5,10,25]$ \%) for \textit{Set1}. Each panel shows the molecule
      with the largest decrease or increase in $\Delta E$ for the
      different percentages.}
    \label{sifig:dist_err_set1_large}
\end{figure}

\noindent
Changes in the predicted energy ($\Delta E_{\rm pred}$) for
\textit{Set1} (Figure \ref{sifig:dist_err_set1_large}) underscore
variations induced by the percentage of samples added.  Across all
variants, except for \set{1a}-Etha, there is an overall mean decrease
in predicted energy (i.e., $\Delta E_{\rm pred}>0$). Notably,
\set{1a}-Acet initially exhibits positive $\Delta E_{\rm pred}$ values
after adding a few samples (i.e., 1\% and 5\%), shifting towards zero
thereafter. Similarly, \set{1a}-Etha consistently displays $\Delta
E_{\rm pred}<0$ across different augmentation levels. For
\set{1b}-Acet, there is a constant positive $\Delta E_{\rm pred}$
centered at approximately 0.5 eV for all percentages tested. The case
of \set{1b}-Etha is particularly intriguing, with varying positions of
$\Delta E_{\rm pred}$ centre.  Notably, at 5\% augmentation, the
distribution shows the largest shift with a centre at 0.7 eV, while at
25\%, the centre shifts to negative values at approximately -0.2
eV. The chemical structures with significant decreases or increases in
predicted energy lack a clear trend. In \set{1a}, decreased predicted
energy is associated with structures featuring an oxazole ring or
multiple triple bonds, while increased $E_{\rm pred}$ is observed for
compounds with a \chemfig{C=N-OH} moiety or multiple cyanide
(\chemfig{C~N}) fragments. Conversely, in \set{1b}, $\Delta E_{\rm
  pred}>0$ is observed for molecules with one carbon centre
substituted by four \chemfig{CH_2-C~CH} or the \chemfig{C=N-OH}
fragment, while negative values are seen for molecules with a
\chemfig{C=O} fragment or a formyl-acetamide fragment
\chemfig{O=C-NH-C=O}.\\

\begin{figure}
    \centering
    \includegraphics[width=0.9\textwidth]{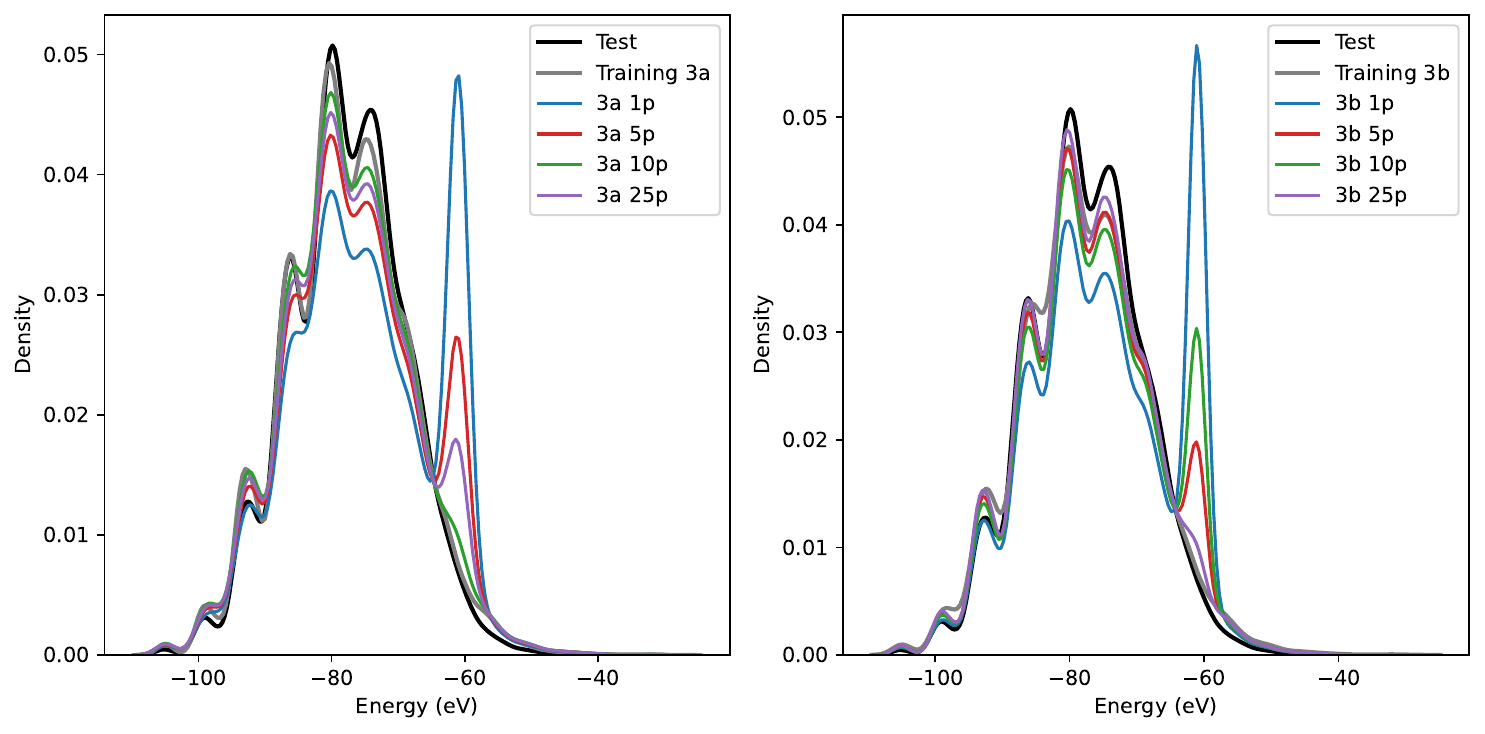}
    \caption{Energy distribution for the testing, initial training
      dataset and the enhanced datasets by different percentages of
      added molecules for \textit{Set3}.}
    \label{sifig:ener_dist_set3_large}
\end{figure}

\begin{figure}
    \centering
    \includegraphics[width=0.9\textwidth]{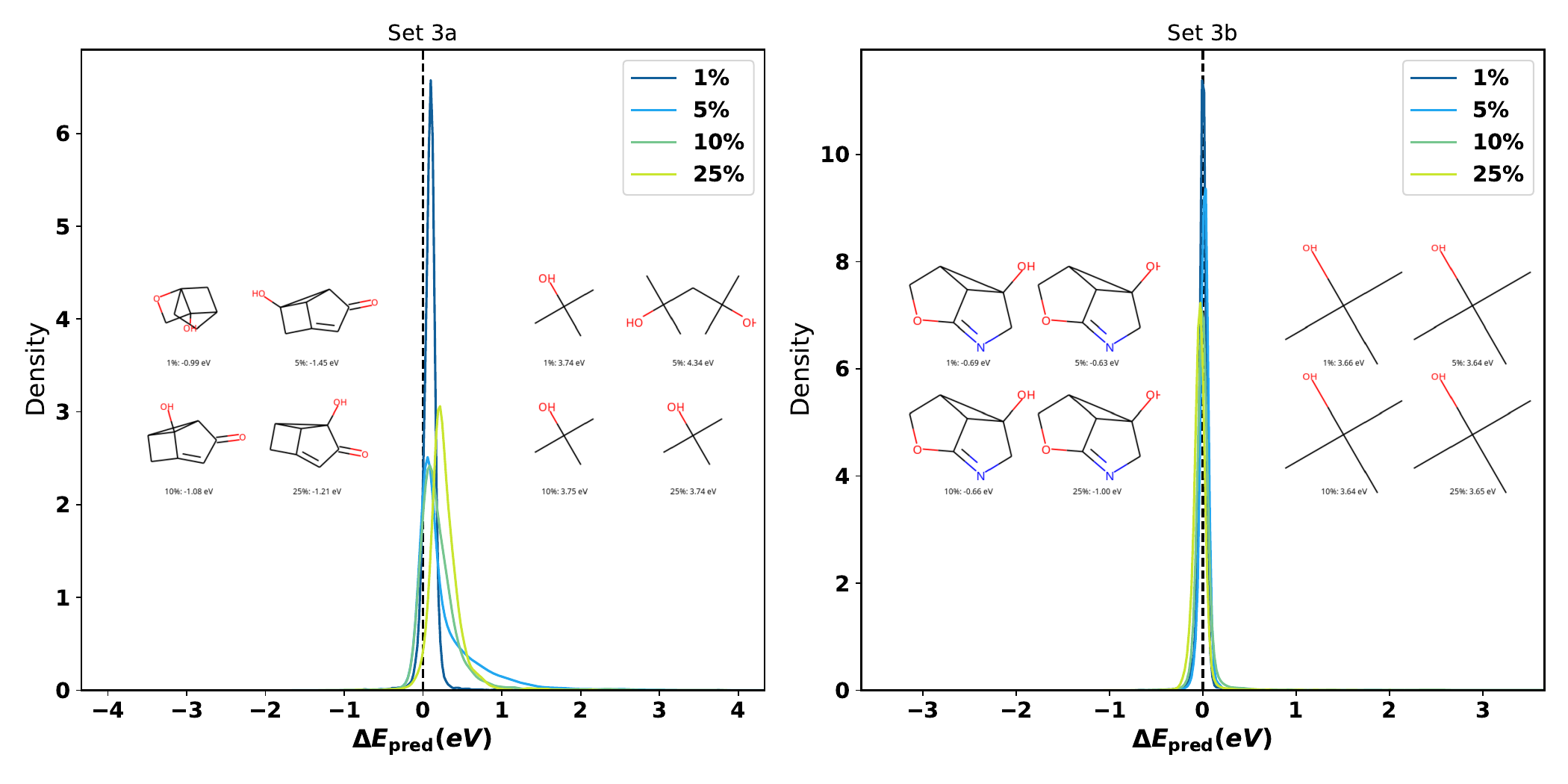}
    \caption{Distribution of change in predicted energy to the
      percentages of samples ($\Delta E = E_{0}- E_{i}$, here $i\in
      \{1,5,10,25\}$\%) for the datasets of \textit{Set3}. Each panel shows the
      molecule with the largest decrease or increase in $\Delta E$ for
      the different percentages.}
    \label{sifig:dist_err_set3_large}
\end{figure}

Results for the distributions of $\Delta E_{\rm pred}$ of \set{2} generally shift towards
positive values for most tested scenarios, except for \set{2a} at low
percentages (1\% and 5\%). Regarding chemical structures, they closely
resemble those observed for temperature effect, characterized by the
presence of numerous heteroatoms (O, N) and \chemfig{C=O}
fragments.  Moving to \set{3}, changes in $\Delta E_{\rm pred}$ are illustrated in Figure
\ref{sifig:dist_err_set3_large}. In \set{3a}, the tails of $P(\Delta
E_{\rm pred})$ shift towards positive values, accompanied by an
increase in the width of $P(\Delta E_{\rm pred})$. These changes
appear to align with the observed trend in energy distribution (see
Figure \ref{sifig:ener_dist_set3_large}) rather than the number of
added samples. Conversely, \set{3b} exhibits a $P(\Delta E_{\rm
  pred})$ centered at 0 eV, with alterations primarily observed in the
distribution's height. The structures of molecules displaying large
$\Delta E_{\rm pred}$ remain consistent with those observed for the temperature effect.\\

\begin{figure}
    \centering
    \includegraphics[width=0.9\textwidth]{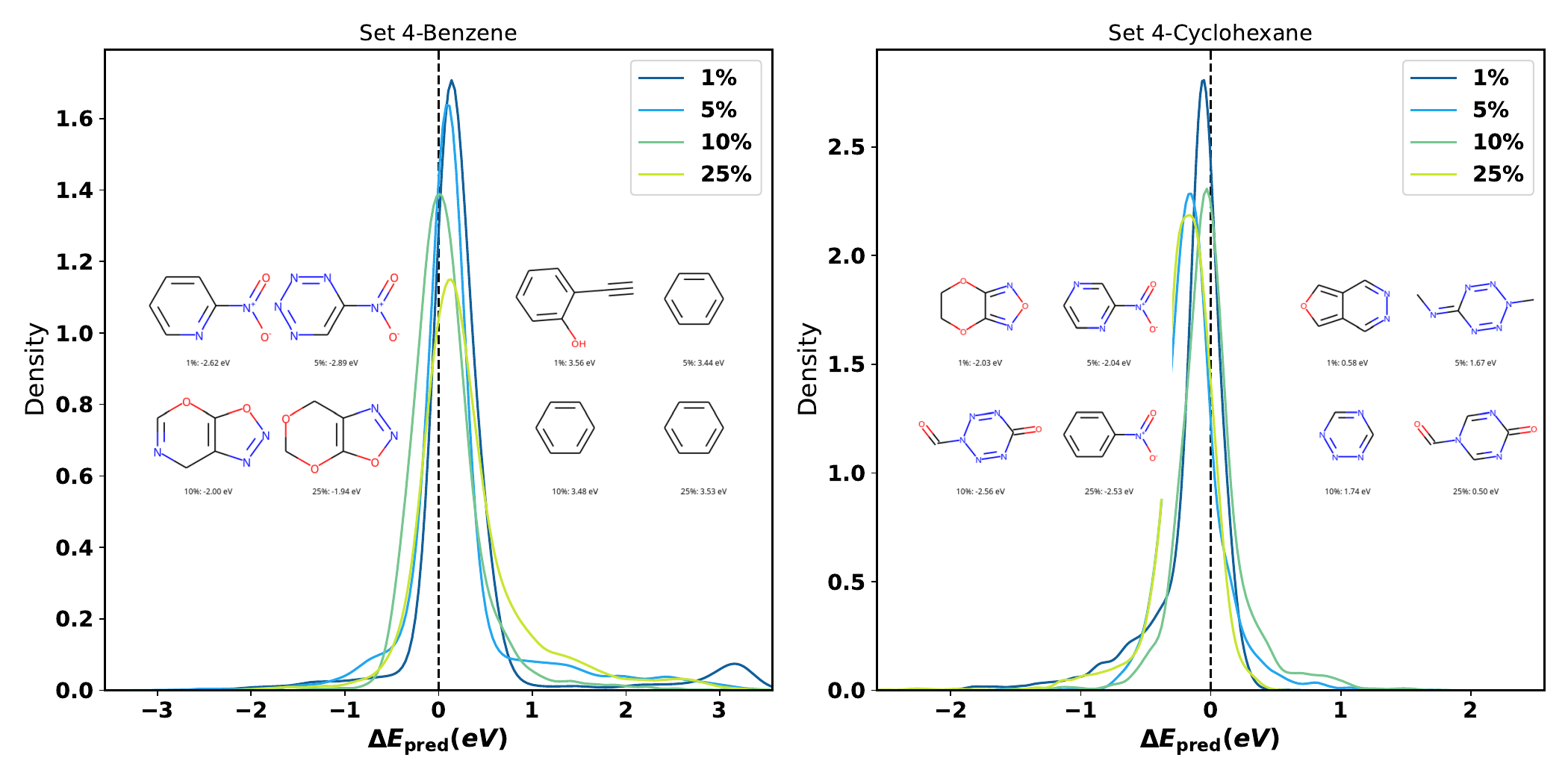}
    \caption{Distribution of change in predicted energy to the
      percentages of samples ($\Delta E = E_{0}- E_{i}$, here $i\in
      \{1,5,10,25\}$\%) for the datasets of \textit{Set4}. Each panel shows the
      molecule with the largest decrease or increase in $\Delta E$ for
      the different percentages.}
    \label{sifig:dist_err_set4_large}
\end{figure}

Lastly,  the distributions of
$P(\Delta E_{\rm pred})$ for \textit{Set4} (see Figure
\ref{sifig:dist_err_set4_large}) reveal contrasting outcomes for
augmentation with benzene and cyclohexane. Benzene enrichment results
in reduced energy predictions with positive values of $\Delta E_{\rm
  pred}$, centering the distribution's mass at larger positive values
for small addition percentages. In contrast, \textit{Set4} enriched
with cyclohexane shows distributions centered at negative values, with
the mass centering at more negative values for small addition
percentages. Molecules exhibiting significant changes in $ E_{\rm
  pred}$ commonly feature fused rings with heteroatoms and nitro
(\chemfig[atom sep=2em]{O-[,,,,lddbond={+}]N-[,,,,lddbond={+}]O})
fragments.

\clearpage

\section{Supplementary Tables}
\begin{table}[h]
\caption{Statistical summary of the performance of the initially
  generated databases on its test set used for training.}
\begin{tabular}{c|c|c}
Subset & MAE(kcal/mol) & RMSE(kcal/mol) \\ \hline
1a      & 0.3918        & 0.6908         \\
1b      & 0.4264        & 0.8564         \\
2a      & 0.4636        & 0.7792         \\
2b      & 0.5044        & 0.8868         \\
2c      & 0.517         & 0.8725         \\
3a      & 0.4379        & 0.6599         \\
3b      & 0.4138        & 0.7649         \\
4       & 0.5181        & 0.9275        
\end{tabular}
\label{sitab:init_perf}
\end{table}

\begin{table}[h]
\caption{Number of samples added to the database as a percentage of
  the total number of samples used for training the different
  databases. Note: For the 25\% of \textit{Set3b}, only the number of
  converged molecules was used.}
\label{sitab:number_of_samples}
\begin{tabular}{c|c|c|c|c}
Dataset & 1 \% & 5 \% & 10 \% & 25 \% \\ \hline
1 and 2    &  250      &  1250 & 2500 & 6250 \\
3a     & 87      & 435 & 870 & 2175  \\
3b      & 206   & 1080 & 2060 & 5138* \\
4      & 125    & 625  & 1250 & 3125 \\
\end{tabular}
\end{table}
\clearpage

\section{Supplementary Figures}

\begin{figure}
    \centering
    \includegraphics{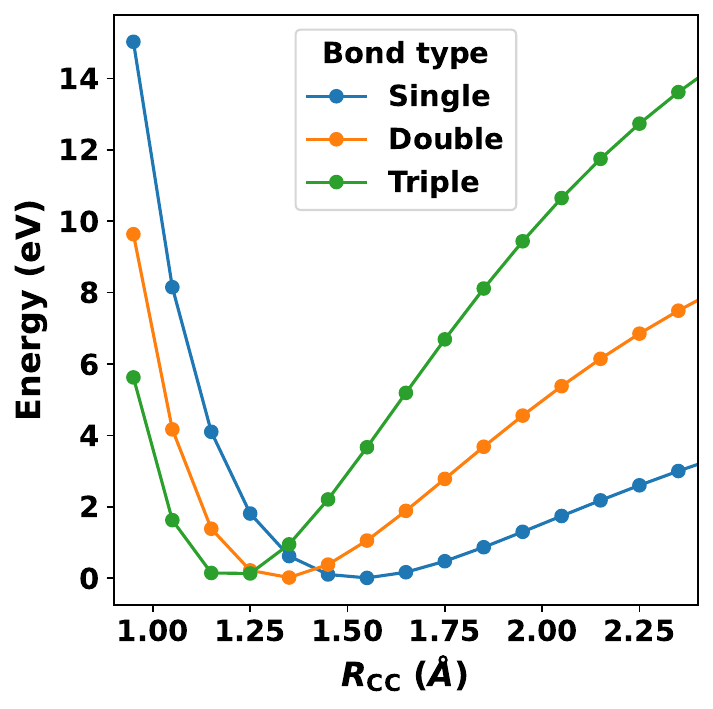}
    \caption{1D potential energy plots for C-C bond on the minimum examples (i.e. ethane, ethylene, and acetylene) at B3LYP/6-31G(2df,p) level. The zero of energy was defined as the energy of the equilibrium geometry for the corresponding molecule.}
    \label{sifig:bonds}
\end{figure}

\begin{figure}[h]
    \centering
    \includegraphics[width=0.9\textwidth]{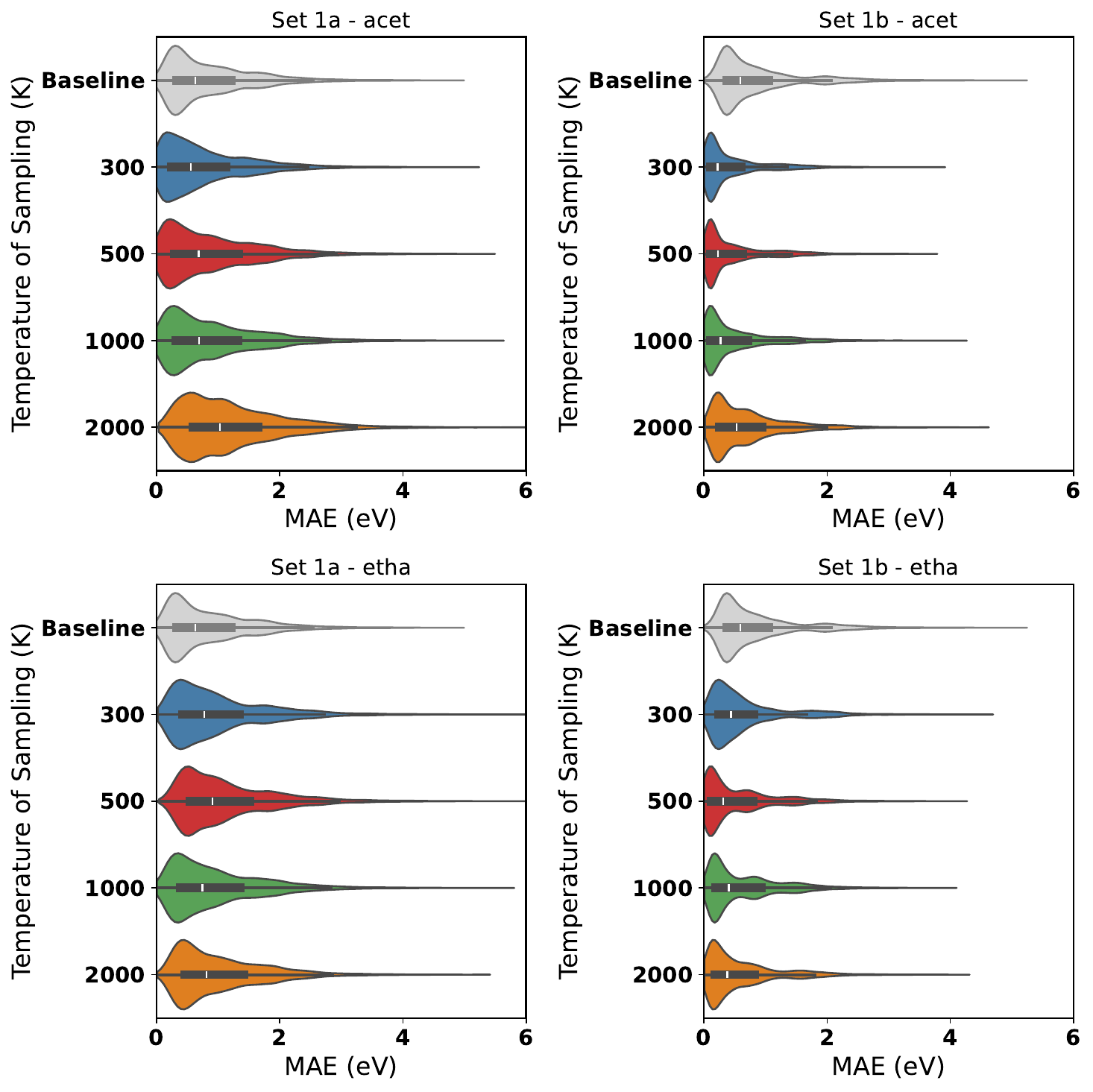}
    \caption{Violin plot of the MAE for the datasets of \textit{Set1} at
      different temperatures.}
    \label{sifig:violin_set1_temp}
\end{figure}

\begin{figure}
    \centering
    \includegraphics[width=0.9\textwidth]{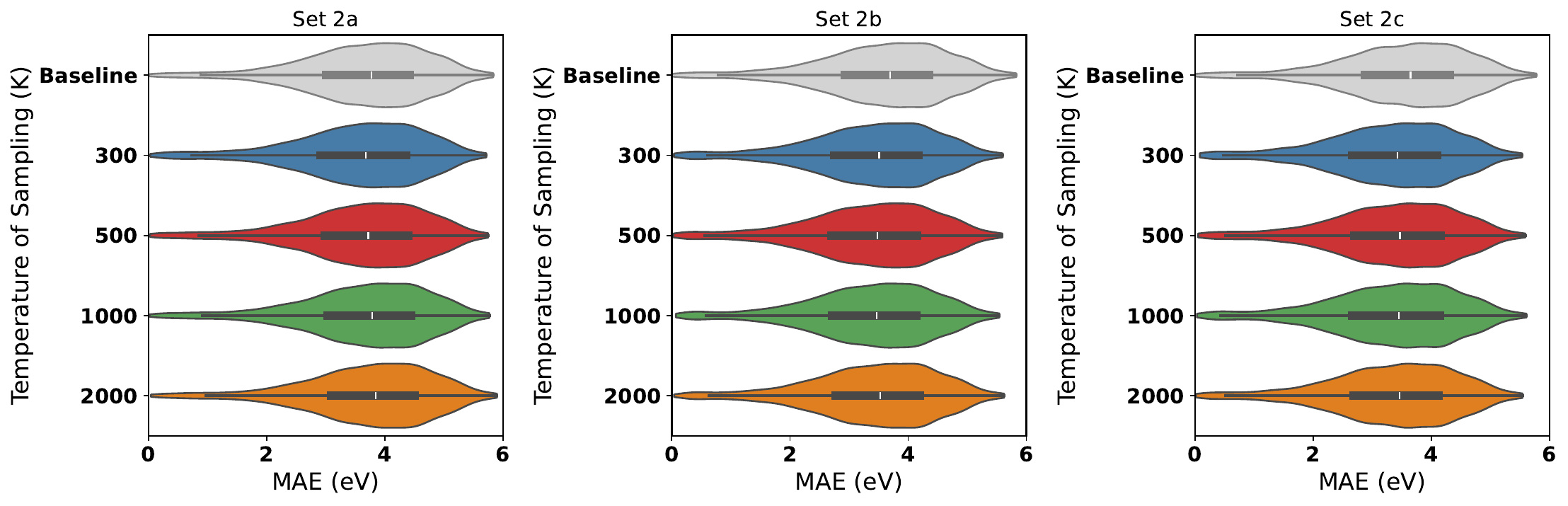}
    \caption{Violin plot of the MAE for the datasets of \textit{Set2} at
      different temperatures.}
    \label{sifig:violin_set2_temp}
\end{figure}

\begin{figure}
    \centering
    \includegraphics[width=0.9\textwidth]{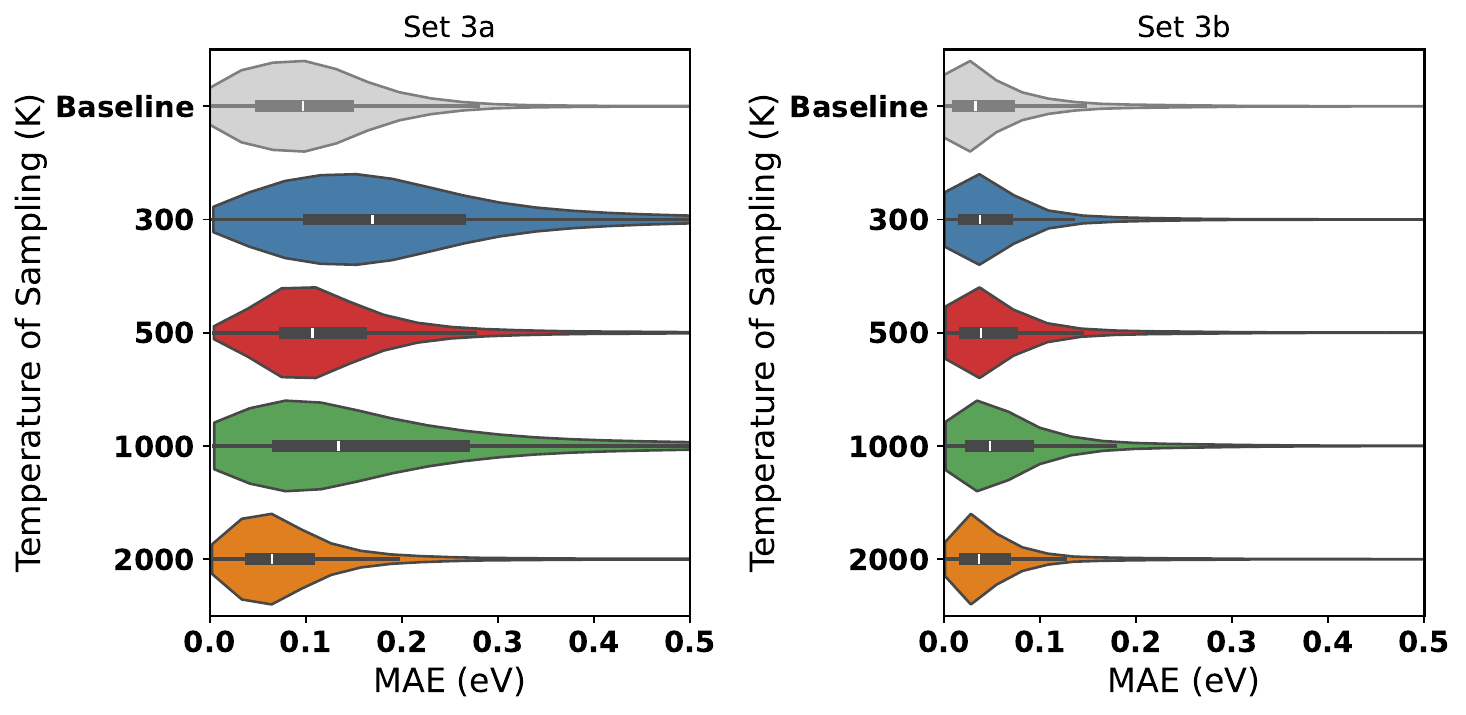}
    \caption{Violin plot of the MAE for the datasets of \textit{Set3} at
      different temperatures.}
    \label{sifig:violin_set3_temp}
\end{figure}

\begin{figure}
    \centering
    \includegraphics[width=0.9\textwidth]{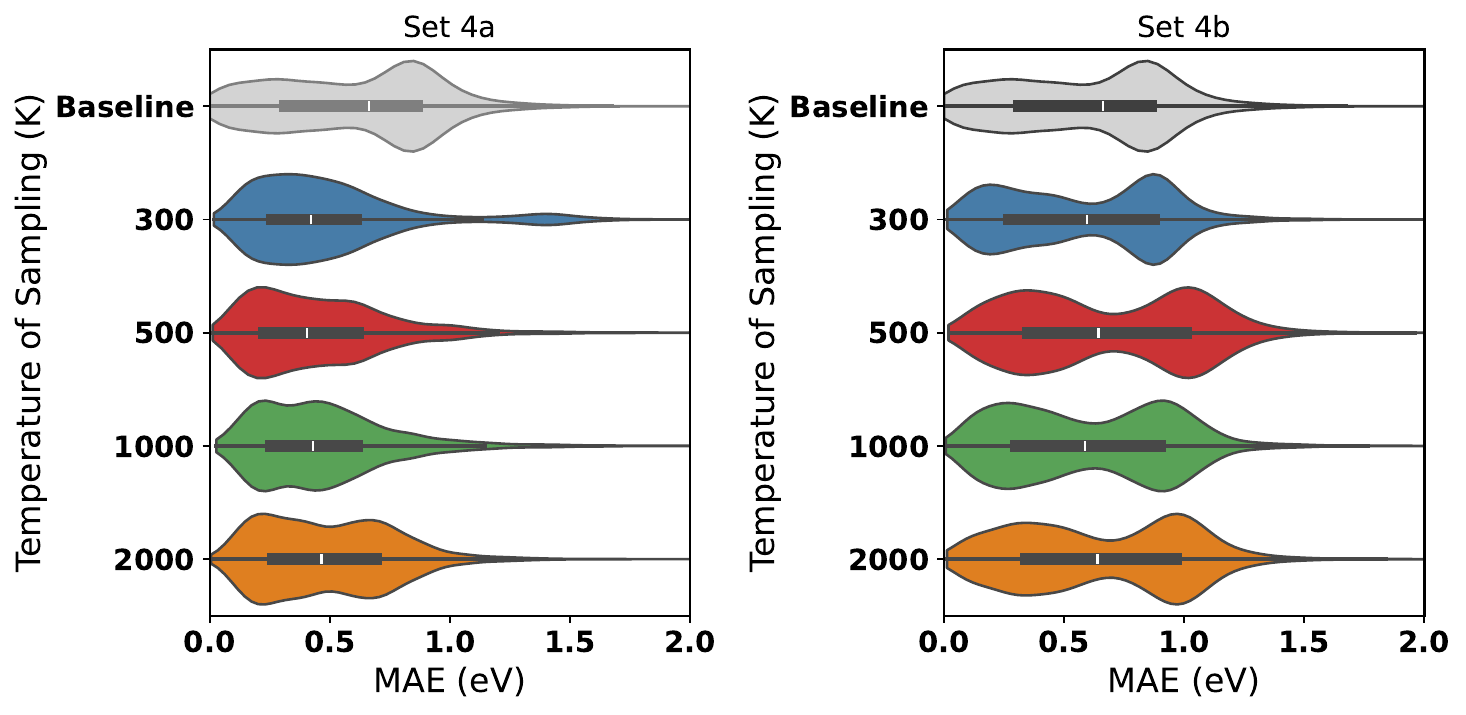}
    \caption{Violin plot of the MAE for the datasets of \textit{Set4} at
      different temperatures.}
    \label{sifig:violin_set4_temp}
\end{figure}

\begin{figure}
    \centering
    \includegraphics[width=0.9\textwidth]{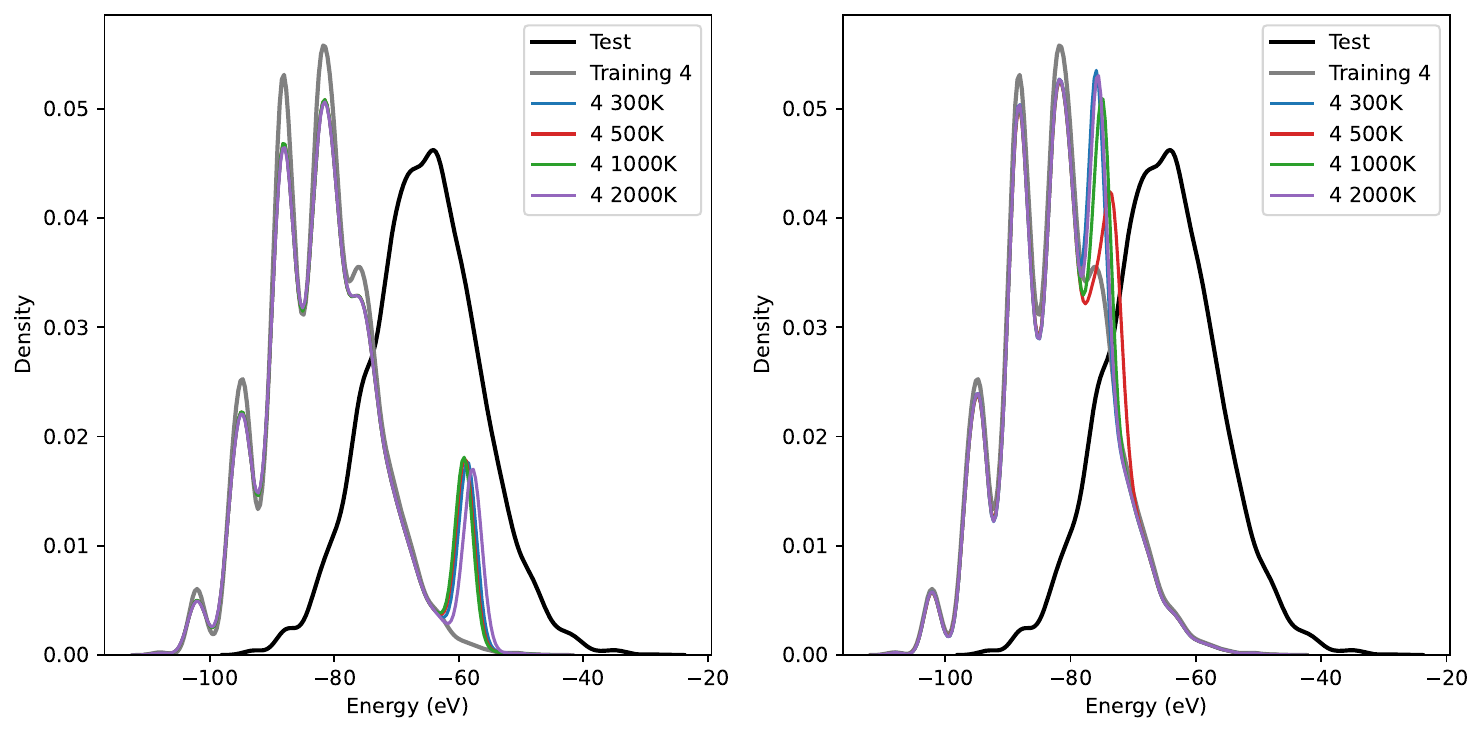}
    \caption{Energy distribution for the testing, initial training
      dataset and the enhanced datasets by temperature for \textit{Set4}.}
    \label{sifig:ener_dist_set4_temp}
\end{figure}

\begin{figure}[h]
    \centering
    \includegraphics[width=0.9\textwidth]{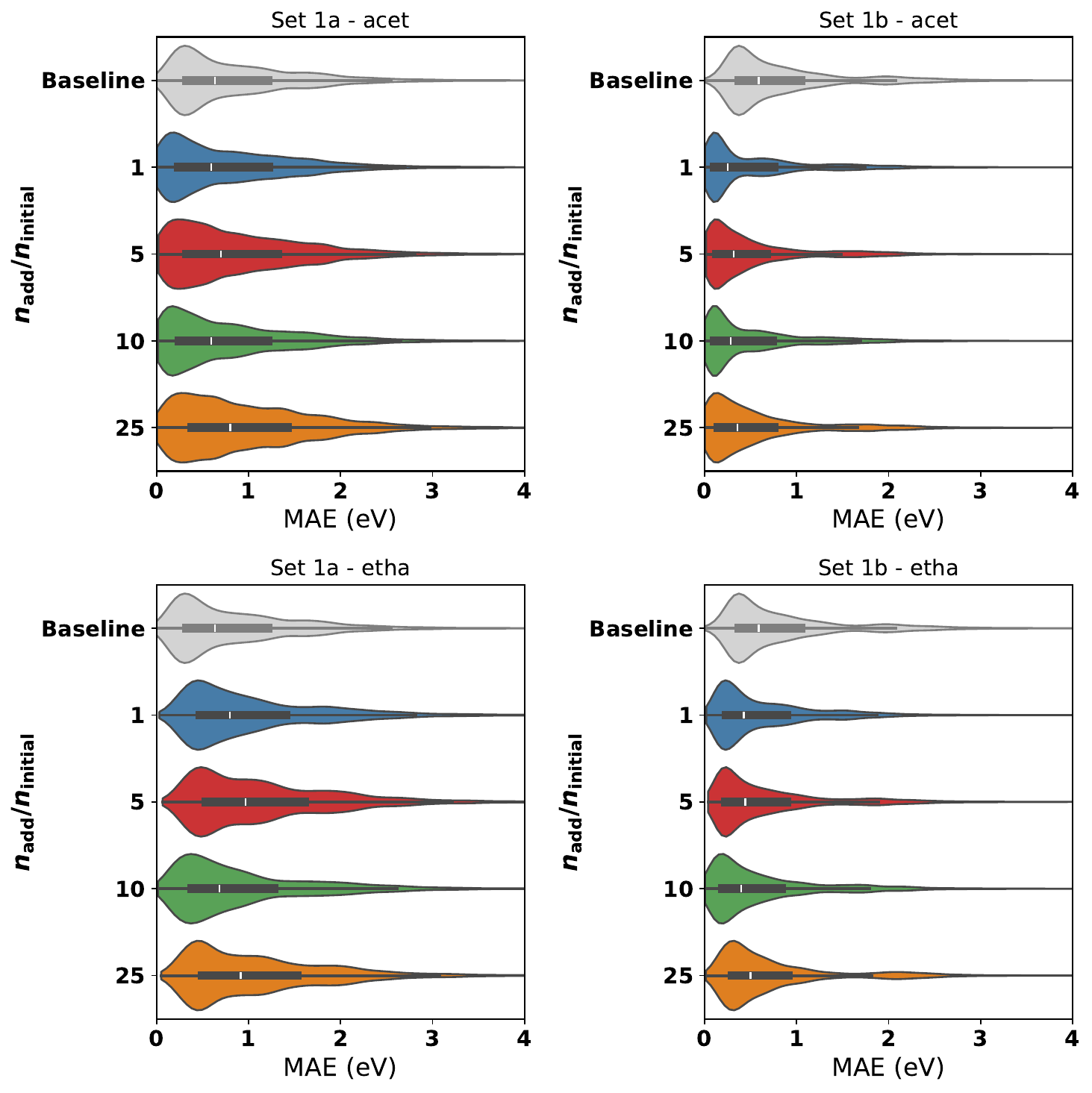}
    \caption{Violin plot of the MAE for the datasets of \textit{Set1} for
      different fractions of added molecules.}
    \label{sifig:violin_set1_large}
\end{figure}

\begin{figure}
    \centering
    \includegraphics[width=0.9\textwidth]{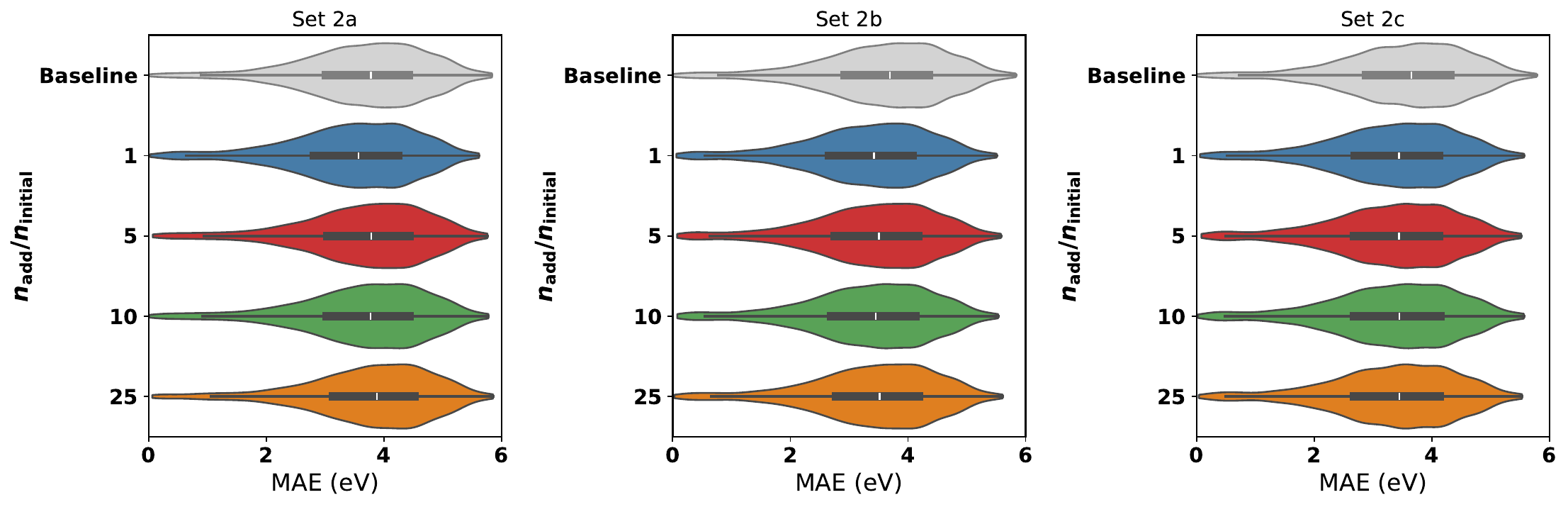}
    \caption{Violin plot of the MAE for the datasets of \textit{Set2} for
      different fractions of added molecules.}
    \label{sifig:violin_set2_large}
\end{figure}

\begin{figure}
    \centering
    \includegraphics[width=0.9\textwidth]{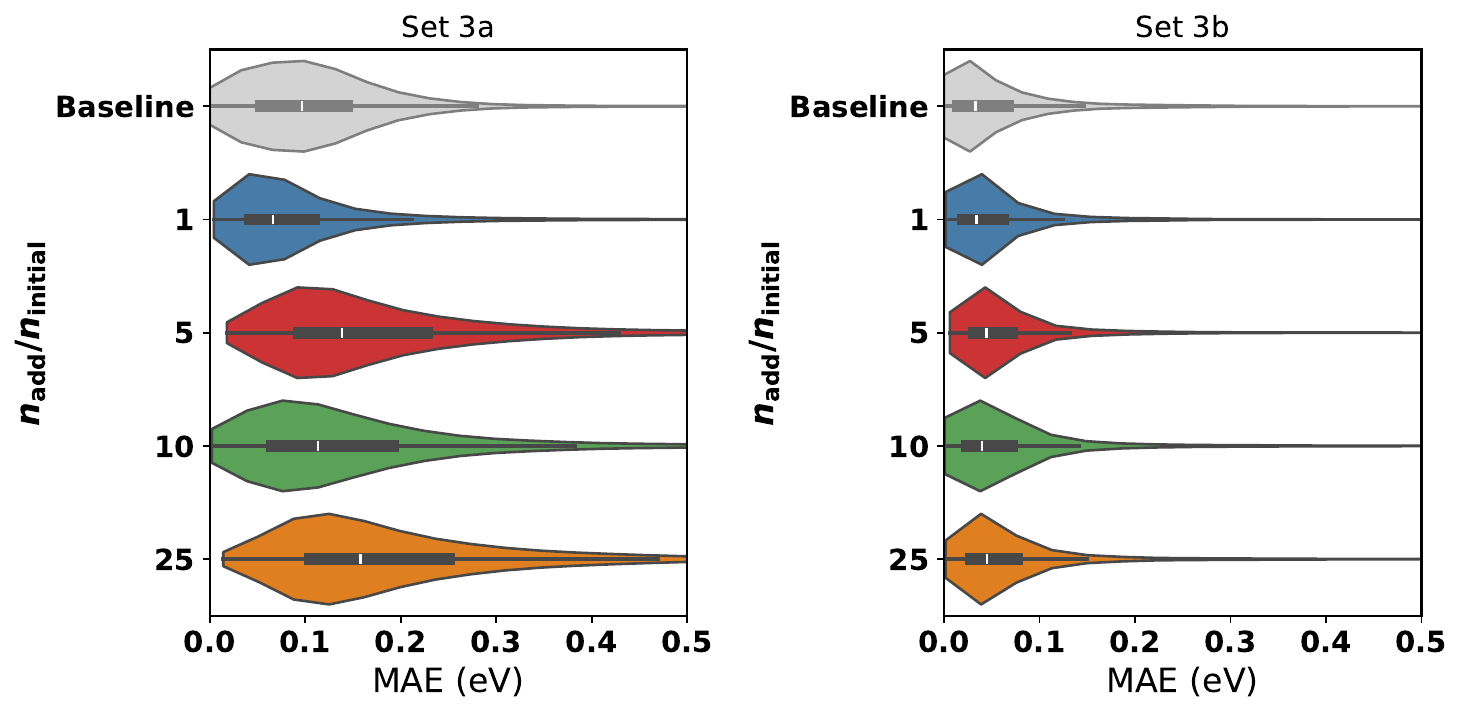}
    \caption{Violin plot of the MAE for the datasets of \textit{Set3} for
      different fractions of added molecules.}
    \label{sifig:violin_set3_large}
\end{figure}

\begin{figure}
    \centering
    \includegraphics[width=0.9\textwidth]{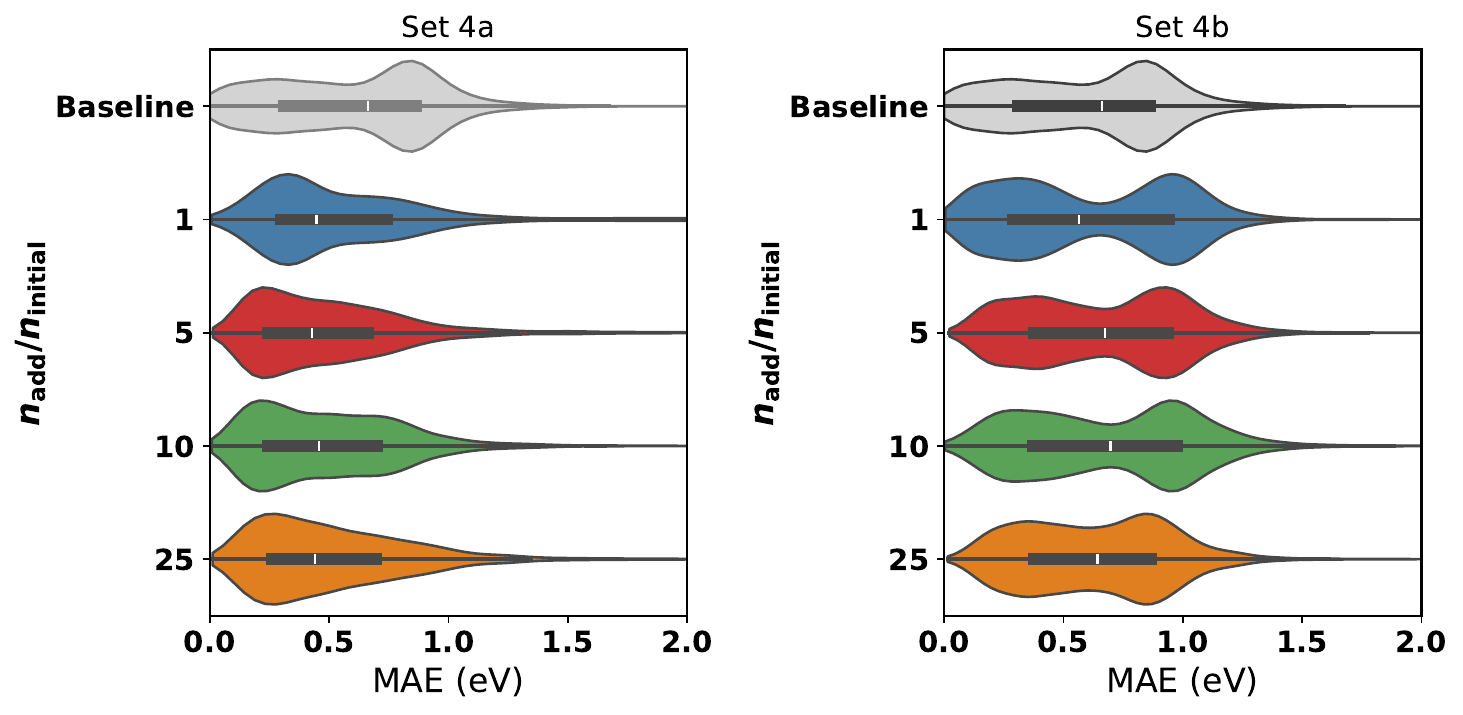}
    \caption{Violin plot of the MAE for the datasets of \textit{Set4} for
      different fractions of added molecules.}
    \label{sifig:violin_set4_large}
\end{figure}

\begin{figure}
    \centering
    \includegraphics[width=0.9\textwidth]{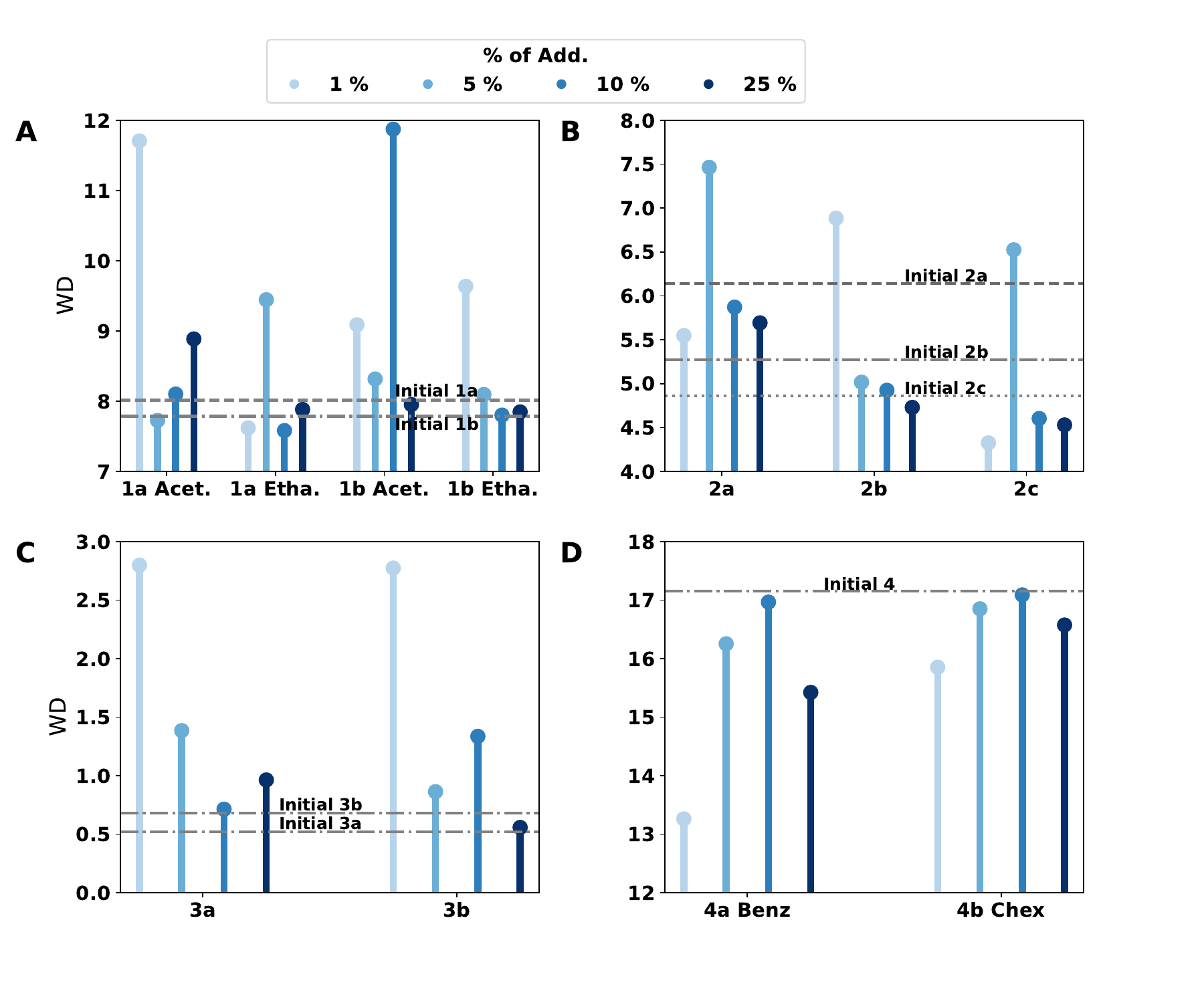}
    \caption{Wasserstein distance between the enhanced and target energy distributions for all the restricted databases studied in this work. A grey line(s) represents the initial distance between training and target distribution in each panel. The scale of the different axes is not uniform to better exemplify the changes in the distances. }
    \label{sifig:wd_perc}
\end{figure}

\begin{figure}
    \centering
    \includegraphics[width=0.9\textwidth]{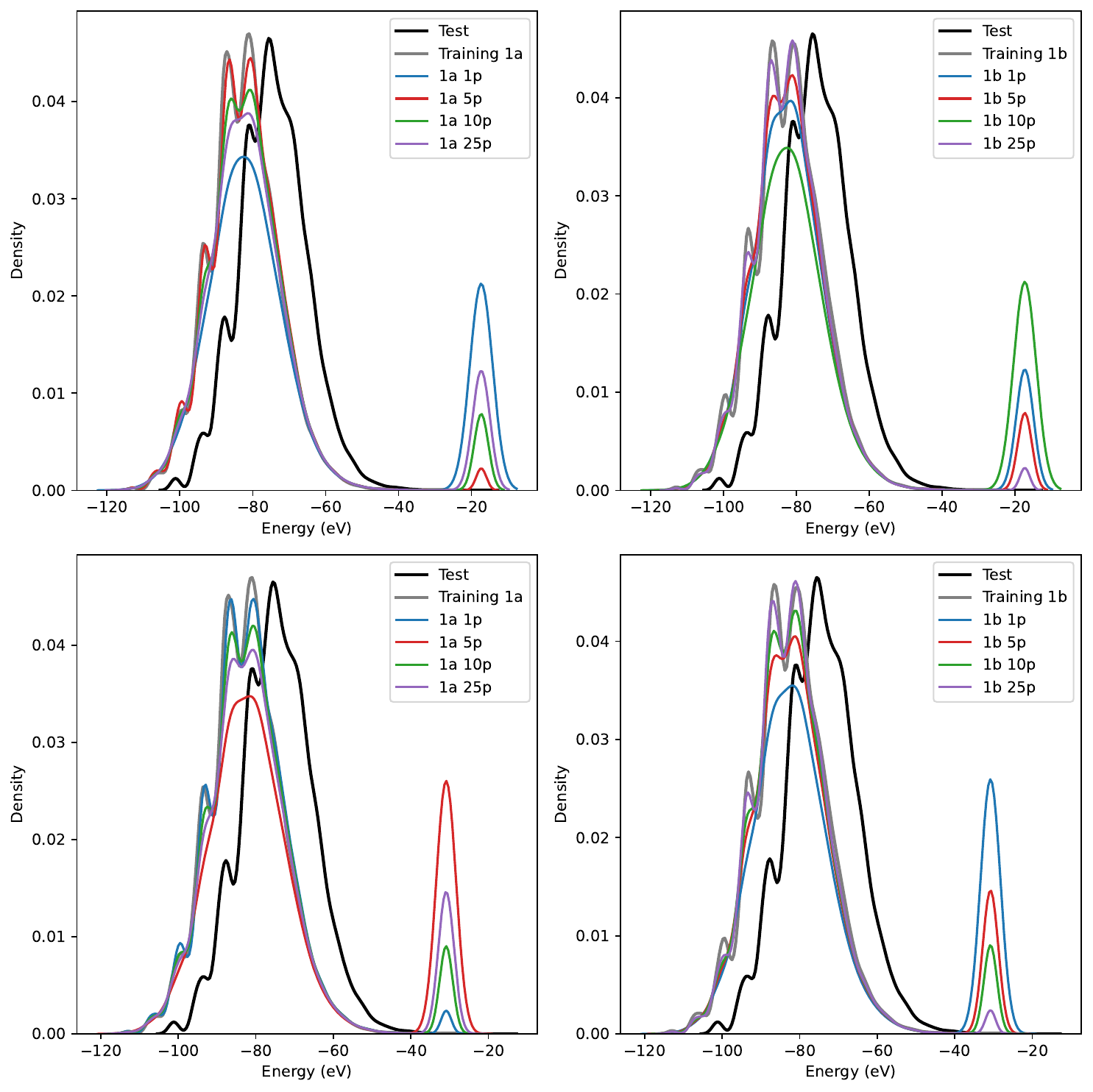}
    \caption{Energy distribution for the testing, initial training
      dataset and the enhanced datasets by different percentages of
      added molecules for \textit{Set1}.}
    \label{sifig:ener_dist_set1_large}
\end{figure}

\begin{figure}
    \centering
    \includegraphics[width=0.9\textwidth]{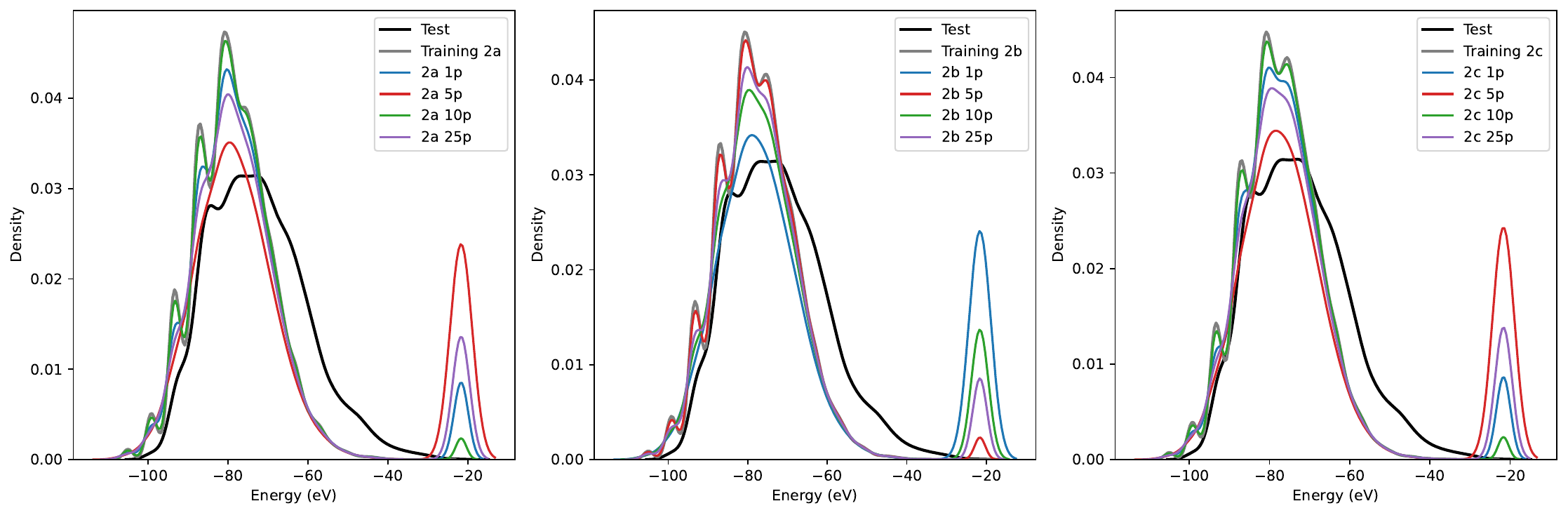}
    \caption{Energy distribution for the testing, initial training
      dataset and the enhanced datasets by different percentages of
      added molecules for \textit{Set2}.}
    \label{sifig:ener_dist_set2_large}
\end{figure}

\begin{figure}
    \centering
    \includegraphics[width=0.9\textwidth]{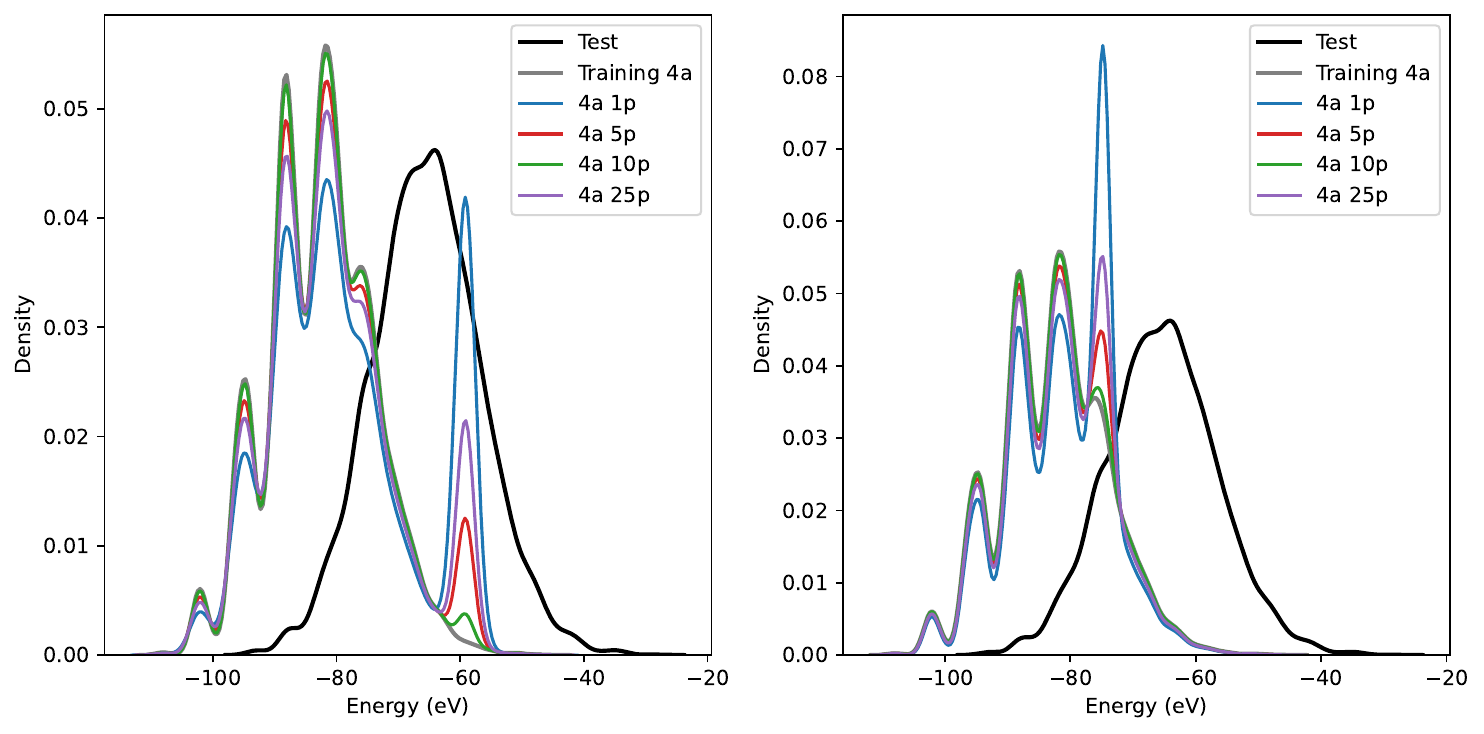}
    \caption{Energy distribution for the testing, initial training
      dataset and the enhanced datasets by different percentages of
      added molecules for \textit{Set4}.}
    \label{sifig:ener_dist_set4_large}
\end{figure}

\begin{figure}
    \centering
    \includegraphics[scale=0.65]{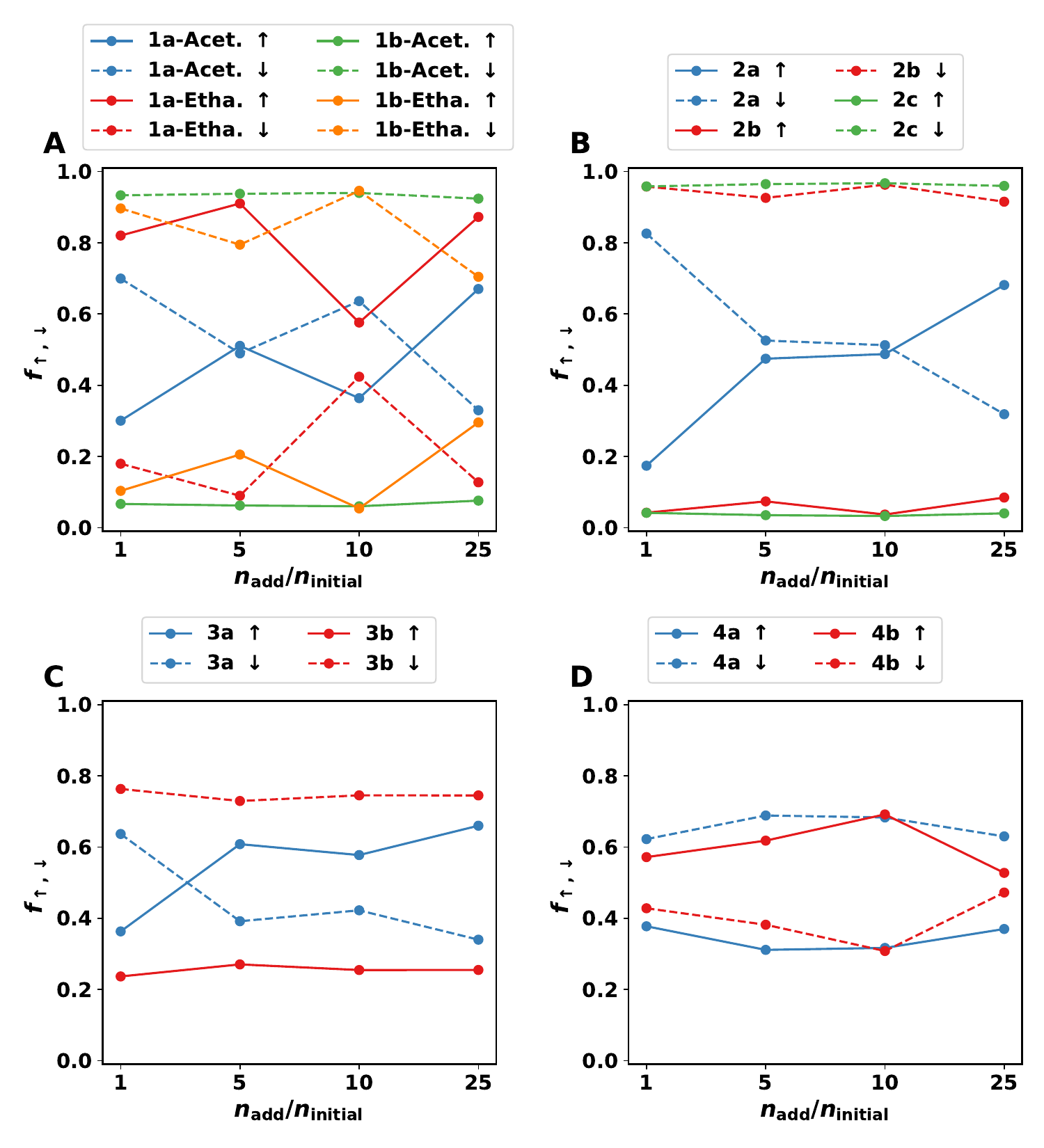}
    \caption{Fraction of samples for which error increases or
      decreases with respect to the fraction of added samples used of
      the different datasets.}
    \label{fig:inc_dec_large}
\end{figure}

\begin{figure}
    \centering
    \includegraphics[width=0.9\textwidth]{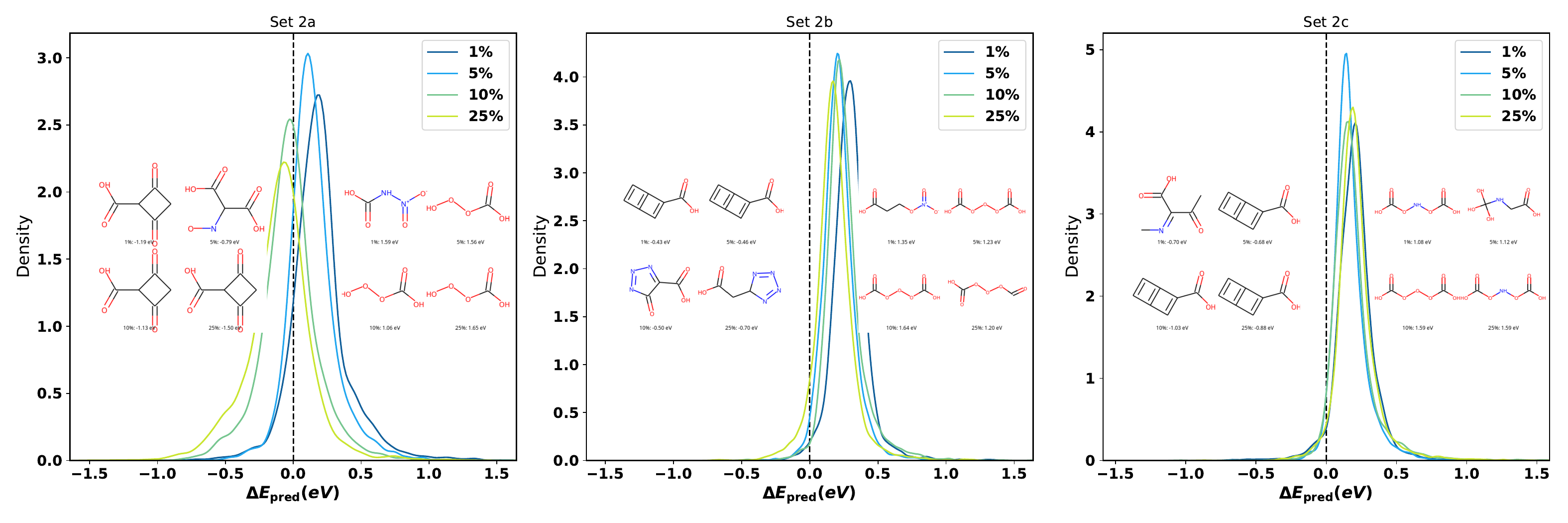}
    \caption{Distribution of change in predicted energy to the
      percentages of samples added ($\Delta E = E_{0}- E_{i}$, here
      $i\in \{1,5,10,25\}$\%) for the datasets of \textit{Set2}. Each panel
      shows the molecule with the largest decrease or increase in
      $\Delta E$ for the different percentages.}
    \label{sifig:dist_err_set2_large}
\end{figure}

\clearpage

%% file: submit.bbl
\providecommand{\latin}[1]{#1}
\makeatletter
\providecommand{\doi}
  {\begingroup\let\do\@makeother\dospecials
  \catcode`\{=1 \catcode`\}=2 \doi@aux}
\providecommand{\doi@aux}[1]{\endgroup\texttt{#1}}
\makeatother
\providecommand*\mcitethebibliography{\thebibliography}
\csname @ifundefined\endcsname{endmcitethebibliography}
  {\let\endmcitethebibliography\endthebibliography}{}
\begin{mcitethebibliography}{58}
\providecommand*\natexlab[1]{#1}
\providecommand*\mciteSetBstSublistMode[1]{}
\providecommand*\mciteSetBstMaxWidthForm[2]{}
\providecommand*\mciteBstWouldAddEndPuncttrue
  {\def\EndOfBibitem{\unskip.}}
\providecommand*\mciteBstWouldAddEndPunctfalse
  {\let\EndOfBibitem\relax}
\providecommand*\mciteSetBstMidEndSepPunct[3]{}
\providecommand*\mciteSetBstSublistLabelBeginEnd[3]{}
\providecommand*\EndOfBibitem{}
\mciteSetBstSublistMode{f}
\mciteSetBstMaxWidthForm{subitem}{(\alph{mcitesubitemcount})}
\mciteSetBstSublistLabelBeginEnd
  {\mcitemaxwidthsubitemform\space}
  {\relax}
  {\relax}

\bibitem[Kirkpatrick and Ellis(2004)Kirkpatrick, and
  Ellis]{kirkpatrick2004chemical}
Kirkpatrick,~P.; Ellis,~C. Chemical space. \emph{Nature} \textbf{2004},
  \emph{432}, 823--824\relax
\mciteBstWouldAddEndPuncttrue
\mciteSetBstMidEndSepPunct{\mcitedefaultmidpunct}
{\mcitedefaultendpunct}{\mcitedefaultseppunct}\relax
\EndOfBibitem
\bibitem[Von~Lilienfeld(2013)]{von2013first}
Von~Lilienfeld,~O.~A. First principles view on chemical compound space: Gaining
  rigorous atomistic control of molecular properties. \emph{Int. J. Quantum
  Chem.} \textbf{2013}, \emph{113}, 1676--1689\relax
\mciteBstWouldAddEndPuncttrue
\mciteSetBstMidEndSepPunct{\mcitedefaultmidpunct}
{\mcitedefaultendpunct}{\mcitedefaultseppunct}\relax
\EndOfBibitem
\bibitem[von Lilienfeld \latin{et~al.}(2020)von Lilienfeld, M{\"u}ller, and
  Tkatchenko]{von2020exploring}
von Lilienfeld,~O.~A.; M{\"u}ller,~K.-R.; Tkatchenko,~A. Exploring chemical
  compound space with quantum-based machine learning. \emph{Nat. Rev. Chem.}
  \textbf{2020}, \emph{4}, 347--358\relax
\mciteBstWouldAddEndPuncttrue
\mciteSetBstMidEndSepPunct{\mcitedefaultmidpunct}
{\mcitedefaultendpunct}{\mcitedefaultseppunct}\relax
\EndOfBibitem
\bibitem[Huang and von Lilienfeld(2021)Huang, and von Lilienfeld]{huang:2021}
Huang,~B.; von Lilienfeld,~O.~A. Ab initio machine learning in chemical
  compound space. \emph{Chem. Rev.} \textbf{2021}, \emph{121},
  10001--10036\relax
\mciteBstWouldAddEndPuncttrue
\mciteSetBstMidEndSepPunct{\mcitedefaultmidpunct}
{\mcitedefaultendpunct}{\mcitedefaultseppunct}\relax
\EndOfBibitem
\bibitem[Coley(2021)]{coley2021defining}
Coley,~C.~W. Defining and exploring chemical spaces. \emph{Trends Chem.}
  \textbf{2021}, \emph{3}, 133--145\relax
\mciteBstWouldAddEndPuncttrue
\mciteSetBstMidEndSepPunct{\mcitedefaultmidpunct}
{\mcitedefaultendpunct}{\mcitedefaultseppunct}\relax
\EndOfBibitem
\bibitem[Medina-Franco \latin{et~al.}(2022)Medina-Franco,
  Ch{\'a}vez-Hern{\'a}ndez, L{\'o}pez-L{\'o}pez, and
  Sald{\'\i}var-Gonz{\'a}lez]{medina2022chemical}
Medina-Franco,~J.~L.; Ch{\'a}vez-Hern{\'a}ndez,~A.~L.; L{\'o}pez-L{\'o}pez,~E.;
  Sald{\'\i}var-Gonz{\'a}lez,~F.~I. Chemical multiverse: an expanded view of
  chemical space. \emph{Mol. Inform.} \textbf{2022}, \emph{41}, 2200116\relax
\mciteBstWouldAddEndPuncttrue
\mciteSetBstMidEndSepPunct{\mcitedefaultmidpunct}
{\mcitedefaultendpunct}{\mcitedefaultseppunct}\relax
\EndOfBibitem
\bibitem[Restrepo(2022)]{restrepo2022chemical}
Restrepo,~G. Chemical space: limits, evolution and modelling of an object
  bigger than our universal library. \emph{Digit. Discov.} \textbf{2022},
  \emph{1}, 568--585\relax
\mciteBstWouldAddEndPuncttrue
\mciteSetBstMidEndSepPunct{\mcitedefaultmidpunct}
{\mcitedefaultendpunct}{\mcitedefaultseppunct}\relax
\EndOfBibitem
\bibitem[Gorse(2006)]{gorse2006diversity}
Gorse,~A.-D. Diversity in medicinal chemistry space. \emph{Curr Top Med Chem}
  \textbf{2006}, \emph{6}, 3--18\relax
\mciteBstWouldAddEndPuncttrue
\mciteSetBstMidEndSepPunct{\mcitedefaultmidpunct}
{\mcitedefaultendpunct}{\mcitedefaultseppunct}\relax
\EndOfBibitem
\bibitem[Sandonas \latin{et~al.}(2023)Sandonas, Hoja, Ernst,
  V{\'a}zquez-Mayagoitia, DiStasio, and Tkatchenko]{sandonas2023freedom}
Sandonas,~L.~M.; Hoja,~J.; Ernst,~B.~G.; V{\'a}zquez-Mayagoitia,~{\'A}.;
  DiStasio,~R.~A.; Tkatchenko,~A. “Freedom of design” in chemical compound
  space: towards rational in silico design of molecules with targeted
  quantum-mechanical properties. \emph{Chem. Sci.} \textbf{2023}, \emph{14},
  10702--10717\relax
\mciteBstWouldAddEndPuncttrue
\mciteSetBstMidEndSepPunct{\mcitedefaultmidpunct}
{\mcitedefaultendpunct}{\mcitedefaultseppunct}\relax
\EndOfBibitem
\bibitem[Fallani \latin{et~al.}(2023)Fallani, Medrano~Sandonas, and
  Tkatchenko]{fallani2023enabling}
Fallani,~A.; Medrano~Sandonas,~L.; Tkatchenko,~A. Enabling Inverse Design in
  Chemical Compound Space: Mapping Quantum Properties to Structures for Small
  Organic Molecules. \emph{arXiv e-prints} \textbf{2023}, arXiv--2309\relax
\mciteBstWouldAddEndPuncttrue
\mciteSetBstMidEndSepPunct{\mcitedefaultmidpunct}
{\mcitedefaultendpunct}{\mcitedefaultseppunct}\relax
\EndOfBibitem
\bibitem[K{\"a}ser \latin{et~al.}(2023)K{\"a}ser, Vazquez-Salazar, Meuwly, and
  T{\"o}pfer]{MM.rev:2023}
K{\"a}ser,~S.; Vazquez-Salazar,~L.~I.; Meuwly,~M.; T{\"o}pfer,~K. Neural
  network potentials for chemistry: concepts, applications and prospects.
  \emph{Digit. Discov.} \textbf{2023}, \emph{2}, 28--58\relax
\mciteBstWouldAddEndPuncttrue
\mciteSetBstMidEndSepPunct{\mcitedefaultmidpunct}
{\mcitedefaultendpunct}{\mcitedefaultseppunct}\relax
\EndOfBibitem
\bibitem[Kulichenko \latin{et~al.}(2024)Kulichenko, Nebgen, Lubbers, Smith,
  Barros, Allen, Habib, Shinkle, Fedik, Li, \latin{et~al.}
  others]{kulichenko2024data}
Kulichenko,~M.; Nebgen,~B.; Lubbers,~N.; Smith,~J.~S.; Barros,~K.;
  Allen,~A.~E.; Habib,~A.; Shinkle,~E.; Fedik,~N.; Li,~Y.~W., \latin{et~al.}
  Data generation for machine learning interatomic potentials and beyond.
  \emph{Chem. Rev.} \textbf{2024}, \emph{124}, 13681--13714\relax
\mciteBstWouldAddEndPuncttrue
\mciteSetBstMidEndSepPunct{\mcitedefaultmidpunct}
{\mcitedefaultendpunct}{\mcitedefaultseppunct}\relax
\EndOfBibitem
\bibitem[Huang and von Lilienfeld(2020)Huang, and von
  Lilienfeld]{huang2020quantum}
Huang,~B.; von Lilienfeld,~O.~A. Quantum machine learning using
  atom-in-molecule-based fragments selected on the fly. \emph{Nat. Chem.}
  \textbf{2020}, 1--7\relax
\mciteBstWouldAddEndPuncttrue
\mciteSetBstMidEndSepPunct{\mcitedefaultmidpunct}
{\mcitedefaultendpunct}{\mcitedefaultseppunct}\relax
\EndOfBibitem
\bibitem[Shaik \latin{et~al.}(2013)Shaik, Rzepa, and Hoffmann]{shaik2013one}
Shaik,~S.; Rzepa,~H.~S.; Hoffmann,~R. One molecule, two atoms, three views,
  four bonds? \emph{Angew. Chem. Int. Ed.} \textbf{2013}, \emph{52},
  3020--3033\relax
\mciteBstWouldAddEndPuncttrue
\mciteSetBstMidEndSepPunct{\mcitedefaultmidpunct}
{\mcitedefaultendpunct}{\mcitedefaultseppunct}\relax
\EndOfBibitem
\bibitem[Vazquez-Salazar \latin{et~al.}(2021)Vazquez-Salazar, Boittier, Unke,
  and Meuwly]{vazquezsalazar2021}
Vazquez-Salazar,~L.~I.; Boittier,~E.~D.; Unke,~O.~T.; Meuwly,~M. Impact of the
  Characteristics of Quantum Chemical Databases on Machine Learning Prediction
  of Tautomerization Energies. \emph{J. Chem. Theory Comput.} \textbf{2021},
  \emph{17}, 4769--4785\relax
\mciteBstWouldAddEndPuncttrue
\mciteSetBstMidEndSepPunct{\mcitedefaultmidpunct}
{\mcitedefaultendpunct}{\mcitedefaultseppunct}\relax
\EndOfBibitem
\bibitem[Hoja \latin{et~al.}(2021)Hoja, Medrano~Sandonas, Ernst,
  Vazquez-Mayagoitia, DiStasio~Jr, and Tkatchenko]{hoja2021qm7}
Hoja,~J.; Medrano~Sandonas,~L.; Ernst,~B.~G.; Vazquez-Mayagoitia,~A.;
  DiStasio~Jr,~R.~A.; Tkatchenko,~A. QM7-X, a comprehensive dataset of
  quantum-mechanical properties spanning the chemical space of small organic
  molecules. \emph{Sci. Data} \textbf{2021}, \emph{8}, 43\relax
\mciteBstWouldAddEndPuncttrue
\mciteSetBstMidEndSepPunct{\mcitedefaultmidpunct}
{\mcitedefaultendpunct}{\mcitedefaultseppunct}\relax
\EndOfBibitem
\bibitem[Smith \latin{et~al.}(2017)Smith, Isayev, and Roitberg]{smith2017ani}
Smith,~J.~S.; Isayev,~O.; Roitberg,~A.~E. ANI-1, A data set of 20 million
  calculated off-equilibrium conformations for organic molecules. \emph{Sci.
  Data} \textbf{2017}, \emph{4}, 3192--3203\relax
\mciteBstWouldAddEndPuncttrue
\mciteSetBstMidEndSepPunct{\mcitedefaultmidpunct}
{\mcitedefaultendpunct}{\mcitedefaultseppunct}\relax
\EndOfBibitem
\bibitem[Quinonero-Candela \latin{et~al.}(2008)Quinonero-Candela, Sugiyama,
  Schwaighofer, and Lawrence]{quinonero2008dataset}
Quinonero-Candela,~J.; Sugiyama,~M.; Schwaighofer,~A.; Lawrence,~N.~D.
  \emph{Dataset shift in machine learning}; Mit Press, 2008\relax
\mciteBstWouldAddEndPuncttrue
\mciteSetBstMidEndSepPunct{\mcitedefaultmidpunct}
{\mcitedefaultendpunct}{\mcitedefaultseppunct}\relax
\EndOfBibitem
\bibitem[Banerjee \latin{et~al.}(2018)Banerjee, Dehnbostel, and
  Preissner]{banerjee2018prediction}
Banerjee,~P.; Dehnbostel,~F.~O.; Preissner,~R. Prediction is a balancing act:
  Importance of sampling methods to balance sensitivity and specificity of
  predictive models based on imbalanced chemical data sets. \emph{Front. Chem.}
  \textbf{2018}, 362\relax
\mciteBstWouldAddEndPuncttrue
\mciteSetBstMidEndSepPunct{\mcitedefaultmidpunct}
{\mcitedefaultendpunct}{\mcitedefaultseppunct}\relax
\EndOfBibitem
\bibitem[Hemmerich \latin{et~al.}(2020)Hemmerich, Asilar, and
  Ecker]{hemmerich2020cover}
Hemmerich,~J.; Asilar,~E.; Ecker,~G.~F. COVER: conformational oversampling as
  data augmentation for molecules. \emph{J. Cheminf.} \textbf{2020}, \emph{12},
  18\relax
\mciteBstWouldAddEndPuncttrue
\mciteSetBstMidEndSepPunct{\mcitedefaultmidpunct}
{\mcitedefaultendpunct}{\mcitedefaultseppunct}\relax
\EndOfBibitem
\bibitem[Korkmaz(2020)]{korkmaz2020deep}
Korkmaz,~S. Deep learning-based imbalanced data classification for drug
  discovery. \emph{J. Chem. Inf. Model.} \textbf{2020}, \emph{60},
  4180--4190\relax
\mciteBstWouldAddEndPuncttrue
\mciteSetBstMidEndSepPunct{\mcitedefaultmidpunct}
{\mcitedefaultendpunct}{\mcitedefaultseppunct}\relax
\EndOfBibitem
\bibitem[Shenoy \latin{et~al.}(2023)Shenoy, Tossou, Noutahi, Mary, Beaini, and
  Ding]{shenoy2023role}
Shenoy,~N.; Tossou,~P.; Noutahi,~E.; Mary,~H.; Beaini,~D.; Ding,~J. Role of
  Structural and Conformational Diversity for Machine Learning Potentials.
  NeurIPS 2023 AI for Science Workshop. 2023\relax
\mciteBstWouldAddEndPuncttrue
\mciteSetBstMidEndSepPunct{\mcitedefaultmidpunct}
{\mcitedefaultendpunct}{\mcitedefaultseppunct}\relax
\EndOfBibitem
\bibitem[Hamakawa and Miyao(2025)Hamakawa, and
  Miyao]{hamakawa2025understanding}
Hamakawa,~Y.; Miyao,~T. Understanding Conformation Importance in Data-Driven
  Property Prediction Models. \emph{J. Chem. Inf. Model.} \textbf{2025},
  \emph{\it In Press}\relax
\mciteBstWouldAddEndPuncttrue
\mciteSetBstMidEndSepPunct{\mcitedefaultmidpunct}
{\mcitedefaultendpunct}{\mcitedefaultseppunct}\relax
\EndOfBibitem
\bibitem[Unke and Meuwly(2019)Unke, and Meuwly]{unke2019physnet}
Unke,~O.~T.; Meuwly,~M. PhysNet: a neural network for predicting energies,
  forces, dipole moments, and partial charges. \emph{J. Chem. Theory Comput.}
  \textbf{2019}, \emph{15}, 3678--3693\relax
\mciteBstWouldAddEndPuncttrue
\mciteSetBstMidEndSepPunct{\mcitedefaultmidpunct}
{\mcitedefaultendpunct}{\mcitedefaultseppunct}\relax
\EndOfBibitem
\bibitem[Reiser \latin{et~al.}(2022)Reiser, Neubert, Eberhard, Torresi, Zhou,
  Shao, Metni, van Hoesel, Schopmans, Sommer, \latin{et~al.}
  others]{reiser2022graph}
Reiser,~P.; Neubert,~M.; Eberhard,~A.; Torresi,~L.; Zhou,~C.; Shao,~C.;
  Metni,~H.; van Hoesel,~C.; Schopmans,~H.; Sommer,~T., \latin{et~al.}  Graph
  neural networks for materials science and chemistry. \emph{Commun. Mater.}
  \textbf{2022}, \emph{3}, 93\relax
\mciteBstWouldAddEndPuncttrue
\mciteSetBstMidEndSepPunct{\mcitedefaultmidpunct}
{\mcitedefaultendpunct}{\mcitedefaultseppunct}\relax
\EndOfBibitem
\bibitem[Corso \latin{et~al.}(2024)Corso, Stark, Jegelka, Jaakkola, and
  Barzilay]{corso2024graph}
Corso,~G.; Stark,~H.; Jegelka,~S.; Jaakkola,~T.; Barzilay,~R. Graph neural
  networks. \emph{Nat. Rev. Methods Primers} \textbf{2024}, \emph{4}, 17\relax
\mciteBstWouldAddEndPuncttrue
\mciteSetBstMidEndSepPunct{\mcitedefaultmidpunct}
{\mcitedefaultendpunct}{\mcitedefaultseppunct}\relax
\EndOfBibitem
\bibitem[Schuett \latin{et~al.}(2018)Schuett, Sauceda, Kindermans, Tkatchenko,
  and Mueller]{schnet:2018}
Schuett,~K.~T.; Sauceda,~H.~E.; Kindermans,~P.~J.; Tkatchenko,~A.;
  Mueller,~K.~R. Schnet - a Deep Learning Architecture for Molecules and
  Materials. \emph{J. Chem. Phys.} \textbf{2018}, \emph{148}, 241722\relax
\mciteBstWouldAddEndPuncttrue
\mciteSetBstMidEndSepPunct{\mcitedefaultmidpunct}
{\mcitedefaultendpunct}{\mcitedefaultseppunct}\relax
\EndOfBibitem
\bibitem[Sch{\"u}tt \latin{et~al.}(2021)Sch{\"u}tt, Unke, and
  Gastegger]{schuett2021painn}
Sch{\"u}tt,~K.; Unke,~O.; Gastegger,~M. Equivariant message passing for the
  prediction of tensorial properties and molecular spectra. International
  Conference on Machine Learning. 2021; pp 9377--9388\relax
\mciteBstWouldAddEndPuncttrue
\mciteSetBstMidEndSepPunct{\mcitedefaultmidpunct}
{\mcitedefaultendpunct}{\mcitedefaultseppunct}\relax
\EndOfBibitem
\bibitem[Batzner \latin{et~al.}(2022)Batzner, Musaelian, Sun, Geiger, Mailoa,
  Kornbluth, Molinari, Smidt, and Kozinsky]{batzner2022nequip}
Batzner,~S.; Musaelian,~A.; Sun,~L.; Geiger,~M.; Mailoa,~J.~P.; Kornbluth,~M.;
  Molinari,~N.; Smidt,~T.~E.; Kozinsky,~B. E(3)-equivariant graph neural
  networks for data-efficient and accurate interatomic potentials. \emph{Nat.
  Comm.} \textbf{2022}, \emph{13}, 1--11\relax
\mciteBstWouldAddEndPuncttrue
\mciteSetBstMidEndSepPunct{\mcitedefaultmidpunct}
{\mcitedefaultendpunct}{\mcitedefaultseppunct}\relax
\EndOfBibitem
\bibitem[Batatia \latin{et~al.}(2022)Batatia, Kovacs, Simm, Ortner, and
  Cs{\'a}nyi]{batatia2022mace}
Batatia,~I.; Kovacs,~D.~P.; Simm,~G.; Ortner,~C.; Cs{\'a}nyi,~G. MACE: Higher
  order equivariant message passing neural networks for fast and accurate force
  fields. \emph{Advances in Neural Information Processing Systems}
  \textbf{2022}, \emph{35}, 11423--11436\relax
\mciteBstWouldAddEndPuncttrue
\mciteSetBstMidEndSepPunct{\mcitedefaultmidpunct}
{\mcitedefaultendpunct}{\mcitedefaultseppunct}\relax
\EndOfBibitem
\bibitem[Vazquez-Salazar \latin{et~al.}(2022)Vazquez-Salazar, Boittier, and
  Meuwly]{vazquez2022uncertainty}
Vazquez-Salazar,~L.~I.; Boittier,~E.~D.; Meuwly,~M. Uncertainty quantification
  for predictions of atomistic neural networks. \emph{Chem. Sci.}
  \textbf{2022}, \emph{13}, 13068--13084\relax
\mciteBstWouldAddEndPuncttrue
\mciteSetBstMidEndSepPunct{\mcitedefaultmidpunct}
{\mcitedefaultendpunct}{\mcitedefaultseppunct}\relax
\EndOfBibitem
\bibitem[Soleimany \latin{et~al.}(2021)Soleimany, Amini, Goldman, Rus, Bhatia,
  and Coley]{soleimany2021evidential}
Soleimany,~A.~P.; Amini,~A.; Goldman,~S.; Rus,~D.; Bhatia,~S.~N.; Coley,~C.~W.
  Evidential deep learning for guided molecular property prediction and
  discovery. \emph{ACS Cent. Sci.} \textbf{2021}, \emph{7}, 1356--1367\relax
\mciteBstWouldAddEndPuncttrue
\mciteSetBstMidEndSepPunct{\mcitedefaultmidpunct}
{\mcitedefaultendpunct}{\mcitedefaultseppunct}\relax
\EndOfBibitem
\bibitem[Amini \latin{et~al.}(2020)Amini, Schwarting, Soleimany, and
  Rus]{amini2020}
Amini,~A.; Schwarting,~W.; Soleimany,~A.; Rus,~D. Deep Evidential Regression.
  Advances in Neural Information Processing Systems. 2020; pp
  14927--14937\relax
\mciteBstWouldAddEndPuncttrue
\mciteSetBstMidEndSepPunct{\mcitedefaultmidpunct}
{\mcitedefaultendpunct}{\mcitedefaultseppunct}\relax
\EndOfBibitem
\bibitem[Kingma and Ba(2014)Kingma, and Ba]{kingma2014adam}
Kingma,~D.~P.; Ba,~J. Adam: A method for stochastic optimization. \emph{arXiv
  preprint arXiv:1412.6980} \textbf{2014}, \relax
\mciteBstWouldAddEndPunctfalse
\mciteSetBstMidEndSepPunct{\mcitedefaultmidpunct}
{}{\mcitedefaultseppunct}\relax
\EndOfBibitem
\bibitem[Ramakrishnan \latin{et~al.}(2014)Ramakrishnan, Dral, Rupp, and
  Von~Lilienfeld]{ramakrishnan2014quantum}
Ramakrishnan,~R.; Dral,~P.~O.; Rupp,~M.; Von~Lilienfeld,~O.~A. Quantum
  chemistry structures and properties of 134 kilo molecules. \emph{Sci. Data}
  \textbf{2014}, \emph{1}, 140022\relax
\mciteBstWouldAddEndPuncttrue
\mciteSetBstMidEndSepPunct{\mcitedefaultmidpunct}
{\mcitedefaultendpunct}{\mcitedefaultseppunct}\relax
\EndOfBibitem
\bibitem[Landrum \latin{et~al.}(2013)Landrum, \latin{et~al.}
  others]{landrum2013rdkit}
Landrum,~G., \latin{et~al.}  RDKit: A software suite for cheminformatics,
  computational chemistry, and predictive modeling. 2013\relax
\mciteBstWouldAddEndPuncttrue
\mciteSetBstMidEndSepPunct{\mcitedefaultmidpunct}
{\mcitedefaultendpunct}{\mcitedefaultseppunct}\relax
\EndOfBibitem
\bibitem[Klein(2017)]{Klein}
Klein,~D. \emph{Organic Chemistry}, 3rd ed.; John Wiley and Sons, 2017\relax
\mciteBstWouldAddEndPuncttrue
\mciteSetBstMidEndSepPunct{\mcitedefaultmidpunct}
{\mcitedefaultendpunct}{\mcitedefaultseppunct}\relax
\EndOfBibitem
\bibitem[Glavatskikh \latin{et~al.}(2019)Glavatskikh, Leguy, Hunault, Cauchy,
  and Da~Mota]{glavatskikh2019dataset}
Glavatskikh,~M.; Leguy,~J.; Hunault,~G.; Cauchy,~T.; Da~Mota,~B. Dataset’s
  chemical diversity limits the generalizability of machine learning
  predictions. \emph{J. Cheminf.} \textbf{2019}, \emph{11}, 69\relax
\mciteBstWouldAddEndPuncttrue
\mciteSetBstMidEndSepPunct{\mcitedefaultmidpunct}
{\mcitedefaultendpunct}{\mcitedefaultseppunct}\relax
\EndOfBibitem
\bibitem[Frisch \latin{et~al.}(2016)Frisch, Trucks, Schlegel, Scuseria, Robb,
  Cheeseman, Scalmani, Barone, Petersson, Nakatsuji, \latin{et~al.}
  others]{gaussian16}
Frisch,~M.; Trucks,~G.; Schlegel,~H.; Scuseria,~G.; Robb,~M.; Cheeseman,~J.;
  Scalmani,~G.; Barone,~V.; Petersson,~G.; Nakatsuji,~H., \latin{et~al.}
  Gaussian 16. 2016\relax
\mciteBstWouldAddEndPuncttrue
\mciteSetBstMidEndSepPunct{\mcitedefaultmidpunct}
{\mcitedefaultendpunct}{\mcitedefaultseppunct}\relax
\EndOfBibitem
\bibitem[Diwekar and David(2015)Diwekar, and David]{diwekar2015probability}
Diwekar,~U.; David,~A. \emph{BONUS Algorithm for Large Scale Stochastic
  Nonlinear Programming Problems}; Springer, 2015; pp 27--34\relax
\mciteBstWouldAddEndPuncttrue
\mciteSetBstMidEndSepPunct{\mcitedefaultmidpunct}
{\mcitedefaultendpunct}{\mcitedefaultseppunct}\relax
\EndOfBibitem
\bibitem[Cover and Thomas(2006)Cover, and Thomas]{cover2006elements}
Cover,~T.~M.; Thomas,~J.~A. \emph{Elements of Information Theory}; Wiley Series
  in Telecommunications and Signal Processing; John Wiley \& Sons, 2006\relax
\mciteBstWouldAddEndPuncttrue
\mciteSetBstMidEndSepPunct{\mcitedefaultmidpunct}
{\mcitedefaultendpunct}{\mcitedefaultseppunct}\relax
\EndOfBibitem
\bibitem[Lin(1991)]{lin1991divergence}
Lin,~J. Divergence measures based on the Shannon entropy. \emph{IEEE Trans.
  Inf. Theory} \textbf{1991}, \emph{37}, 145--151\relax
\mciteBstWouldAddEndPuncttrue
\mciteSetBstMidEndSepPunct{\mcitedefaultmidpunct}
{\mcitedefaultendpunct}{\mcitedefaultseppunct}\relax
\EndOfBibitem
\bibitem[Nielsen(2019)]{nielsen2019jensen}
Nielsen,~F. On the Jensen--Shannon symmetrization of distances relying on
  abstract means. \emph{Entropy} \textbf{2019}, \emph{21}, 485\relax
\mciteBstWouldAddEndPuncttrue
\mciteSetBstMidEndSepPunct{\mcitedefaultmidpunct}
{\mcitedefaultendpunct}{\mcitedefaultseppunct}\relax
\EndOfBibitem
\bibitem[Virtanen \latin{et~al.}(2020)Virtanen, Gommers, Oliphant, Haberland,
  Reddy, Cournapeau, Burovski, Peterson, Weckesser, Bright, \latin{et~al.}
  others]{virtanen2020scipy}
Virtanen,~P.; Gommers,~R.; Oliphant,~T.~E.; Haberland,~M.; Reddy,~T.;
  Cournapeau,~D.; Burovski,~E.; Peterson,~P.; Weckesser,~W.; Bright,~J.,
  \latin{et~al.}  SciPy 1.0: fundamental algorithms for scientific computing in
  Python. \emph{Nat. Meth.} \textbf{2020}, \emph{17}, 261--272\relax
\mciteBstWouldAddEndPuncttrue
\mciteSetBstMidEndSepPunct{\mcitedefaultmidpunct}
{\mcitedefaultendpunct}{\mcitedefaultseppunct}\relax
\EndOfBibitem
\bibitem[Villani(2009)]{villani2009wasserstein}
Villani,~C. \emph{Optimal transport: old and new}; Springer, 2009; pp
  93--111\relax
\mciteBstWouldAddEndPuncttrue
\mciteSetBstMidEndSepPunct{\mcitedefaultmidpunct}
{\mcitedefaultendpunct}{\mcitedefaultseppunct}\relax
\EndOfBibitem
\bibitem[Weng(2019)]{weng2019gan}
Weng,~L. From gan to wgan. \emph{arXiv preprint arXiv:1904.08994}
  \textbf{2019}, \relax
\mciteBstWouldAddEndPunctfalse
\mciteSetBstMidEndSepPunct{\mcitedefaultmidpunct}
{}{\mcitedefaultseppunct}\relax
\EndOfBibitem
\bibitem[Panaretos and Zemel(2019)Panaretos, and
  Zemel]{panaretos2019statistical}
Panaretos,~V.~M.; Zemel,~Y. Statistical aspects of Wasserstein distances.
  \emph{Annu. Rev. Stat. Appl.} \textbf{2019}, \emph{6}, 405--431\relax
\mciteBstWouldAddEndPuncttrue
\mciteSetBstMidEndSepPunct{\mcitedefaultmidpunct}
{\mcitedefaultendpunct}{\mcitedefaultseppunct}\relax
\EndOfBibitem
\bibitem[Ramdas \latin{et~al.}(2017)Ramdas, Garc{\'\i}a~Trillos, and
  Cuturi]{ramdas2017wasserstein}
Ramdas,~A.; Garc{\'\i}a~Trillos,~N.; Cuturi,~M. On wasserstein two-sample
  testing and related families of nonparametric tests. \emph{Entropy}
  \textbf{2017}, \emph{19}, 47\relax
\mciteBstWouldAddEndPuncttrue
\mciteSetBstMidEndSepPunct{\mcitedefaultmidpunct}
{\mcitedefaultendpunct}{\mcitedefaultseppunct}\relax
\EndOfBibitem
\bibitem[Vazquez-Salazar \latin{et~al.}(2025)Vazquez-Salazar, K{\"a}ser, and
  Meuwly]{vazquez2025outlier}
Vazquez-Salazar,~L.~I.; K{\"a}ser,~S.; Meuwly,~M. Outlier-detection for
  reactive machine learned potential energy surfaces. \emph{npj Comp. Mat.}
  \textbf{2025}, \emph{11}, 33\relax
\mciteBstWouldAddEndPuncttrue
\mciteSetBstMidEndSepPunct{\mcitedefaultmidpunct}
{\mcitedefaultendpunct}{\mcitedefaultseppunct}\relax
\EndOfBibitem
\bibitem[Torrie and Valleau(1977)Torrie, and Valleau]{torrie1977nonphysical}
Torrie,~G.~M.; Valleau,~J.~P. Nonphysical sampling distributions in Monte Carlo
  free-energy estimation: Umbrella sampling. \emph{J Comp. Phys.}
  \textbf{1977}, \emph{23}, 187--199\relax
\mciteBstWouldAddEndPuncttrue
\mciteSetBstMidEndSepPunct{\mcitedefaultmidpunct}
{\mcitedefaultendpunct}{\mcitedefaultseppunct}\relax
\EndOfBibitem
\bibitem[K{\"a}stner(2011)]{kastner2011umbrella}
K{\"a}stner,~J. Umbrella sampling. \emph{WIREs Comput. Mol. Sci} \textbf{2011},
  \emph{1}, 932--942\relax
\mciteBstWouldAddEndPuncttrue
\mciteSetBstMidEndSepPunct{\mcitedefaultmidpunct}
{\mcitedefaultendpunct}{\mcitedefaultseppunct}\relax
\EndOfBibitem
\bibitem[Dukler \latin{et~al.}(2019)Dukler, Li, Lin, and
  Mont{\'u}far]{dukler2019wasserstein}
Dukler,~Y.; Li,~W.; Lin,~A.; Mont{\'u}far,~G. Wasserstein of Wasserstein loss
  for learning generative models. International conference on machine learning.
  2019; pp 1716--1725\relax
\mciteBstWouldAddEndPuncttrue
\mciteSetBstMidEndSepPunct{\mcitedefaultmidpunct}
{\mcitedefaultendpunct}{\mcitedefaultseppunct}\relax
\EndOfBibitem
\bibitem[Kingma and Welling(2019)Kingma, and Welling]{kingma2019introduction}
Kingma,~D.~P.; Welling,~M. An introduction to variational autoencoders.
  \emph{Foundations and Trends in Machine Learning} \textbf{2019}, \emph{12},
  307--392\relax
\mciteBstWouldAddEndPuncttrue
\mciteSetBstMidEndSepPunct{\mcitedefaultmidpunct}
{\mcitedefaultendpunct}{\mcitedefaultseppunct}\relax
\EndOfBibitem
\bibitem[Schwalbe-Koda and G{\'o}mez-Bombarelli(2020)Schwalbe-Koda, and
  G{\'o}mez-Bombarelli]{Schwalbe-Koda2020}
Schwalbe-Koda,~D.; G{\'o}mez-Bombarelli,~R. \emph{Machine Learning Meets
  Quantum Physics}; Springer, 2020; pp 445--467\relax
\mciteBstWouldAddEndPuncttrue
\mciteSetBstMidEndSepPunct{\mcitedefaultmidpunct}
{\mcitedefaultendpunct}{\mcitedefaultseppunct}\relax
\EndOfBibitem
\bibitem[Esders \latin{et~al.}(2025)Esders, Schnake, Lederer, Kabylda,
  Montavon, Tkatchenko, and M\"uller]{esders2025analyzing}
Esders,~M.; Schnake,~T.; Lederer,~J.; Kabylda,~A.; Montavon,~G.;
  Tkatchenko,~A.; M\"uller,~K.-R. Analyzing Atomic Interactions in Molecules as
  Learned by Neural Networks. \emph{J. Chem. Theory Comput.} \textbf{2025},
  \emph{21}, 714--729\relax
\mciteBstWouldAddEndPuncttrue
\mciteSetBstMidEndSepPunct{\mcitedefaultmidpunct}
{\mcitedefaultendpunct}{\mcitedefaultseppunct}\relax
\EndOfBibitem
\bibitem[Gardner \latin{et~al.}(2023)Gardner, Beaulieu, and
  Deringer]{gardner2023synthetic}
Gardner,~J.~L.; Beaulieu,~Z.~F.; Deringer,~V.~L. Synthetic data enable
  experiments in atomistic machine learning. \emph{Digit. Discov.}
  \textbf{2023}, \emph{2}, 651--662\relax
\mciteBstWouldAddEndPuncttrue
\mciteSetBstMidEndSepPunct{\mcitedefaultmidpunct}
{\mcitedefaultendpunct}{\mcitedefaultseppunct}\relax
\EndOfBibitem
\bibitem[Reizinger \latin{et~al.}(2024)Reizinger, Bizeul, Juhos, Vogt,
  Balestriero, Brendel, and Klindt]{reizinger2024cross}
Reizinger,~P.; Bizeul,~A.; Juhos,~A.; Vogt,~J.~E.; Balestriero,~R.;
  Brendel,~W.; Klindt,~D. Cross-Entropy Is All You Need To Invert the Data
  Generating Process. \emph{arXiv preprint arXiv:2410.21869} \textbf{2024},
  \relax
\mciteBstWouldAddEndPunctfalse
\mciteSetBstMidEndSepPunct{\mcitedefaultmidpunct}
{}{\mcitedefaultseppunct}\relax
\EndOfBibitem
\end{mcitethebibliography}
